\newif\ifshort
\shortfalse 

\documentclass{ecai}
\usepackage[utf8]{inputenc}
\usepackage{amsmath,amssymb,graphicx,amsthm,dsfont,mathtools}
\usepackage{color,soul}
\usepackage{xspace}
\usepackage{enumerate}
\usepackage{mathtools}
\usepackage{bbm}
\usepackage{multicol}
\usepackage{booktabs}
\usepackage{paralist}
\usepackage{graphicx}
\usepackage{latexsym}
\usepackage{subcaption}
\usepackage[noend,ruled,linesnumbered]{algorithm2e} 
\usepackage{xifthen}

\usepackage[textsize=tiny,textwidth=1.5cm,linecolor=green!70!black, backgroundcolor=green!10, bordercolor=black,disable=true]{todonotes}
\newcommand{\todoH}[1]{\todo[linecolor=yellow!70!black, backgroundcolor=yellow!10]{H: #1}}
\newcommand{\todoIH}[1]{\todo[inline,linecolor=yellow!70!black, backgroundcolor=yellow!10]{H: #1}}

\newcommand{\todoIC}[1]{\todo[inline,linecolor=blue!70!black, backgroundcolor=blue!10]{C:#1}}
\usepackage{mdframed}

\usepackage[colorlinks,citecolor=blue!80!black,linkcolor=red!60!black,pagebackref]{hyperref} 
\renewcommand*{\backref}[1]{}
\renewcommand*{\backrefalt}[4]{%
	\ifcase #1%
	\marginpar{\tiny no cite}
	\or
	$\rightarrow$~p.~#2.%
	\else
	$\rightarrow$~pp.~#2.%
	\fi
}

\usepackage{tikz}
\usetikzlibrary{decorations,arrows,petri,topaths,backgrounds,shapes,positioning,fit,calc,decorations.pathreplacing,patterns,intersections,decorations.pathmorphing,matrix,angles,quotes}

\tikzstyle{thickline} = [line width=1.8pt]
\tikzstyle{gl} = [draw, gray]
\tikzstyle{pl} = [draw, orange]
\tikzstyle{ml} = [draw, blue]
\tikzstyle{sett} = [draw, thick, purple]
\tikzstyle{ele} = [draw, thick, green!70!black]
\tikzstyle{approvingvoters} = [thickline, black]
\tikzstyle{profileframe} = [step=1,black,thin]

\tikzset{voterlabel/.style={inner sep=1pt, font=\small}}
\tikzset{alternativelabel/.style={inner sep=1pt, font=\small}}

\makeatletter
\newcommand{\gettikzxy}[3]{%
	\tikz@scan@one@point\pgfutil@firstofone#1\relax
	\edef#2{\the\pgf@x}%
	\edef#3{\the\pgf@y}%
}
\makeatother

\usepackage{cleveref}
\usepackage{hyperref}

\newtheorem{corollary}{Corollary}
\newtheorem{lemma}{Lemma}

\newtheorem{observation}{Observation}
\newtheorem{proposition}{Proposition}

\newtheorem{thm}{Theorem}
\newtheorem{claim}{Claim}

\theoremstyle{definition}
\newtheorem{definition}{Definition}

\crefname{table}{Table}{Tables}
\crefname{figure}{Figure}{Figures}
\crefname{theorem}{Theorem}{Theorems}
\crefname{definition}{Definition}{Definitions}
\crefname{corollary}{Corollary}{Corollaries}
\crefname{observation}{Observation}{Observations}
\crefname{lemma}{Lemma}{Lemmas}
\crefname{example}{Example}{Examples}
\crefname{reduction}{Reduction}{Reductions}
\crefname{construction}{Construction}{Constructions}
\crefname{subsection}{Subsection}{Subsections}
\crefname{section}{Section}{Sections}
\crefname{proposition}{Proposition}{Propositions}
\crefname{algorithm}{Algorithm}{Algorithms}
\crefname{claim}{Claim}{Claims}
\crefname{thm}{Theorem}{Theorems}

\tikzstyle{alter} = [draw, circle, minimum size=4ex, inner sep=1pt, text centered, align=center]

\tikzstyle{nn} = [draw, circle, inner sep=.7pt,fill=black]
\tikzstyle{dnode} = [nn, rectangle, blue, fill=blue]

\newcommand{\decprob}[3]{%
\smallskip   
    \begin{minipage}{0.94\linewidth}%
      \textsc{#1}\\
      \textbf{Input:} #2\\
      \textbf{Question:} #3
    \end{minipage}%

}
\newcommand{\delVect}{\mathcal{\bar{V}}}

\newcommand{\misr}{\ensuremath{\mathsf{\rho}}}
\newcommand{\costdiff}{\mathsf{diff}}
\newcommand{\repre}{\ensuremath{\sigma}}
\newcommand{\happyrepre}{\ensuremath{\hat{\sigma}}}
\newcommand{\partialrepre}{\repre}
\newcommand{\VV}{\ensuremath{\mathsf{A}}}
\newcommand{\committee}{\ensuremath{W}}
\newcommand{\committeeR}{\committee_R}
\newcommand{\solution}{(\committee,\repre)}
\newcommand{\dominate}{\ensuremath{\sqsupset}}
\newcommand{\dominators}{\ensuremath{\mathsf{Dom}}}
\newcommand{\subordinates}{\ensuremath{\mathsf{Sub}}}
\newcommand{\incomp}{\ensuremath{\mathsf{Incom}}}
\newcommand{\earlier}{\ensuremath{\mathsf{Earlier}}}
\newcommand{\later}{\ensuremath{\mathsf{Later}}}
\newcommand{\inbetw}{\ensuremath{\mathsf{Inbet}}}
\newcommand{\inbetwS}{\ensuremath{\widehat{\inbetw}}}
\newcommand{\kUpper}{\ensuremath{\hat{\mathsf{k}}}}
\newcommand{\kLower}{\ensuremath{\check{\mathsf{k}}}}
\newcommand{\na}{\ensuremath{n_a}}
\newcommand{\nb}{\ensuremath{n_b}}
\newcommand{\nd}{\ensuremath{n^*}}
\newcommand{\promise}{\ensuremath{(c', i', j')}}
\newcommand{\upperB}{\ensuremath{\lceil n/ \csize\rceil}}
\newcommand{\lowerB}{\ensuremath{\lfloor n/ \csize \rfloor}}
\newcommand{\MNO}{\ensuremath{\MNT}}
\newcommand{\MNT}{\ensuremath{\mathsf{MNT}}}
\newcommand{\NT}{\ensuremath{\mathsf{NT}}}
\newcommand{\rank}{\ensuremath{\mathsf{rk}}}
\newcommand{\RR}{\ensuremath{\mathcal{R}}}

\newcommand{\aaa}{\ensuremath{\mathcal{C}}}
\newcommand{\vvv}{\ensuremath{\mathcal{V}}}

\newcommand{\ppp}{\ensuremath{\mathcal{P}}}
\newcommand{\profiletuple}{\ensuremath{(\aaa, \vvv, \RR)}}
\newcommand{\firsttabletuple}{\ensuremath{a,b,i,j,\kUpper,\kLower,\na,\nb,\nd,B,c',i',j'}}
\newcommand{\bound}{\ensuremath{\beta}}
\newcommand{\CC}{\textsc{CC}}
\newcommand{\lin}{\textsc{Linear}}
\newcommand{\app}{\textsc{Approval}}
\newcommand{\CCsum}{\CC-\summw}
\newcommand{\CCmax}{\CC-\maxmw}
\newcommand{\linCCsum}{\lin-\CC-\summw}
\newcommand{\appCCsum}{\app-\CC-\summw}

\newcommand{\Msum}{\Monroe-\summw}
\newcommand{\Mmax}{\Monroe-\maxmw}

\newcommand{\appMsum}{\app-\Monroe-\summw}

\newcommand{\Mmw}{\Monroe-\mw}
\newcommand{\CCmw}{\CC-\mw}

\newcommand{\Monroe}{\textsc{Monroe}}

\newcommand{\mw}{\textsc{MW}}

\newcommand{\summw}{\textsc{MW}\textsuperscript{+}}
\newcommand{\maxmw}{\textsc{MW}\textsuperscript{$\max$}}

\newcommand{\SP}{SP\xspace}
\newcommand{\SC}{SC\xspace}
\newcommand{\ii}{\ensuremath{\hat{i}}}
\newcommand{\jj}{\ensuremath{\hat{j}}}

\newcommand{\FPT}{\text{\normalfont{FPT}}\xspace}

\newcommand{\myemph}[1]{{\color{green!40!black}\emph{#1}}}

\newcommand{\delv}{t}
\newcommand{\dela}{t}

\newcommand{\csize}{k}
\newcommand{\delV}{\hat{V}}
\newcommand{\reV}{V_R}
\newcommand{\Alltypes}{\mathcal{T}}

\newcommand{\correct}{correct}

\newcommand{\good}{good}
\newcommand{\hide}[1]{}
\newcommand{\newR}[1]{\textbf{#1}}

\DeclareMathOperator{\cost}{cost}
\DeclareMathOperator{\type}{type}
\DeclareMathOperator{\dom}{\subordinates}
\DeclareMathOperator{\dominating}{\dominators}

\DeclareMathOperator{\usableset}{\mathcal{U}}
\DeclareMathOperator{\usablesetS}{\hat{\mathcal{U}}}
\DeclareMathOperator{\dyn}{\mathcal{D}}
\DeclareMathOperator{\Sig}{Sig}
\DeclareMathOperator*{\argmax}{arg\,max}
\DeclareMathOperator*{\argmin}{arg\,min}
\DeclareMathOperator{\conflictsetNO}{CONF}
\newcommand{\optimal}{maximally good}

\newcommand{\hassignable}{h-assignable}
\newcommand{\hassignment}{happy-assignment}

\usepackage{color}

\usepackage{comment}

\usepackage{algpseudocode}
\newcommand*\leftmost[1]{\operatorname{fi}(#1)}
\newcommand*\rightmost[1]{\operatorname{la}(#1)}

\newcommand*\typenr{y}
\DeclareMathOperator{\size}{size}
\DeclareMathOperator{\level}{level}

\newcommand{\appsymb}{\ensuremath{\star}}

\ifshort

\usepackage{etoolbox} %
\newcommand{\toappendix}[1]{%
  \gappto{\appendixtext}{
    {#1}
   }
}

\newcommand{\appendixproof}[2]{%
  \gappto{\appendixtext}{
    \subsection{Proof of \cref{#1}}\label{proof:#1}
    #2
    }
}

\newcommand{\appendixproofwithsketch}[3]{%
#2
  \gappto{\appendixtext}{
    \subsection{Proof of \cref{#1}}\label{proof:#1}
    #3
    }
}

\newcommand{\appendixsection}[1]{%
  \gappto{\appendixtext}{
    \section{Additional Material for Section~\ref{#1}}
    \label{appsec:#1}
  }
}

\else %

\usepackage{etoolbox} %
\newcommand{\toappendix}[1]{#1}

\newcommand{\appendixproof}[2]{#2}

\newcommand{\appendixsection}[1]{}

\newcommand{\appendixproofwithsketch}[3]{%
#3
}

\fi

\newcommand{\mypara}[1]{

  \smallskip
  \noindent \textbf{#1}
}
\begin{document}

\begin{frontmatter}

\title{Efficient Algorithms for Monroe and CC Rules in Multi-Winner Elections with \mbox{(Nearly) Structured Preferences}}

\author[A]{\fnms{Jiehua}~\snm{Chen}\orcid{0000-0002-8163-1327}}
\author[A]{\fnms{Christian}~\snm{Hatschka}\orcid{0000-0002-0881-8259}}
\author[A]{\fnms{Sofia}~\snm{Simola}\orcid{0000-0001-7941-0018}} %

\address[A]{TU Wien, Austria}

\begin{abstract}
  We investigate winner determination for two popular proportional representation systems: the Monroe and Chamberlin-Courant (abbrv.\ \CC) systems.
  Our study focuses on (nearly) single-peaked resp. single-crossing preferences.
  We show that for single-crossing approval preferences, winner determination of the Monroe rule is polynomial, and for both rules, winner determination mostly admits FPT algorithms with respect to the number of voters to delete to obtain single-peaked or single-crossing preferences.
  Our results answer some complexity questions from the literature~\cite{LacknerSkownron23ABC,SYFE2015,MisraSonarVaid2017MW}.
\end{abstract}

\end{frontmatter}
\section{Introduction}
\looseness=-1
\allowdisplaybreaks
In multi-winner elections, a preference profile consists of a set of alternatives and voters, each with preferences over the alternatives. The objective is to select a committee of fixed size k, representing voters' preferences optimally. Two well-known voting rules for multi-winner elections are the Monroe rule and the Chamberlin-Courant (CC) rule~\cite{chamberlin1983representative,monroe1995fully}, both aiming to proportionally represent different voter groups.
The Monroe rule minimizes the overall misrepresentation (i.e., voters' dissatisfaction) while ensuring each committee member represents roughly the same number of voters.
In contrast, the CC rule also seeks to minimize the overall misrepresentation, but does not require equal representation among committee members.

Both the Monroe and CC rules are popular proportional voting rules that have been studied extensively from both social choice and computational perspectives~\cite{Dummett1984,potthoff1998proportional,brams2009mathematics,procaccia2008complexity,lu2011budgeted,BetzlerMW2013,SYFE2015,SFS2015,FELABS2017,MisraSonarVaid2017MW,ConstantinescuElkind21MWSC,Sornat2022,ChenRoy2022}.
Unfortunately, the winner determination problems for these rules, \Mmw\ and \CCmw, are NP-hard~\cite{lu2011budgeted,procaccia2008complexity}, making it difficult to find an optimal solution.
However, the NP-hardness reduction may result in instances with unstructured preferences.
Consequently, researchers have explored the computational complexity of instances with nice preference structures, such as single-peaked (SP)~\cite{Black1948}, single-crossing (SC)~\cite{Mirrlees1971}, or nearly SP or SC preferences~\cite{BreCheWoe2016,ELP17NSP} (also see \cref{sec:prelim} for the definition), as oftentimes the preferences of voters align with these structures. %

Betzler et al.~\cite{BetzlerMW2013} designed polynomial-time algorithms for both \Mmw\ and \CCmw\ when voters have either approval or linear preferences, which are SP. Skowron et al.~\cite{SYFE2015} continued this line of research on linear and SC preferences, demonstrating that in this case \CCmw\ is also polynomial-time solvable,
while \Mmw\ with linear and SC preferences remains NP-hard. 
They left open the complexity of \Mmw\ for SC approval preferences or when minimizing the maximum misrepresentation.
Elkind and Lackner~\cite{EL15} give polynomial-time algorithms for weighted proportional approval voting (weighted PAV) rules for SC approval preferences, which includes the CC rule.
Recently, Constantinescu and Elkind~\cite{ConstantinescuElkind21MWSC} and Sornat et al.~\cite{Sornat2022} improved existing positive results by providing more efficient algorithms for SP and nearly SP preferences. 
As for nearly structured preferences, Misra et al.~\cite{MisraSonarVaid2017MW} examined the parameterized complexity of CC-MW concerning the distance to SPness or SCness.
The distance is measured by the number of voters (resp.\ alternatives) to delete to obtain SPness or SCness.
They provide fixed-parameter (FPT) algorithms for the distance measure of deleting alternatives, but leave open the question of whether the same holds for deleting voters; we answer this question positively.
For other distance measure, Skowron et al.~\cite{SYFE2015} show that their algorithm for \CCmw can be modified into an FPT algorithm wrt.~the single-crossing width.
The \Monroe\ and \CC\ rules have also been studied on other extensions of SPness.
Peters et al.~\cite{peters2022preferences} study \CCmw\ on preferences that are SP on trees.
Peters and Lackner~\cite{peters2020preferences} study Ordered Weight Average voting, which is an extension of the CC rule, on preferences that are SP on a circle.
Godziszewski et al.~\cite{godziszewski2021analysis} study multi-winner elections, including \CCsum\ when the voters and alternatives can be embedded in 2-Euclidean space.

\mypara{Our contributions.}
In this paper, we contribute to the algorithmic research of the Monroe and CC rules under (nearly) structure preferences, and provide several efficient algorithms.
All algorithms are based on dynamic programming (DP).
Foremost, in \cref{sec:monroe-SC}, we develop a novel polynomial-time algorithm for \Mmw\ with SC approval preferences, using DP; this answers an open question~\cite{SYFE2015}.
Note that under SC preferences, a key building block in standard DP algorithms for \CCmw\ is the continuous block property that an optimal solution may satisfy.
A major challenge when developing algorithms for \Mmw\ however is the absence of such continuous block property even under SC approval preferences. %
We overcome this challenge by introducing a structural concept called \emph{\optimal\ voter intervals} and show that they one-to-one correspond to the alternatives in an optimal solution. 
We observe that there exists an optimal solution where we can greedily assign to each alternative in the committee a voter interval (which may be potentially larger than the voter set that it represents). %
This approach allows us to effectively combine partial solutions and solve the \Mmw\ problem with SC approval preferences efficiently.
\begin{table*}[t!]
  \caption{Overview of the complexity for Monroe and CC (columns) with (nearly) SP/SC preferences (rows). Results in bold text are new. Note that ``linear'' means that the voters have linear preferences and the misrepresentation is based on Borda.
    \summw\ refers to the problem of minimizing the \emph{sum} of misrepresentations of all voters, while \maxmw\ refers to the problem of minimizing the \emph{maximum} misrepresentation of all voters. Next to the results one can see the source, or the corresponding corollary or theorem.}\label{tab:overview}
\resizebox{2\columnwidth}{!}{%
  \begin{tabular}{lllcllcllllcllll}
    \toprule
    Preference structure & \multicolumn{2}{c}{\CC-\summw} && \multicolumn{2}{c}{\CC-\maxmw} && \multicolumn{4}{c}{\Monroe-\summw} && \multicolumn{4}{c}{\Monroe-\maxmw}\\\cline{2-3} \cline{5-6} \cline{8-11}\cline{13-16}
                         &\multicolumn{2}{c}{approval/linear} && \multicolumn{2}{c}{approval/linear} 
                         && \multicolumn{2}{c}{approval} & \multicolumn{2}{c}{linear} && \multicolumn{2}{c}{approval} & \multicolumn{2}{c}{linear} \\
    Single-peaked  & P & \cite{BetzlerMW2013}&& P & \cite{BetzlerMW2013}&& P & \cite{BetzlerMW2013} & ? &&& P & \cite{BetzlerMW2013}& P & \cite{BetzlerMW2013}\\
    Single-crossing & P & \cite{SYFE2015,EL15}&& P & \cite{SYFE2015,EL15}&& \newR{P} & (T\ref{thm:MonroeP})& NP-hard & \cite{SYFE2015} && \newR{P} & (T\ref{thm:MonroeP})& ? & \\\bottomrule
    $t$-voters SP & \newR{FPT} & (T\ref{thm:nearSPvoter})&& \newR{FPT} & (C\ref{corr:max})&& \newR{FPT} & (T\ref{thm:MonroenearSCvot})& ? &&& \newR{FPT} & (T\ref{thm:MonroenearSCvot})& \newR{FPT} & (T\ref{thm:MonroenearSCvot})\\
    $t$-voters SC & \newR{FPT} & (T\ref{thm:Approval-CC-SC-t-voters:FPT})&& \newR{FPT} & (C\ref{corr:max})&& \newR{FPT} &(T\ref{thm:MonroenearSCvot})& NP-hard &\cite{SYFE2015}&& \newR{FPT} &(T\ref{thm:MonroenearSCvot})& ?&\\\bottomrule
    $t$-alternatives SP & FPT & \cite{MisraSonarVaid2017MW}&& FPT & \cite{MisraSonarVaid2017MW}&& \newR{XP} & (T\ref{thm:MonroenearSCalt})& ? &&& \newR{XP} & (T\ref{thm:MonroenearSCalt})& \newR{XP} & (T\ref{thm:MonroenearSCalt})\\
    $t$-alternatives SC & FPT & \cite{MisraSonarVaid2017MW}&& FPT & \cite{MisraSonarVaid2017MW}&& \newR{XP} &(T\ref{thm:MonroenearSCalt})& NP-hard &\cite{SYFE2015}&& \newR{XP} &(T\ref{thm:MonroenearSCalt})& ?&\\
  \end{tabular}}
\end{table*}

Then, in \cref{sec:FPTSPCC}, we show that \CCmw\ and \Mmw\ (except one case) are fixed-parameter tractable (FPT) for the distance measure of deleting $\delv$ voters to achieve SPness or SCness, i.e., the corresponding problems can be solved in $f(\delv)\cdot (n+m)^{O(1)}$ time, where $n$ and $m$ denote the number of voters and alternatives, respectively.
The results for the CC rule answer an open issue by Misra et al. %
The basic idea is to guess in FPT time how different deleted voters are going to be represented by the same alternatives and in which order these representing alternatives. %
We then combine the guessed structure into a DP which moves along the SC order and a specific order of the alternatives and finds a partial solution that additionally covers this guessed structure in FPT time.
Finally, we present straightforward polynomial-time algorithms for the Monroe rule when only a constant number~$\dela$ of alternatives need to be deleted to achieve SPness (resp. SCness), demonstrating that the corresponding problems are in XP with respect to~$\dela$.
See \cref{tab:overview} for an overview.

\looseness=-1

\section{Preliminaries}\label{sec:prelim}
\ifshort(Full) proofs for results marked by \appsymb\ are deferred to the appendix. %
\fi
Given a non-negative integer~$t\in \mathds{N}$, let $[t]$ denote the set~$\{1,\ldots, t\}$. We assume basic knowledge of parameterized complexity and refer to the textbook by Cygan et al.~\cite{CyFoKoLoMaPiPiSa2015} for more details.
For a more general introduction to multi-winner elections, see for example the book chapter by Faliszewski et al.~\cite{faliszewski2017multiwinner}.
For further methods and topics on multi-winner elections with approval preferences we refer to a recent book by Lackner and Skowron~\cite{LacknerSkownron23ABC}. 

\mypara{Preference profiles and structured preferences.}
A \myemph{preference profile} (\myemph{profile} in short) is a triple~$(\aaa, \vvv, \RR)$, where $\aaa$ denotes a set~$\aaa$ of $m$ alternatives, $\vvv$ denotes a set of $n$ voters with $\vvv=\{v_1,\ldots, v_n\}$,
and $\RR$ is a collection~$\RR=(\succeq_1,\ldots, \succeq_n)$ of preference orders such that each~$\succeq_i$, is either a linear order or a subset of $\aaa$
and shall represents the preferences of voter~$v_i$ over~$\aaa$, $i\in [n]$.
For instance, for $\aaa=\{1,2,3,4\}$, a preference order can be $3\succ 1 \succ 2 \succ 4$ or a subset $\{2,3\}$.
The former means that $3$ is preferred to $1$, $1$ to $2$, and $2$ to $4$,
while the latter means that $2$ and $3$ are approved while $1$ and $4$ not. 
Note that an approval set can also be considered as a dichotomous weak order, so having approval set~$\{2,3\}$ is equivalent to $\{2,3\}\succ \{1,4\}$, and we say that $2$ is tied with $3$, while $1$ is tied with $4$.
For the sake of clarity, we will call a preference profile an \myemph{approval profile} if the preference orders are approval sets; otherwise it is a \myemph{linear preference profile}.
Given a preference order~$\succ$ and two alternatives~$x$ and $y$, we write \myemph{$x\succeq y$} to mean that $x$ is preferred to or %
tied with $y$.
 For instance, for $3\succ 1\succ 2$, we have that $3\succeq 1$,
 and for the approval set~$\{2,3\}$, we have that $2\succeq 3$ and $3 \succeq 2$. 
 Given a voter~$v_i \in \vvv$ and an alternative~$a \in \aaa$, we define the rank of~$a$ in the preferences of~$v_i$ as $\rank_i(a) = |\{b\in \aaa \mid b \succ_i a\}|$.
For approval preferences, we also use \myemph{$\VV(a)$} to denote the set of voters who each approve of~$a$.

\begin{definition}[Single-peaked and single-crossing]
  Let $\ppp$ be a preference profile (with either linear or approval preferences).
  We say that~$\ppp$ is \myemph{single-peaked} (\SP) wrt.\ a linear order~$\rhd$ of the alternatives~$\aaa$ if for each voter $v_i \in \vvv$ and three distinct alternatives~$a,b,c \in \aaa$ with
  $a\rhd b \rhd c$ or $c\rhd b \rhd a$ it holds that  
 ``$a \succeq_i b$'' implies ``$b\succeq_i c$''. 
  Accordingly, we say that $\ppp$ is \myemph{SP} if there exists a linear order~$\rhd$ on~$\aaa$ such that $\ppp$ is SP wrt.~$\rhd$.

  We say that~$\ppp$ with \emph{linear preferences} is \myemph{single-crossing} (\SC) wrt.\ a linear order~$\rhd$ of the voters~$\vvv$ if for each pair~$\{x,y\}\subseteq \aaa$ of alternatives and each three voters~$v_i, v_j, v_k\in \vvv$ with $v_i\rhd v_j \rhd v_k$ it holds that
  ``$x\succ_i y$ and $x \succ_k y$'' implies ``$x\succ_j y$''.
  We say that~$\ppp$ with \emph{approval preferences} is \myemph{single-crossing} (\SC) wrt.\ a linear order~$\rhd$ of the voters~$\vvv$ if for each alternative~$x$ the set of voters approving of~$x$ form an interval in $\rhd$.
\end{definition}
It is polynomial-time solvable to check whether a given preference profile is single-peaked~\cite{Bartholdi1986} (resp.\ single-crossing~\cite{BCW12}). 
We say that a preference profile~$\ppp$ is \myemph{$\dela$-alternatives nearly SP} (resp.\ \myemph{nearly SC}) if it is possible to delete at most $\dela$ alternatives from $\ppp$ (and update the preference orders accordingly) to obtain an SP (resp.\ SC) profile.
We define \myemph{$\delv$-voters nearly SP}  (resp. \myemph{nearly SC}) profiles analogously.
Determining the smallest $\dela$ for a given profile to be $\dela$-alternatives nearly SP or $\delv$-voters nearly SC can be done in polynomial-time, while it is NP-hard to determine the smallest $\delv$ for a given preference profile to be $\delv$-voters nearly SP or $\dela$-alternatives nearly SC~\cite{BreCheWoe2016,ELP17NSP}. 
Nevertheless, the latter two problems are \FPT with respect to $\delv$~\cite{elkind2014detecting}.

\mypara{Multi-winner election with proportional misrepresentations.}
Let $\ppp$ denote a profile and $\csize \in \mathbb{N}$ be a number.
An \myemph{assignment} $\repre$ is a function that maps voters to alternatives, i.e. $\repre\colon \vvv\rightarrow\aaa$.
We say that an assignment~$\repre\colon \vvv\to\aaa$ is a \myemph{$\csize$-assignment} if $|\repre(V)|=\csize$, and that it is \myemph{proportional} if each assigned alternative represents roughly the same portion of the voters, i.e., $\lowerB \le |\{v\in V\mid \repre^{-1}(a)\}|\le \lceil{n/k\rceil}$ holds for all~$a\in \repre(V)$. A \myemph{partial assignment} $\repre$ is a partial function that maps voters to alternatives.

A \myemph{misrepresentation function $\misr \colon \vvv \times \aaa \to \mathbb{R}_{\geq 0}$} specifies the dissatisfaction of a voter towards an alternative. For linear preference profiles we consider Borda misrepresentation function, where $\misr(v_i, a) = \rank_i(a)$. For approval profiles we use the approval misrepresentation function, where $\misr(v_i, a) = 0$ if $v_i$ approves $a$ and $1$ otherwise.
There are two common ways of measuring the overall misrepresentations, one is the sum of misrepresentations of the voters and the other the maximum misrepresentation among all voters.

For a fixed number~$\csize$, the \Monroe{} rule finds a proportional $\csize$-assignment that minimizes either the sum or maximum of all misrepresentations.
Under the CC rule, we do not have such a requirement--each alternative may represent an arbitrary number of voters.
Consequently, we obtain four multi-winner (\mw) election problems.

\decprob{\Msum (\text{\normalfont{resp.}} \Mmax)}
{A preference profile~\ppp, misrepresentation function~$\misr$, a committee size $\csize$.}
{Find a  \emph{proportional} 
  $\csize$-assignment $\repre\colon \vvv \to \aaa$ with minimum~$\sum_{v_i \in \vvv} \misr(v_i, \repre(v_i))$ (resp.\ $\max_{v_i \in \vvv} \misr(v_i, \repre(v_i))$).}
If we drop the proportionality requirement, we obtain the \CCsum\ and \CCmax\ problems.
For the sake of brevity, we also use the same names to refer to the decision variants of these problems where a misrepresentation bound is given as input. 
Moreover, for each problem~$\Pi\in \{$\CC-\summw, \CC-\maxmw, \Monroe-\summw, \Monroe-\maxmw$\}$, we use \lin-$\Pi$ and \app-$\Pi$ to refer to the problem~$\Pi$ with linear preferences and approval preferences, respectively.
For instance, \linCCsum\ denotes the problem \CC-\summw\ with linear preferences.

Let $I=(\profiletuple, \misr, \csize)$ denote an instance of multi-winner election. %
\ifshort We say that a tuple~$(\committee, \repre)$ is a \myemph{solution of $I$} if $\repre$ is a $\csize$-assignment and $\repre(V)=\committee$.
\else
We say that a tuple~$(\committee, \repre)$ is a \myemph{partial solution of $I$} if $\repre$ is a partial assignment and $\repre(V)\subseteq \committee$ with $|\committee| \le \csize$. 
We say that a partial solution~$(\committee, \repre)$ is a \myemph{solution of $I$} if $\repre$ is a indeed a $\csize$-assignment and $\repre(V)=\committee$. \fi\todoH{Did we ever use partial solution outside of section 3? We have a different definition for partial solution there. S: Removed partial solution from short version, left in long. Removed Monroe partial solution completely.}
Under the \Monroe\ rule, we additionally require that a solution is proportional.
 Finally, if $\repre$ minimizes the corresponding misrepresentation measure, then we call~$(\committee, \repre)$ an \myemph{optimal solution}.

As already noted by Betzler et al.~\cite{BetzlerMW2013}, the decision variant of \lin-\maxmw\ can be considered as a restriction of the decision variant \app-\summw\ by letting each voter approve of those alternatives for whom the misrepresentation is within the bound~$\bound$. %
\begin{proposition}[\cite{BetzlerMW2013}]\label{obs:reduction}
\CCmax\ (resp.\ \Mmax) can be reduced to \appCCsum (resp.\ \appMsum) in linear time.
\end{proposition}	
This reduction preserves the SP property, as was shown by Betzler et al., however the SC property is not necessarily satisfied. %

\section{Monroe for approval and SC preferences}\label{sec:monroe-SC}

In this section, we consider profiles with SC approval preferences and show that under the \Monroe\ rule we can find an optimal solution in polynomial-time, using dynamic programming~(DP). 
In \cref{sub:additional} we describe some structural definitions that are in \cref{sub:technical} used to show that there is always an optimal solution where the happy voters of the alternatives are ordered as certain types of voter intervals. This ordering allows us to build and combine partial solutions from the bottom up. Using our knowledge of this ordering, we will build and describe the DP in \cref{sec:dynprogtable}.

\subsection{Additional definitions}\label{sub:additional}
To ease notation, we assume that the voters in~$\vvv$ are named $1,2,\ldots,n$ such that $1>\cdots >n$ is an SC order.
Before we show the DP approach, we first introduce necessary concepts and notations. 

Let $\ppp=\profiletuple$ be an SC approval preference profile.
We call an assignment $\repre$ an \myemph{$[\ii,\jj]$-assignment} if it is an assignment that maps the voter interval $[\ii,\jj]$ to the set of alternatives, i.e. $\repre\colon [\ii,\jj]\rightarrow\aaa$. 

Given an $[\ii,\jj]$-assignment $\repre$ for $\ppp$, we derive a partial assignment called \myemph{$[\ii,\jj]$-\hassignment}~\myemph{\happyrepre} from~\repre\ which only considers happy voters. Therefore, $\happyrepre(v)=\repre(v)$ if $\misr(v,\repre(v))=0$ and voters that are not happy under $\repre$ are left unassigned. If $[\ii,\jj]=[n]$ we will just refer to it as a \hassignment.

In the context of SC approval profiles, we say that \myemph{$(W,\partialrepre)$} is a \myemph{partial solution wrt.~to an interval of voters $[\ii,\jj]$} if $\partialrepre$ is a partial $[\ii,\jj]$-assignment with $\partialrepre([\ii,\jj])\subseteq W$ and $|\repre^{-1}(a)| \leq \lceil \frac{n}{\csize} \rceil$ for all $a \in \repre(\VV)$.\todo{S: Added this weaker proportionality requirement, is it needed?}
Note that a solution is a partial solution wrt.~$[n]$.
We define the \myemph{misrepresentation of a partial solution} to be $\misr(\partialrepre)=\sum_{i\in[\ii,\jj]}\misr(i,\partialrepre(i))$.
Let $(W,\repre)$ be a (partial) solution. We say that a voter $v$ is \myemph{happy with $c$} if he is assigned to $c$ and has zero misrepresentation for $c$, i.e., $\repre(v)=c$ and $\misr(v,c)=0$.
We say a voter $v$ is \myemph{\hassignable} wrt.~a (partial) solution $(W,\partialrepre)$ if there exists an alternative $a\in\happyrepre(V)$ such that $v\in\VV(a)$. %

Next, we define some relations between alternatives. 
\begin{definition}[Domination, incomparable sets, earlier sets]\label{def:dom-incom-earlier}
  Let $\ppp$ be an SC approval profile and let $\rhd$ be an SC order.
  For an arbitrary alternative~$a\in \aaa$, we use \myemph{$\leftmost{a}$} and \myemph{$\rightmost{a}$} to denote the first and the last voter in the order~$\rhd$ that approve of $a$. %

  Let $a$ and $b$ be two alternatives.
  We say that $a$ \myemph{dominates} $b$ if $\VV(a) \supset \VV(b)$,
  and call $a$ (resp.\ $b$) the \myemph{dominator} (resp.\ \myemph{subordinate}) of $b$ (resp.\ $a$).
  We use \myemph{$\dominators(a)$} (resp.\ \myemph{$\subordinates(a)$}) to denote the set consisting of all dominators (resp.\ subordinates) of $a$, i.e.,
  $\dominators(a)=\{b\in \aaa \mid \VV(a) \subset \VV(b)\}$ and
  $\subordinates(a) = \{b\in \aaa \mid \VV(b) \subset \VV(a)\}$.
  We say that $a$ and $b$ are \myemph{comparable} if $b\in\dominators(a)\cup \subordinates(a) \cup \{a\}$, and they are \myemph{incomparable} if not. We will use the set $\incomp(a)=\aaa \setminus (\dominators(a)\cup \subordinates(a) \cup \{a\})+$ to denote the set of alternatives that are incomparable to $a$.%
  
  We say that $a$ \myemph{starts earlier} (resp. \myemph{starts later}) than $b$ if $a\in\incomp(b)$ and $\leftmost{a}<\leftmost{b}$ (resp. $\leftmost{b} < \leftmost{a}$),
  and use \myemph{$\earlier(a)$} (resp. \myemph{$\later(a)$}) to denote the set consisting of all alternatives that start earlier than $a$, i.e., $\earlier(a) = \{ b \in \incomp(a) \mid \leftmost{b} < \leftmost{a}\}$ (resp. $\later(a) = \{ b \in \incomp(a) \mid \leftmost{a} < \leftmost{b}\}$).
\end{definition}

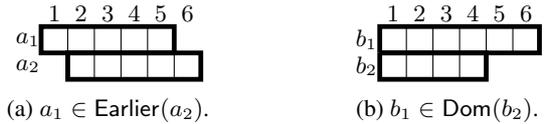
\begin{figure}[t!]
  \captionsetup[subfigure]{justification=centering}
  \centering
  \begin{subfigure}[T]{.495\linewidth}
    \centering
    \begin{tikzpicture}[scale=0.35]
      \draw[profileframe] (0,1) grid (5,0);
      \draw[profileframe] (1,0) grid (6,-1);
      \draw[approvingvoters] (0,0) -- (5,0) -- (5,1) -- (0,1) -- (0,0) -- cycle;
      \draw[approvingvoters] (1,0) -- (6,0) -- (6,-1) -- (1,-1) -- (1,0) -- cycle;

      \foreach \n  in {1, ..., 6} {
        \node[voterlabel] at (\n-0.5, 1.5) {$\n$};
      }
       \foreach \n  in {1, 2} {
        \node[alternativelabel] at (-0.5, -\n+1.5) {$a_{\n}$};
      }
    \end{tikzpicture}
    \caption{$a_1\in \earlier(a_2)$.}
  \end{subfigure}~~
  \begin{subfigure}[T]{.495\linewidth}
    \centering
    \begin{tikzpicture}[scale=0.35]
      \draw[step=1.0,black,thin] (0,1) grid (6,0);
      \draw[step=1.0,black,thin] (0,0) grid (4,-1);
      \draw[approvingvoters] (0,0) -- (6,0) -- (6,1) -- (0,1) -- (0,0);
      \draw[approvingvoters] (0,0) -- (4,0) -- (4,-1) -- (0,-1) -- (0,0);
      \foreach \n  in {1, ..., 6} {
        \node[voterlabel] at (\n-0.5, 1.5) {$\n$};
      }
       \foreach \n  in {1, 2} {
        \node[alternativelabel] at (-0.5, -\n+1.5) {$b_{\n}$};
      }
    \end{tikzpicture}
    \caption{$b_1\in \dominators(b_2)$.}
  \end{subfigure}
  \caption{Illustration of relations between two alternatives in an SC approval profile (see \cref{def:dom-incom-earlier}).
    Throughout all figures, the columns correspond to the voters such that the left-to-right order is SC, while the rows correspond to the alternatives.
    The squares in each row specify which voters approve the corresponding alternative.
    Due to the SC property, the squares in each row are consecutive.}     \label{fig:alternative-relation}
\end{figure}

\Cref{fig:alternative-relation} illustrates the relation between two alternatives.

\todoIH{Intuition. C: done}
The next definition gives a grouping of the alternatives based on how ``dominated'' they are and provides an ordering of the alternatives that is relevant for the DP. An alternative that is undominated is on level $1$, an alternative that is only dominated by level $1$ alternatives is on level $2$ and so on.
\begin{definition}[Levels and canonical ordering]\label{def:levels}
  Let $A\subseteq \aaa$ be a set of alternatives. We define the level sets inductively:
  \begin{align*}
    \level_A(1)\coloneqq&\{c\in A\mid\dominators(c)=\emptyset\}\\
    \level_A(i+1)\coloneqq&\{c\in A\mid ( \level_A(i)\cap\dominating(c)\neq\emptyset)\\&\wedge (\dominating(c)\cap A\subseteq\level_A(1)\cup\ldots\cup\level_A(i))\}.
  \end{align*}
  If $A$ is clear from context, we will omit it from the subscript.
The \myemph{canonical} ordering of~$A$ is defined recursively as follows:
It starts with the alternatives in $\level(1)$ ordered with alternatives that start earlier being earlier in the ordering, then $\level(2)$ ordered based on the first voter that approves them and so on. 
\end{definition}

For the ease of notation, throughout the whole section, we assume that $W=\{a_1,\ldots,a_t\}$ so that the order~$a_1,\ldots,a_t$ respects the canonical ordering defined above.

We now go over desirable properties of a solution, based on which we then derive a corresponding nice structure of voter intervals to compute an optimal solution efficiently. 
The first property assumes that we can select the best alternatives among all that satisfy a group of voters, whereas the second property assumes that happy voters are assigned in an intuitive way.
\begin{definition}[Monotone and neatly ordered solution]\label{def: mono}
Let $\solution$ be a partial solution wrt.\ a voter interval $[\ii,\jj]$.
 We say that $(W,\partialrepre)$ is \myemph{monotone wrt.\ a set of alternatives $A\subseteq\aaa$}, if it is inclusion-wise maximal with relation to the domination relation. Formally it means:
  \begin{compactenum}[(i)]
    \item For every~$c\in W$ it holds that $\dominators(c)\cap A\subseteq W$ and
    \item for all~$a,b\in A$ with $a\in\dominators(b)$ it holds that if there exists a voter $u\in[\ii,\jj]$ who is happy with $b$, then there exists a voter $v\in[\ii,\jj]$ who is happy with $a$.
    \end{compactenum}
    We say a solution is \myemph{monotone} if it is monotone wrt.\ \aaa.\todoIC{Not needed maybe?}

  Let $a,b\in\happyrepre([\ii,\jj])$ be a pair of alternatives.
  We say that $a$ \myemph{precedes} $b$ in $\solution$ if $\max(\happyrepre^{-1}(a))<\min(\happyrepre^{-1}(b))$. Note that we identify the voter by its index in the single-crossing order, and take $\max$ and $\min$ over those.
  We say that $a$ and $b$ are \myemph{neatly ordered (\NT)} in $\solution$ if the following two statements hold, where we assume, without loss of generality, that
  $a \in \earlier(b) \cup \dominators(b)$:
  \begin{compactenum}[(i)]
    \item If $a \in \earlier(b)$, then $a$ precedes~$b$.
    \item If $a \in \dominators(b)$, then every voter~$i\in \happyrepre^{-1}(a)$ has
    $i < \leftmost{b}$ or $i > \max(\happyrepre^{-1}(b))$.
  \end{compactenum}
  We say a solution is neatly ordered (\NT) if every pair $a,b \in\happyrepre([\ii,\jj])$ is \NT.
  We use \MNO\ to abbreviate monotone and neatly ordered.
\end{definition} 
\begin{figure}[t!]
  \centering
 \begin{subfigure}[T]{.5\linewidth}
    \centering
	\begin{tikzpicture}[scale=0.35]
		\draw[profileframe] (1,0) grid (7,1);
		\draw[profileframe] (0,0) grid (8,-1);
		\draw[approvingvoters] (1,0) -- (7,0) -- (7,1) -- (1,1) -- (1,0);
		\foreach \x in {0,...,3}
		\draw [pattern=north west lines, pattern color=blue](1+\x,0) -- (2+\x,0) -- (2+\x,1) -- (1+\x,1) -- (1+\x,0);
		\draw[approvingvoters] (0,0) -- (8,0) -- (8,-1) -- (0,-1) -- (0,0);
		\draw [pattern= north west lines,pattern color=red](0,0) -- (1,0) -- (1,-1) -- (0,-1) -- (0,0);
		\foreach \x in {0,...,2}
		\filldraw [pattern=north west lines, pattern color=red](5+\x,0) -- (6+\x,0) -- (6+\x,-1) -- (5+\x,-1) -- (5+\x,0);
		\foreach \n  in {1, ..., 8} {
		          \node[voterlabel] at (\n-0.5, 1.5) {$\n$};
		        }
		        \foreach \n  in {1,...,2} {
		          \node[alternativelabel] at (-0.5, -\n+1.5) {$a_{\n}$};
		      }
		      
	\end{tikzpicture}
	\caption{The pair $(a_1,a_2)$ is \NT.}\label{fig:NO1}
\end{subfigure}~~
\begin{subfigure}[T]{.5\linewidth}
    \centering
\begin{tikzpicture}[scale=0.35]
	\draw[profileframe] (0,1) grid (5,0);
	\draw[profileframe] (3,-1) grid (8,0);
	\draw[approvingvoters] (0,0) -- (5,0) -- (5,1) -- (0,1) -- (0,0);
	\foreach \x in {0,...,3}
	\draw [pattern=north west lines, pattern color=blue](0+\x,0) -- (1+\x,0) -- (1+\x,1) -- (0+\x,1) -- (0+\x,0);
	\draw[approvingvoters] (8,-1) -- (3,-1) -- (3,0) -- (8,0) -- (8,-1);
	\foreach \x in {0,...,3}
	\filldraw [pattern=north west lines, pattern color=red](4+\x,-1) -- (5+\x,-1) -- (5+\x,0) -- (4+\x,0) -- (4+\x,-1);
	\foreach \n  in {1, ..., 8} {
			          \node[voterlabel] at (\n-0.5, 1.5) {$\n$};
			        }
			        \foreach \n  in {1,...,2} {
			          \node[alternativelabel] at (-0.5, -\n+1.5) {$b_{\n}$};
			      }
\end{tikzpicture}
	\caption{The pair $(b_1,b_2)$ is \NT.}\label{fig:NO2}
\end{subfigure}\\
\begin{subfigure}[t]{.3\linewidth}
	\centering
		\begin{tikzpicture}[scale=0.35]
		\draw[profileframe] (0,1) grid (5,0);
		\draw[profileframe] (1,-1) grid (6,0);
		\draw[approvingvoters] (0,0) -- (5,0) -- (5,1) -- (0,1) -- (0,0);
		\foreach \x in {0,1}
		\filldraw [pattern=north west lines, pattern color=blue](3+\x,0) -- (4+\x,0) -- (4+\x,1) -- (3+\x,1) -- (3+\x,0);
		\draw[approvingvoters] (1,0) -- (6,0) -- (6,-1) -- (1,-1) -- (1,0);
		\foreach \x in {0,1}
		\filldraw [pattern=north west lines, pattern color=red](1+\x,0) -- (2+\x,0) -- (2+\x,-1) -- (1+\x,-1) -- (1+\x,0);
		\foreach \n  in {1, ..., 6} {
							          \node[voterlabel] at (\n-0.5, 1.5) {$\n$};
							        }
							        \foreach \n  in {1,...,2} {
							          \node[alternativelabel] at (-0.5, -\n+1.5) {$c_{\n}$};}
	\end{tikzpicture}
	\caption{The pair $(c_1,c_2)$ is not neatly ordered due to violating~$(i)$.}\label{NotNO1}
\end{subfigure}\hfill
	\begin{subfigure}[t]{.3\linewidth}
		\centering
		\begin{tikzpicture}[scale=0.35]
			\draw[profileframe] (0,1) grid (6,0);
			\draw[profileframe] (0,-1) grid (4,0);
			\draw[approvingvoters] (0,0) -- (6,0) -- (6,1) -- (0,1) -- (0,0);
			\foreach \x in {0,1}
			\filldraw [pattern=north west lines, pattern color=blue](0+\x,0) -- (1+\x,0) -- (1+\x,1) -- (0+\x,1) -- (0+\x,0);
			\draw[approvingvoters] (0,0) -- (4,0) -- (4,-1) -- (0,-1) -- (0,0);
			\foreach \x in {0,1}
			\filldraw [pattern=north west lines, pattern color=red](2+\x,0) -- (3+\x,0) -- (3+\x,-1) -- (2+\x,-1) -- (2+\x,0);
			\foreach \n  in {1, ..., 6} {
			\node[voterlabel] at (\n-0.5, 1.5) {$\n$};
										        }
			\foreach \n  in {1,...,2} {
			\node[alternativelabel] at (-0.6, -\n+1.5) {$d_{\n}$};}
		\end{tikzpicture}
		\caption{The pair $(d_1,d_2)$ is not neatly ordered due to violating~$(ii)$.}\label{NotNO2}
	\end{subfigure}\hfill	
	\begin{subfigure}[t]{.3\linewidth}
	\centering
	\begin{tikzpicture}[scale=0.35]
	\draw[profileframe] (0,1) grid (6,-0);
	\draw[profileframe] (1,-1) grid (5,-0);
		\draw[approvingvoters] (0,0) -- (6,0) -- (6,1) -- (0,1) -- (0,0);
		\foreach \x in {0,1}
		\filldraw [pattern=north west lines, pattern color=blue](2+\x,0) -- (3+\x,0) -- (3+\x,1) -- (2+\x,1) -- (2+\x,0);
		\draw[approvingvoters] (1,0) -- (5,0) -- (5,-1) -- (1,-1) -- (1,0);
		\filldraw [pattern=north west lines, pattern color=red](1,0) -- (2,0) -- (2,-1) -- (1,-1) -- (1,0);
		\filldraw [pattern=north west lines, pattern color=red](4,0) -- (5,0) -- (5,-1) -- (4,-1) -- (4,0);
\foreach \n  in {1, ..., 6} {
			\node[voterlabel] at (\n-0.5, 1.5) {$\n$};
										        }
			\foreach \n  in {1,...,2} {
			\node[alternativelabel] at (-0.55, -\n+1.5) {$e_{\n}$};}
	\end{tikzpicture}
	\caption{The pair $(e_1,e_2)$ is not neatly ordered due to violating~$(ii)$.}\label{NotNO3}
	\end{subfigure}
	\caption{Illustration of neatly ordered and not neatly ordered (partial) solutions. The hatched parts indicate the assignment.} \label{fig:sketchesno}
\end{figure}
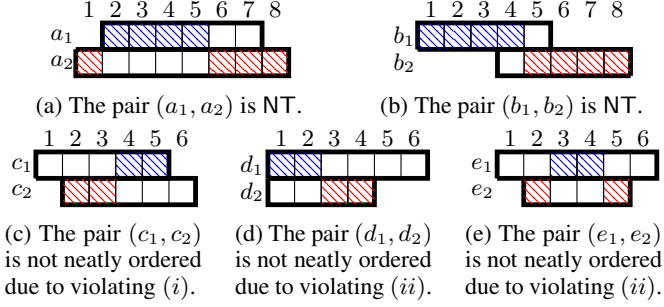
See \cref{{fig:sketchesno}} for an illustration of monotonicity and neat ordering.
Next, we define the voter interval structure which helps us find an optimal solution; also see \cref{fig:sketchesgood} for an illustration.
\todoIH{Update the definition regarding the partial solution.}
\begin{definition}[Good and \optimal\ intervals]\label{def:good-optimal-intervals}
  Let $\solution$ be a (partial) solution and $a\in W$ an alternative.
  A voter interval $[i,j]$ is called \myemph{good} for $a$ with respect to~$\solution$ if it satisfies the following:
\begin{compactenum}[(i)]
  \item \label{good:3} If $\happyrepre^{-1}(a)=\emptyset$, then $[i,j]=\emptyset$.
  \item \label{good:1} $\happyrepre^{-1}(a) \subseteq [i,j]\subseteq \VV(a)$.
  \item \label{good:2} For all $\ell\in[i,j]\cap\happyrepre^{-1}(W)$, we have $\happyrepre(l)\cap\incomp(a)=\emptyset$. %
  \item \label{good:6} If $\solution$ is a partial sol.\ wrt. $[\ii,\jj]$, then $[i,j]\subseteq[\ii,\jj]$. 
\end{compactenum}
Note that the third requirement above states that every alternative that satisfies some voter from the range is comparable with $a$. 

Let $\mathcal{I} \coloneqq ([i_1,j_1],\ldots,[i_t,j_t])$ be a collection of voter intervals; recall that $W=\{a_1,\ldots,a_t\}$, where $a_1,\ldots,a_t$ are ordered according to the canonical ordering.
We say that $\mathcal{I}$ is \myemph{\good} for $\solution$ if for all $\ell\in[1,t]$, it holds that $[i_\ell,j_\ell]$ is good for $a_\ell$ and for all pairs $\{a_r,a_s\}$ with $a_r,a_s\in W$ it holds that
\begin{compactenum}[(i)]
  \setcounter{enumi}{4}
  \item \label{good:4}if $a_s\in\incomp(a_r)$, then $[i_s,j_s]\cap[i_r,j_r]=\emptyset$ and
  \item \label{good:5}if $a_s\in\dom(a_r)$, then $[i_s,j_s]\subseteq [i_r,j_r]$ or\\ $[i_s,j_s]\cap[i_r,j_r]=\emptyset$.
\end{compactenum}
We now define a signature to compare two \good\ collections.
Let $\mathcal{I}\coloneqq \{[i_1,j_1],\ldots,[i_t,j_t]\}$ be a \good\ collection of intervals with respect to~$\solution$, then the signatures of $\mathcal{I}$ is defined as
$\Sig(\mathcal{I})=(j_1,\ldots,j_t,-i_1,\ldots,-i_t)$.
We say that $\mathcal{I}$ is \myemph{\optimal} if $\Sig(\mathcal{I})$ is lexicographically maximal.
\end{definition}
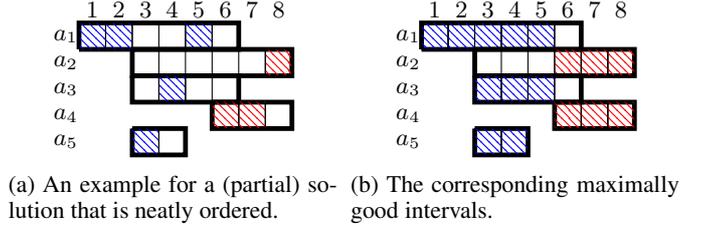
\begin{figure}[t!]
  \centering
 \begin{subfigure}[T]{.5\linewidth}
    \centering
	\begin{tikzpicture}[scale=0.35]
		\draw[profileframe] (0,1) grid (6,0);
		\draw[profileframe] (2,0) grid (8,-1);
		\draw[profileframe] (2,-1) grid (6,-2);
		\draw[profileframe] (5,-2) grid (8,-3);
		\draw[profileframe] (2,-3) grid (4,-4);
		\draw[approvingvoters] (0,0) -- (6,0) -- (6,1) -- (0,1) -- (0,0);
		\foreach \x in {-1,...,0}
		\draw [pattern=north west lines, pattern color=blue](1+\x,0) -- (2+\x,0) -- (2+\x,1) -- (1+\x,1) -- (1+\x,0);
		\draw [pattern=north west lines, pattern color=blue](4,1) -- (5,1) -- (5,0) -- (4,0) -- (4,1);
		\draw[approvingvoters] (2,0) -- (8,0) -- (8,-1) -- (2,-1) -- (2,0);
		\draw [pattern= north west lines,pattern color=red](7,0) -- (8,0) -- (8,-1) -- (7,-1) -- (7,0);
		\draw[approvingvoters] (2,-1) -- (6,-1) -- (6,-2) -- (2,-2) -- (2,-1);
		\foreach \x in {2,...,2}
				\draw [pattern=north west lines, pattern color=blue](1+\x,-1) -- (2+\x,-1) -- (2+\x,-2) -- (1+\x,-2) -- (1+\x,-1);
		\draw[approvingvoters] (5,-2) -- (8,-2) -- (8,-3) -- (5,-3) -- (5,-2);
		\foreach \x in {4,...,5}
						\draw [pattern=north west lines, pattern color=red](1+\x,-2) -- (2+\x,-2) -- (2+\x,-3) -- (1+\x,-3) -- (1+\x,-2);
		\draw[approvingvoters] (2,-3) -- (4,-3) -- (4,-4) -- (2,-4) -- (2,-3);
		\foreach \x in {1,...,1}
			\draw [pattern=north west lines, pattern color=blue](1+\x,-3) -- (2+\x,-3) -- (2+\x,-4) -- (1+\x,-4) -- (1+\x,-3);
		\foreach \n  in {1, ..., 8} {
		          \node[voterlabel] at (\n-0.5, 1.5) {$\n$};
		        }
		        \foreach \n  in {1,...,5} {
		          \node[alternativelabel] at (-0.5, -\n+1.5) {$a_{\n}$};
		      }
		      
	\end{tikzpicture}
	\caption{An example for a (partial) solution that is neatly ordered.}\label{fig:good1}
\end{subfigure}~~
\begin{subfigure}[T]{.5\linewidth}
    \centering
\begin{tikzpicture}[scale=0.35]
	\draw[profileframe] (0,1) grid (6,0);
		\draw[profileframe] (2,0) grid (8,-1);
		\draw[profileframe] (2,-1) grid (6,-2);
		\draw[profileframe] (5,-2) grid (8,-3);
		\draw[profileframe] (2,-3) grid (4,-4);
		\draw[approvingvoters] (0,0) -- (6,0) -- (6,1) -- (0,1) -- (0,0);
		\foreach \x in {-1,...,3}
		\draw [pattern=north west lines, pattern color=blue](1+\x,0) -- (2+\x,0) -- (2+\x,1) -- (1+\x,1) -- (1+\x,0);
		\draw[approvingvoters] (2,0) -- (8,0) -- (8,-1) -- (2,-1) -- (2,0);
		\foreach \x in {4,...,6}
						\draw [pattern=north west lines, pattern color=red](1+\x,-0) -- (2+\x,0) -- (2+\x,-1) -- (1+\x,-1) -- (1+\x,0);
		\draw[approvingvoters] (2,-1) -- (6,-1) -- (6,-2) -- (2,-2) -- (2,-1);
		\foreach \x in {1,...,3}
				\draw [pattern=north west lines, pattern color=blue](1+\x,-1) -- (2+\x,-1) -- (2+\x,-2) -- (1+\x,-2) -- (1+\x,-1);
		\draw[approvingvoters] (5,-2) -- (8,-2) -- (8,-3) -- (5,-3) -- (5,-2);
		\foreach \x in {4,...,6}
						\draw [pattern=north west lines, pattern color=red](1+\x,-2) -- (2+\x,-2) -- (2+\x,-3) -- (1+\x,-3) -- (1+\x,-2);
		\draw[approvingvoters] (2,-3) -- (4,-3) -- (4,-4) -- (2,-4) -- (2,-3);
		\foreach \x in {1,...,2}
			\draw [pattern=north west lines, pattern color=blue](1+\x,-3) -- (2+\x,-3) -- (2+\x,-4) -- (1+\x,-4) -- (1+\x,-3);
		\foreach \n  in {1, ..., 8} {
		          \node[voterlabel] at (\n-0.5, 1.5) {$\n$};
		        }
		        \foreach \n  in {1,...,5} {
		          \node[alternativelabel] at (-0.5, -\n+1.5) {$a_{\n}$};
		      }
		      
\end{tikzpicture}
	\caption{The corresponding \optimal\ intervals.}\label{fig:good2}
\end{subfigure}
\caption{Illustration of a \optimal\ collection of intervals corresponding to a solution; see \cref{def:good-optimal-intervals}. The hatched parts indicate the assignment.} \label{fig:sketchesgood}
\end{figure}
Next, we define what it means to be inbetween two alternatives and what it means to be usable.
\begin{definition}[Inbetween and usable alternatives]\label{def: U}
  We start by defining subsets of incomparable alternatives that are only approved by voters from a given range: 
  \myemph{$\inbetw(a,b)$} $= \{c \in \incomp(a)\cap \incomp(b) \mid \VV(c) \subseteq [\leftmost{a}, \rightmost{b}]\}$ and
  \myemph{$\inbetwS(a,b)$} consists of all alternatives from $\inbetw(a,b)$ that are not dominated by any other alternative in $\inbetw(a,b)$.
  Formally, $\inbetwS(a,b) = \level_{\inbetw(a,b)}(1)$.
  
  Let $c, c'$ be two (not necessarily distinct) alternatives and $[i,j]$ and $[i',j']$ two voter intervals such that $i' \le i \le j \le j'$, $[i,j]\subseteq \VV(c)$, and $[i', j']\subseteq \VV(c')$.
  We say that a third alternative~$a$ is \myemph{usable} for the six-tuple~$(c,i,j,c',i',j')$ if $a\in \subordinates(c)$, $[i,j]\cap \VV(a) \neq \emptyset$, and one of the following three conditions holds.
  \begin{compactenum}[(i)]
    \item $\leftmost{a} \ge i$, or
    \item $\leftmost{a} \in [i',i-1]$ and for each alternative~$b\in \dominators(a) \cap \subordinates(c') \cap \earlier(c)$, it holds that  $\leftmost{b} < i'$ and there exists an alternative~$b'\in \earlier(c')\cap \dominators(b)$, or
    \item $\leftmost{a} < i'$ and for each alternative~$b\in \dominators(a)\cap \incomp(c)$ it holds that $c\in \earlier(b)$.
  \end{compactenum}
  If an alternative $b$ leads to the second or third condition being violated we say that \myemph{$b$ blocks $a$ from being usable} (wrt.\ the six-tuple~$(c,i,j,c',i',j')$).
  We use \myemph{$\usableset(c,i,j,c',i',j')$} to denote the set consisting of all useful alternatives:
  \begin{align*}
  &\usableset(c,i,j,c',i',j') =\\
  &\begin{cases}
    \{a\in  \subordinates(c) \mid a \text{ is usable for } (c,i,j,c',i',j')\},\\
    \qquad\qquad~~  \text{ if } c' \in \dominators(c)\cup \{c\} \text{ and }[i,j]\subseteq \VV(c) \cap [i', j']\\
    \qquad \qquad~~ \text{ and } [i',j'] \subseteq \VV(c')\\
    \{a\in  \subordinates(c) \mid a \text{ is usable for } (c,i,j,c,i,j)\},\\
    \qquad\qquad~~  \text{ if } c'=i'=j'=0\\
    \emptyset, 
    \qquad \qquad \text{otherwise}.
  \end{cases}
  \end{align*}
  Finally, let \myemph{$\usablesetS(c, i,j,c', i', j')$} denote the set consisting of all alternatives from $\usableset(c,i,j,c',i',j')$ that are not dominated by others in~$\usableset(c,i,j,c',i',j')$, i.e., $\usablesetS(c, i,j,c', i', j') = \level_{\usableset(c, i,j,c', i', j')}(1)$.
\end{definition}
By the above definition it holds that $\inbetw(a,a)=\emptyset$.
Intuitively the alternatives in $\usableset(c,i,j,c',i',j')$ will be the alternatives dominated by $c$ that can be assigned voters on the interval $[i,j]$ in a \MNT\ solution where neither an alternative that is incomparable to $c$ is assigned voters on $[i,j]$ nor an alternative that is incomparable to $c'$ is assigned alternatives on $[i',j']$. \todoIC{Ok? Too vague? Kind of hard to intuitively explain U without the good intervals. S: Changed this, removed the references to tree-like structure since we do not use it anymore, but now it uses MNT.}
\cref{fig:InbetU} illustrates \cref{def: U}.
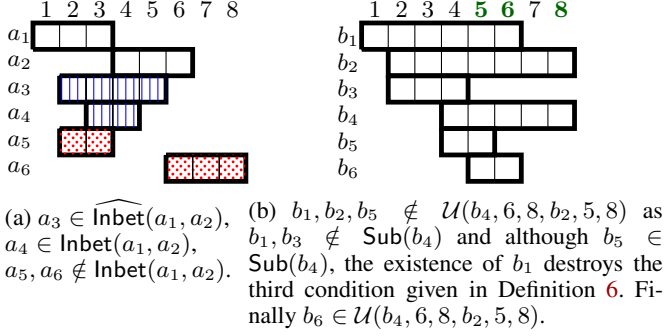
\begin{figure}[t!]
  \centering
  \begin{subfigure}[T]{.35\linewidth}
    \centering
    \begin{tikzpicture}[scale=0.35]
      \draw[profileframe] (0,1) grid (3,0);
      \draw[profileframe] (4,-1) grid (6,0);
      \draw[profileframe] (1,-1) grid (5,-2);
      \draw[profileframe] (2,-3) grid (4,-2);
      \draw[profileframe] (1,-3) grid (3,-4);
      \draw[profileframe] (5,-5) grid (8,-4);
      \draw[approvingvoters] (0,0) -- (3,0) -- (3,1) -- (0,1) -- (0,0);
      \draw[approvingvoters] (3,0) -- (6,0) -- (6,-1) -- (3,-1) -- (3,0);
      \draw[approvingvoters] (1,-1) -- (5,-1) -- (5,-2) -- (1,-2) -- (1,-1);
      \foreach \x in {0,...,3}
      \filldraw [pattern=vertical lines, pattern color=blue](1+\x,-1) -- (2+\x,-1) -- (2+\x,-2) -- (1+\x,-2) -- (1+\x,-1);
      \draw[approvingvoters] (2,-2) -- (4,-2) -- (4,-3) -- (2,-3) -- (2,-2);
      \foreach \x in {1,...,2}
      \filldraw [pattern=vertical lines, pattern color=blue](1+\x,-2) -- (2+\x,-2) -- (2+\x,-3) -- (1+\x,-3) -- (1+\x,-2);
      \draw[approvingvoters] (1,-3) -- (3,-3) -- (3,-4) -- (1,-4) -- (1,-3);
      \foreach \x in {0,...,1}
      \filldraw [pattern=crosshatch dots, pattern color=red](1+\x,-3) -- (2+\x,-3) -- (2+\x,-4) -- (1+\x,-4) -- (1+\x,-3);
      \draw[approvingvoters] (5,-4) -- (8,-4) -- (8,-5) -- (5,-5) -- (5,-4);
      \foreach \x in {4,...,6}
      \filldraw [pattern=crosshatch dots, pattern color=red](1+\x,-4) -- (2+\x,-4) -- (2+\x,-5) -- (1+\x,-5) -- (1+\x,-4);

      \foreach \n  in {1, ..., 8} {
        \node[voterlabel] at (\n-0.5, 1.5) {$\n$};
      }
       \foreach \n  in {1,...,6} {
        \node[alternativelabel] at (-0.5, -\n+1.5) {$a_{\n}$};
      }
    \end{tikzpicture}
    \caption{$a_3\in \inbetwS(a_1,a_2)$,\\
      $a_4\in \inbetw(a_1,a_2)$,\\
      $a_5, a_6\notin$ $\inbetw(a_1,a_2)$.}\label{fig:Inbet}
  \end{subfigure}~~
  \begin{subfigure}[T]{.63\linewidth}
    \centering
    \begin{tikzpicture}[scale=0.35]
        \draw[profileframe] (0,1) grid (6,0);
        \draw[profileframe] (1,-1) grid (8,0);
        \draw[profileframe] (1,-1) grid (4,-2);
        \draw[profileframe] (3,-3) grid (8,-2);
        \draw[profileframe] (3,-3) grid (5,-4);
        \draw[profileframe] (4,-5) grid (6,-4);
        \draw[approvingvoters] (0,0) -- (6,0) -- (6,1) -- (0,1) -- (0,0);
        \draw[approvingvoters] (1,0) -- (8,0) -- (8,-1) -- (1,-1) -- (1,0);
        \draw[approvingvoters] (1,-1) -- (4,-1) -- (4,-2) -- (1,-2) -- (1,-1);
        \draw[approvingvoters] (3,-2) -- (8,-2) -- (8,-3) -- (3,-3) -- (3,-2);
        \draw[approvingvoters] (3,-3) -- (5,-3) -- (5,-4) -- (3,-4) -- (3,-3);
        \draw[approvingvoters] (4,-4) -- (6,-4) -- (6,-5) -- (4,-5) -- (4,-4);
        \foreach \n  in {1, ..., 4} {
          \node[voterlabel] at (\n-0.5, 1.5) {$\n$};
        }
        \node[voterlabel] at (6.5, 1.5) {$7$};
        \foreach \n  in {5,6,8} {
                  \node[voterlabel] at (\n-0.5, 1.5) {\myemph{$\mathbf{\n}$}};
                }
        \foreach \n  in {1,...,6} {
          \node[alternativelabel] at (-0.5, -\n+1.5) {$b_{\n}$};
      }
      
    \end{tikzpicture} 
      \caption{$b_1,b_2, b_5\notin \usableset(b_4,6,8,b_2,5,8)$ as 
$b_1,b_3\notin \subordinates(b_4)$ and although $b_5\in \subordinates(b_4)$, the existence of $b_1$ destroys the third condition given in \cref{def: U}. Finally $b_6\in \usableset(b_4,6,8,b_2,5,8)$.}\label{fig:U}
    \end{subfigure}
    \caption{Illustration of the inbetween and usable alternatives from \cref{def: U}.} \label{fig:InbetU}
\end{figure}

\subsection{Technical lemmas}\label{sub:technical}
In this section we introduce some lemmas that relate to the structure of an optimal solution. The main observation is that any optimal solution can be translated into another optimal solution that is \MNT\ and admits a \optimal\ collection of voter intervals. 

\todoIH{The following proof sketch is not very informative. Also, I would rather call it idea... S: Better?}
\ifshort\begin{lemma}[\appsymb]\else\begin{lemma}\fi\label{lemmaMNO}
  Let $(\profiletuple, \misr, \csize)$ be an instance of \Monroe-\summw, $A \subseteq \aaa$ a subset of alternatives,
  and $(\committee, \repre)$ a partial solution wrt.~$[\ii, \jj]$.
  Then, we can construct another partial solution wrt.~$[\ii, \jj]$ that is \NT and monotone with respect to~$A$ such that the misrepresentation is at most $\misr(\repre)$.
\end{lemma}
\appendixsection{sub:technical}

\appendixproof{lemmaMNO}{
\begin{proof}
  We show how to construct such a partial solution in three steps, with each step supported by a separate claim.
  In the first step, we transform $(\committee, \repre)$ into one~$(\committee, \repre_1)$ which is monotone wrt.~$A$.
  In the second step, we transform $(\committee, \repre_1)$ into one~$(\committee, \repre_2)$ where each two incomparable alternatives are \NT. 
  In the final step, we transform $(\committee, \repre_2)$ into one~$(\committee, \repre_2)$ which is \NT.
  In each step, we also maintain the properties obtained in the previous step and make sure that the overall misrepresentation does not increase.
  For the sake of readability, we call a pair of alternatives that violates the monotonicity condition (wrt.~$A$) an \myemph{non-monotonic} pair; otherwise it is \myemph{monotone}.

  \begin{claim}\label{cl:monotone}
    There exists a partial solution~$(\committee_1, \repre_1)$ wrt.~$[\ii, \jj]$ that is monotone wrt.~$A$ such that $\misr(\repre_1)\le \misr(\repre)$.
  \end{claim}
  \begin{proof} \renewcommand{\qedsymbol}{(of \cref{cl:monotone})~$\diamond$}
    We start with $(\committee, \repre)$, i.e., let $\committee_1=\committee$ and $\repre_1=\repre$, and iteratively find a \emph{non-monotonic} pair of alternatives (wrt.~$\repre$) and make it monotone without introducing new non-monotonic pairs.
    To this end, let $\{a, b\}$ be a non-monotonic pair with $a\in \dominators(b)$
    such that no subordinate~$c$ of $b$ exists for which $\{b,c\}$ is non-monotonic
    and no dominator~$c'$ of $a$ exists for which $\{c',a\}$ is non-monotonic.
    That is, one of the following cases applies:
    \begin{compactenum}[(i)]
      \item $b \in W$ but $a \notin W$ \label{lem:monotone_no1} or
      \item $a, b \in A$ and there is a voter $u \in [\ii, \jj]$ who is happy with $b$ but no voter in $[\ii, \jj]$ is happy with $a$. \label{lem:monotone_no2}
    \end{compactenum}
    For case~\eqref{lem:monotone_no1}, we simply replace $b$ with $a$ in the partial solution, that is,
    let $\committee_1=(\committee\setminus \{b\}) \cup \{a\}$
    and $\repre_1(i) = a$ for all $i\in V$ with $\repre(i) = b$. 
    Clearly, after such modification,~$\{a,b\}$ becomes monotone wrt.~$\repre_1$, while the misrepresentation does not increase since $a$ dominates $b$.
    Since $(\committee,\repre)$ is a partial solution, it satisfies $k \ge |\repre^{-1}(b)| = |\repre^{-1}_1(a)|$.
    This means that $(\committee_1,\repre_1)$ is also a partial solution.
    
    For case~\eqref{lem:monotone_no2}, we exchange the inverse assignment of $a$ and $b$, that is,
    for all voter~$i\in [\ii,\jj]$, let $\repre_1(i) = a$ if $\repre(i) = b$,
    and $\repre_1(i) = b$ if  $\repre(i) = a$.
    Observe that after the reassignment, $\{a,b\}$ becomes monotone wrt.~$\repre_1$ and the misrepresentation does not increase since $a$ dominates~$b$.
    Moreover, since $(\committee,\repre)$ is a partial solution and we exchanged the assignment of voters to $a$ and $b$, the new tuple~$(\committee_1,\repre_1)$ is also a partial solution.

    Finally, due to the choice of $b$, we did not introduce any additional non-monotonic pair that involves $b$, and due to the choice of $a$, neither did we introduce any additional non-monotonic pair that involves $a$.
    Hence, we did not introduce any additional non-monotonic pair. 
    By repeating the above modifications as long as there are non-monotonic pairs,
    we will obtain a new partial solution that is monotone wrt.~$A$.
  \end{proof}

  Next, we show how to obtain an equally good partial solution such that every incomparable pair~$\{a,b\}$ of alternatives is \NT.
  \begin{claim}\label{cl:NT-1}
    There exists a partial solution~$(\committee_2,\repre_2)$ wrt.~$[\ii,\jj]$ that is monotone wrt.~$A$ such that $\misr(\repre_2) \le \misr(\repre)$ and each incomparable pair of alternatives is \NT.
  \end{claim}
  \begin{proof} \renewcommand{\qedsymbol}{(of \cref{cl:NT-1})~$\diamond$}
    By \cref{cl:monotone}, we can assume that $(\committee, \repre)$ is a a partial solution which is monotone wrt.~$A$, and let $\committee_2=\committee$ and $\repre_2=\repre$.
    We will iteratively find an incomparable and non-\NT\ pair of alternatives (wrt.~$\repre_2$) and make it \NT\ without introducing any incomparable and non-\NT\ pair of alternatives.
    To this end, let $\{a, b\}\subseteq \happyrepre_2([\ii,\jj])$ be a non-\NT\ pair of alternatives with $a\in \earlier(b)$, i.e., no alternative~$c\in \earlier(a)\cap \happyrepre_2([\ii,\jj])$ would form a non-\NT\ pair with $b$
    and no alternative~$c \in \happyrepre_2([\ii,\jj])$ with $b \in \earlier(c)$ would form a non-\NT\ pair with $a$.
    Then, we exchange the assignments of two specific voters~$v_a$ and $v_b$ by setting
    $v_a = \min(\happyrepre_2^{-1}(a))$ and $v_b=\max(\happyrepre_2^{-1}(b))$, and then 
    $\repre_2(v_a) = b$ and $\repre_2(v_b) = a$.
    We repeat the above exchange operation until $\{a,b\}$ is \NT, i.e., $\max(\happyrepre_2^{-1}(a)) < \min(\happyrepre_2^{-1}(b))$.
    Note that we can achieve this inequality since in each exchange we decrease $\max(\happyrepre_2^{-1}(a))$ and increase $\min(\happyrepre_2^{-1}(b))$. 
    Since the number of voters assigned to each alternative remains the same,
    tuple~$(\committee_2, \repre_2)$ is a partial solution.
    Since we did not change~$\committee_2$ and the voters that changed their assignments both approve of~$a$ and $b$,
    the obtained tuple~$(\committee_2,\repre_2)$ is monotone wrt.~$A$ such that $\misr(\repre_2) \le \misr(\repre)$.

    Next, we show that we did not introduce any incomparable and non-\NT\ pair of alternatives. 
    Towards a contradiction, suppose that $\{c,d\}\subseteq \repre_2([\ii,\jj])$ with $c\in \earlier(d)$ is \NT\ in $\repre$ but non-\NT\ in $\repre_2$.
    Since only the voters that were assigned to~$a$ or $b$ change their assignments,
    it follows that $\{c,d\}$ involves either~$a$ or~$b$.

    If $\{c,d\}$ involves~$a$, then $d = a$ since otherwise $\{c,d\}=\{a,d\}$ would have been non-\NT\ in $\repre$ as weal (note that $\max(\happyrepre^{-1}_2(a)) < \max(\happyrepre^{-1}(a))$), which contradicts our assumption.
    This implies that $\min(\happyrepre_2^{-1}(a)) < \max(\happyrepre^{-1}(c))$.
    By our choice of~$\{a,b\}$, we infer that $\min(\happyrepre_2^{-1}(a)) < \max(\happyrepre^{-1}(c)) < \min(\happyrepre^{-1}(a))$.
    By construction, it follows that $\min(\happyrepre_2^{-1}(a))$ is a voter who was previously assigned to $b$, which further implies that $\min(\happyrepre^{-1}(b)) \le \min(\happyrepre_2^{-1}(a)) < \max(\happyrepre^{-1}(c))$.
    Since $c\in \earlier(a)$ and $a \in \earlier(b)$,
    we have that $\{c,b\}$ is an incomparable and non-\NT\ pair, a contradiction to our choice of~$\{a,b\}$.
    
    Similarly, if $\{c,d\}$ involves~$b$, then $c=b$ since otherwise $\{c,d\}=\{c,b\}$ would have been non-\NT\ in $\repre$ as well (note that $\min(\happyrepre^{-1}_2(b)) > \min(\happyrepre^{-1}(b))$), which contradicts our assumption.
    This implies that $\min(\happyrepre^{-1}(d)) < \max(\happyrepre_2^{-1}(b))$.
    Since $\{b,d\}=\{c,d\}$ is \NT\ in $\repre$, voter~$\max(\happyrepre_2^{-1}(b))$ must have been assigned to $a$ in $\repre$.
    However, this would mean that $\max(\happyrepre^{-1}(a)) \ge  \max(\happyrepre_2^{-1}(b)) > \min(\happyrepre^{-1}(d))$, so $\{a,d\}$ is non-\NT\ in $\repre$, a contradiction to our choice of $\{a,b\}$ since $b\in \earlier(d)$.

    Summarizing, by repeating the above find an replace procedure, we obtain a new partial solution satisfying the desired properties stated in the claim. 
  \end{proof}

  \begin{claim}\label{cl:NT-2}
    There exists a partial solution~$(\committee_3, \repre_3)$ wrt.~$[\ii,\jj]$
    that is monotone wrt.~$A$ and \NT\ such that $\misr(\repre_2) \le \misr(\repre)$.
  \end{claim}

  \begin{proof}
    \renewcommand{\qedsymbol}{(of \cref{cl:NT-2})~$\diamond$}
    \todoIH{Prove me. Avoid mixed usages of different notations for the same meaning.}
    \todoIC{Done?}
    By \cref{cl:NT-1}, we can assume that $(\committee,\repre)$ is a partial solution which is monotone wrt.~$A$ and each incomparable pair of alternatives is \NT. Let $W_3=W$ and $\repre_3=\repre$. We will iteratively find alternatives that are not \NT\ together with some alternative that dominates them and make it \NT\ with any alternative that dominates them. To this end we define $\conflictsetNO(c, \repre) = \{i \in [\leftmost{c}, \max(\happyrepre^{-1}(c))] \mid \repre(i)\in\dominators( c) \text{ and } \misr(i, c) = 0\}$. Should the alternative $c$ not be assigned any happy voters we will set $\conflictsetNO(c, \repre) = \emptyset$. This is the set of voters whose representing alternatives dominate $c$ and violate neat ordering for $c$ at that voter. If $\conflictsetNO(c, \repre)$ is empty, then $c$ and any alternative that dominates it are \NT.
    
    Let $a \in W_3$ be an alternative such that $\conflictsetNO(a, \repre_3) \neq \emptyset$ and  there is no $b \in W_3$ such that $a\in\dominators(b)$ and $\conflictsetNO(b, \repre_3) \neq \emptyset$. Moreover, there is no $c \in W\cap\earlier(a)$ that satisfies $\conflictsetNO(c, \repre_3) \neq \emptyset$. Informally speaking, we choose an alternative $a$ such that each of its subordinates is \NT\ with respect to its dominators and among alternatives that satisfy this condition we choose the one that starts earliest.  
    
      We now exchange assigned voters in $\repre_3$ such that the following conditions are satisfied:
      \begin{compactenum}[(a)]
        \item misrepresentation is not increased and \label{lem:NOopt}
        \item $|\conflictsetNO(a, \repre_3)|$ is decreased and \label{lem:NOa}
        \item for every $b \in W_3$, if $b\in\subordinates(a)$ or $b \in \earlier(a)$, then $\conflictsetNO(b, \repre_3) = \emptyset$ and\label{lem:NOb}
        \item for every $c, d \in W_3$, if $c$ and $d$ are incomparable, then $\{c, d\}$ is \NT\ in $(W, \repre_3)$. \label{lem:NOc}\\
      \end{compactenum}
    
      \noindent Let $i \coloneqq \max(\conflictsetNO(a, \repre_3))$ and let $b \coloneqq \repre_3(i)$ and let $j \coloneqq \max(\happyrepre_3^{-1}(a))$. By choice of $a$, $\happyrepre_3^{-1}(a) \neq \emptyset$. We exchange voters resulting in the same partial assignment except $\repre_3(i) \coloneqq a$ and $\repre_3(j) \coloneqq b$. For simplicity's sake we will refer to the assignment before this exchange as $\alpha_3$. We proceed to show that Properties \eqref{lem:NOopt}--\eqref{lem:NOc} hold.
    
      For Property \eqref{lem:NOopt} note that $V(a) \subseteq V(b)$ because $b\in\dominating(a)$, so $j$ is happy to be reassigned from $a$ to $b$. For the voter $i$, notice that $\max(\conflictsetNO(a)) \in [\leftmost{a}, \max(\happyrepre^{-1}(a))] \subseteq V(a)$ and thus $i$ is happy to be represented by~$a$. Since only these two voters were reassigned misrepresentation was not increased.\\
    
      For Property \eqref{lem:NOa}, note that $i \notin \conflictsetNO(a, \repre_3)$, because $\repre_3(i) = a$, which does not dominate $a$. Thus we have an element that was removed from the conflict set. Next we show that no elements are added to $\conflictsetNO(a, \repre_3)$. Because $i < j$, we must have that $\max(\happyrepre_3^{-1}(a)) < j$ and moreover $[\leftmost{a}, \max_1(\happyrepre_3^{-1}(a))] \subset [\leftmost{a},j]$. The only voter whose representative is changed to something that dominates $a$ is $j$, but $j = \max(\happyrepre_3^{-1}(a)) > \max(\happyrepre_3^{-1}(a))$ and thus $j$ is not in $\conflictsetNO(a, \repre_3)$. \\

      For Property \eqref{lem:NOb}, first note that by our choice of $a$, every alternative in $c \in \dom(a)$ satisfies $\conflictsetNO(c, \hat{\alpha}_3) = \emptyset$. Similarly, every alternative $c \in W$ such that $c \in \earlier(a)$, satisfies $\conflictsetNO(c, \hat{\alpha}_3) = \emptyset$ before the exchange. %
    
      Assume there is some $c$ that violates Property \eqref{lem:NOb}. Because we only change assigned voters of $a$ and $b$, any new element in $\conflictsetNO(c, \repre_1)$ must be either $i$ or $j$.
    
      We proceed in 2 cases:
    
      \textbf{Case \ref{lem:NOb}.1:} $c\in\subordinates(a)$.
      
      First assume $a \in \happyrepre_3(\conflictsetNO(c, \repre_3))$. By our choice of $a$, we know that $a$ and $c$ are \NT\ in $(W, \alpha_3)$ and moreover that $\conflictsetNO(c, \alpha_3)$ is empty. Thus there must be a voter $i' \in [\leftmost{c}, \max(\happyrepre_3^{-1}(c))]$ such that $\repre_3(i') = a$ and $i'$ is happy with $a$. However, the only voter newly assigned to $a$ is $i$. The voter $i$ was happily assigned to $b$ beforehand though. Thus $c, b$ are not \NT\ in $(W, \alpha_3)$. This contradicts $\conflictsetNO(c, \alpha_3) = \emptyset$.
    
      Next assume $b \in \happyrepre_3(\conflictsetNO(c, \repre_3))$. By our choice of $a$, we know that $b$ and $c$ are \NT\ in $(W, \alpha_3)$ and moreover that $\conflictsetNO(c,\alpha_3)$. Thus there must be a voter $i' \in [\leftmost{c}, \max(\happyrepre_3^{-1}(c))]$ such that $\repre_1(i') = b$ and $i'$ is happy with $b$. However, the only voter newly assigned to $b$ is $j$, which was previously assigned to $a$. Then $a$ and $c$ are not \NT\ in $(\committee_3,\alpha_3)$, however. This contradicts $\conflictsetNO(c, \hat{\alpha}_3)$ being empty.

      \textbf{Case \ref{lem:NOb}.2:} $c \in \earlier(a)$.
      
      Note that $a \in \happyrepre_3(\conflictsetNO(c, \repre_3))$ is not possible, because $c$ and $a$ are incomparable. %

      Lets assume $b \in \happyrepre_3(\conflictsetNO(c, \repre_3))$. By our choice of $c$, we know that $b$ and $c$ are \NT\ in $(W, \alpha_3)$ and moreover that $\conflictsetNO(c, \alpha_3) = \emptyset$. Thus there must be a voter $i' \in [\leftmost{c}, \max(\happyrepre_3^{-1}(c))]$ such that $\repre_3(i') = b$ and $i'$ is happy with $b$. The only voter newly assigned to $b$ is $j$, which was previously assigned to $b$.
    
      Since $\min(\happyrepre_3^{-1}(a)) \leq j < \max(\happyrepre_3^{-1}(c))$, before the exchange and $a$ and $c$ are \NT, we must have that $\leftmost{a} < \leftmost{c}$. This contradicts our assumption that $c \in \earlier(a)$.\\
    
      For Property \eqref{lem:NOc}, first note that because we only swap the representatives of $i$ and $j$, any pair that violates Property \eqref{lem:NOc} must consist of either $a$ or $b$ and an alternative that represents a happy voter in $[i + 1, j - 1]$. %

      Let us proceed in two cases based on whether $\{a,c\}$ or $\{b,c\}$ is not \NT\ in $(W, \repre_3)$

      \textbf{Case \ref{lem:NOc}.1:} \textit{$a$ and $c$ are incomparable and not \NT\ in $(W, \repre_3)$.}
      
      If $a \in \earlier(c)$, then $\max(\happyrepre_3^{-1}(a)) \leq j\leq\min(\happyrepre_3^{-1}(c))$ as  $a, c$ are \NT\ in $(W, \repre_1)$, contradicting our case assumption. Thus we must have that $c \in \earlier(a)$ and $ \max(\happyrepre_3^{-1}(c)) > \min(\happyrepre_3^{-1}(a))$. As the only voter that is newly represented by $a$ is $i$, it follows that $\max(\happyrepre_3^{-1}(c)) >i$.
    
    Let us first assume $b$ and $c$ are incomparable and thus \NT\ in $(W, \repre)$. From $\max(\happyrepre^{-1}(c)) > \min{\happyrepre^{-1}(b)}$ we can deduce that $b \in \earlier(c)$ and thus $\rightmost{a} < \rightmost{b} < \rightmost{c}$. However, this contradicts $c \in \earlier(a)$.
    
    Next let us consider the case where $b\in\dominators(c)$. Because $c \in \earlier(a)$ and $a$ and $c$ are incomparable, $\conflictsetNO(c, \alpha_3) = \emptyset$ and moreover, the pair $\{b,c\}$ is \NT\ in $(W, \alpha_3)$.
    
    However, $i \in [\leftmost{a},\max(\happyrepre_3^{-1}(c))] \subset [\leftmost{c}, \max(\happyrepre^{-1}(c))]$,  and $i$ was happy with $b$. This implies that $b$ and $c$ are not \NT\ in $(W, \alpha_3)$, a contradiction to our original assumption.
    
    Finally, we cannot have that $c\in\dominators(b)$, because then $c\in\dominators(a)$ as well, a contradiction to our case assumption. 

    \textbf{Case \ref{lem:NOc}.2:} \textit{$b$ and $c$ are incomparable and not \NT\ in $(W, \repre_3)$.}

    First assume $c \in \earlier(b)$. Then $\max(\happyrepre_3^{-1}(c))< \min(\happyrepre_3^{-1}(b))$, as the minimum over the voters assigned to $b$ has either increased or stayed the same and therefore $\{b,c\}$ are \NT\ in $(W, \repre_3)$, a contradiction to our case assumption.

    Thus it must hold that $b \in \earlier(c)$. If $\max(\happyrepre_3^{-1}(b)) = \max(\happyrepre_3^{-1}(b))$ then $b$ and $c$ are also \NT\ in $(W, \repre_3)$. Thus it must hold that $\max(\happyrepre_3^{-1}(b)) = j$. If $a$ and $c$ are incomparable, it follows that $a\in\earlier(c)$ because $b\in\earlier(c)$. Then because $a$ and $c$ are \NT\ in $(W,\alpha_3)$, it follows that $\max(\happyrepre_3^{-1}(a))\leq j < \min(\happyrepre^{-1}(c))$ and $b$ and $c$ are also \NT\ in $(W,\repre_3)$, a contradiction to our case assumption. Therefore $a$ and $c$ must be comparable. It follows that $c\in\dominators(a)$, as otherwise $c\in\subordinates(b)$, a contradiction to our case assumption. Because $i = \max(\conflictsetNO(a, \alpha_3))$ and $b = \alpha_3(i)$, there is no voter $i' \in [i+1, \max(\hat{\alpha}_3^{-1}(a))]$ such that $\alpha(i') = c$ and $i'$ is happy with $c$. 
    
    Moreover, because $b$ and $c$ are \NT in $(W,\alpha_3)$, $\max(\hat{\alpha}_3^{-1}(b)) < \min(\hat{\alpha}_3^{-1}(c))$ holds before the exchange. Since $j<\min(\happyrepre_3^{-1}(c)$, it follows that every voter who is happy with $c$ must be after $j$ and therefore $b$ and $c$ are \NT, contradicting our case assumption.\\
    
    Because we do not change $W$ and both $a$ and $b$ have a voter $v \in [\ii, \jj]$ that is happy with them, the solution $(W, \repre_3)$ is still monotone.
    
    Thus we have shown that $\repre_3$ satisfies properties \eqref{lem:NOopt}--\eqref{lem:NOc}. By repeatedly creating such a new assignment, $\repre_3$ will satisfy $\conflictsetNO(c, \repre_3) = \emptyset$ for all $c\in W_3$. By property \eqref{lem:NOopt} its misrepresentation is at least as good as that of $(W, \repre)$ and by Properties \eqref{lem:NOb} and \eqref{lem:NOc} incomparable alternatives remain \NT\ and no alternative ``before" $c$ has $\conflictsetNO(c, \repre^*) \neq \emptyset$. By repeating this process, we will eventually reach an \NT\ solution. 
  \end{proof}

We have shown that given a partial solution $(W,\repre)$ we can find a partial solution that is monotone with respect to $A$ without increasing misrepresentation in \cref{cl:monotone}. Then given this monotone partial solution we can find a partial solution that is monotone with respect to $A$ and \NT\ using \cref{cl:NT-1} and \cref{cl:NT-2}. Therefore we have shown the lemma.\end{proof}

}
\ifshort\begin{observation}[\appsymb]\else\begin{observation}\fi\label{LemmaInt}
  Let $(\committee,\repre)$ be an \NT\ partial solution  wrt.\ $[\ii, \jj]$. Then there exists a unique \optimal\ collection of intervals $\mathcal{I}$.
\end{observation}

\appendixproof{LemmaInt}{
\begin{proof}
  We start with assigning to each alternative $a_l\in W$ the interval $[\min(\happyrepre^{-1}(a_l)),\max(\happyrepre^{-1}(a_l))]$. Note that the resulting collection of intervals is \good\ due to the neat ordering. Therefore there exists a \good\ collection of intervals for an \NT\ solution. 
  Next, by the definition of signatures, we observe that 
  for each two collections of \good\ intervals $\mathcal{I}_1$ and $\mathcal{I}_2$ of $(\committee, \repre)$ with the same signature, it holds that $\mathcal{I}_1=\mathcal{I}_2$. 
  Obviously in the lexicographical ordering any two signatures are comparable. Since the number of 
  signatures is finite (there are $O(n^{2k})$ many), there is a unique lexicographically maximal signature, and hence a unique \optimal\ collection of intervals.
\end{proof}
}

Intuitively, the next lemma guarantees that every voter that could be part of some interval is part of at least one interval.
\todoIH{Add a proof idea for the next two lemmas? Done?}
\ifshort\begin{lemma}[\appsymb]\else\begin{lemma}\fi\label{LemmaIntvotcontainment}
  Let $\solution$ be an \NT\ partial solution wrt.\ $[\ii, \jj]$ and $\mathcal{I}=([i_1,j_1],\ldots,[i_t,j_t])$ the corresponding \optimal\ collection of intervals.
  Then, for each \hassignable\ voter $v\in [\ii, \jj]$
  there exists an alternative $a_x\in \happyrepre([\ii, \jj])$
  such that $v\in[i_x,j_x]$.
\end{lemma}

\appendixproof{LemmaIntvotcontainment}{
\begin{proof}
  Suppose, for the sake of contradiction, that $v\in [\ii, \jj]$ is an \hassignable\ voter who is not part of any interval in~$\mathcal{I}$.
  We aim to show the contradiction that $\mathcal{I}$ is not maximally good for $\solution$ by providing one whose signature is lexicographically larger than $\Sig(\mathcal{I})$.
  To this end, let $u$ be a voter that minimizes $|v-u|$ such that there exists
  an alternative~$a_z$, $z\in [t]$, with $v \in \VV(a_z)$ and $u\in [i_z,j_z]$; we choose an arbitrary one if there are more than one such voter.
  Formally,
  \begin{align*}
    u=\argmin_{u' \in \bigcup\limits_{z\in [t]\colon v\in \VV(a_z)} [i_z,j_z]} |v-u'|.
  \end{align*}
  Then, let us choose an alternative~$a_p$, $p\in [t]$, such that $u\in [i_p,j_p]$, $v\in \VV(a_p)$, and no dominator~$a_{z}\in \dominators(a_p)$ has $u\in [i_{z}, j_{z}]$.
  We only consider the case when $u < v$ as the proof for the case when $v < u$ works similarly.
  Note that, by definition, it holds that $u=j_p$.
  We claim that we can extend the interval of $a_p$ to $[i_p, v]$ and obtain a good collection of intervals whose signature is lexicographically larger than the one for $\mathcal{I}$.
  
  First of all, due to our choice of $u$, we observe the following for all $z\in [t]$:
  \begin{align}\label{eq:lemmaIntvotcontainment}
    \nonumber & \text{if } [i_z, j_z]\cap [u+1, v-1] \neq \emptyset, \\
 &   \text{then } [i_z, j_z] \subseteq [u+1, v-1]    \text{ and } a_z \in \subordinates(a_p). 
  \end{align}
  To show the above, consider an arbitrary~$z\in [t]$ such that $[i_z,j_z]\cap [u+1, v-1] \neq \emptyset$.
  Then, $j_z \le v-1$ since $v$ is not part of any interval of~$\mathcal{I}$.
  Moreover, $i_z \ge u+1$ as otherwise $[i_z,j_z]$ and $[i_p, j_p]$ would have a non-empty intersection (at $u=j_p$), and hence by the definition of good intervals and by our choice of $a_p$ we have that $a_p \in \dominators(a_z)$ and $[i_z, j_z]\subseteq [i_p,j_p] = [i_p, u]$, a contradiction to our choice that $[i_z,j_z]\cap [u+1, v-1]\neq \emptyset$.
  Summarizing, we have that $[i_z,j_z]\subseteq [u+1, v-1]$, achieving the first part of the implication in~\eqref{eq:lemmaIntvotcontainment}.
  Next, by our choice of $u$, we know that $v\notin \VV(a_z)$ since $j_p = u < i_z \le j_z$ and hence $|v-j_z| < |v-u|$.
  This implies that $\rightmost{a_z} < \rightmost{a_p}$.
  Since $j_p = u < u+1 \le i_z$ and $\rightmost{a_z} < \rightmost{a_p}$,
  we infer that $\leftmost{a_z} \ge \leftmost{a_p}$ as otherwise $a_z \in \earlier(a_p)$ which contradicts the neat ordering of $\solution$; note that both $a_z$ and $a_p$ are assigned at least one happy voter. 
  Hence, $a_z \in \subordinates(a_p)$.

  Now, let $\mathcal{I'}$ be identical to $\mathcal{I}$ except that the interval of $a_p$ is changed from $[i_p,j_p]$ to $[i_p, v]$.
  Clearly, $\Sig(I')$ is lexicographically larger than $\Sig(I)$.
  If we can show that $\mathcal{I}'$ is good for $\solution$, then we obtain a contradiction that $\mathcal{I}$ is \optimal\ for $\solution$. 
  In the remainder of the proof, we aim to show that $\mathcal{I}'$ is good for $\solution$, i.e., it satisfies the six conditions given in \cref{def:good-optimal-intervals}.
  By definition, at least one voter is happy with $a_p$, because otherwise $[i_p,j_p]=\emptyset$ would need to hold in $\mathcal{I}$. This however would contradict our initial assumption and therefore condition \eqref{good:3} is satisfied.
  Clearly $[i_p, v]$ satisfies condition~\eqref{good:1} for $a_p$ since interval~$[i_p,j_p]$ is good for $a_p$ and $v\in \VV(a_p)$.
  Moreover, by \eqref{eq:lemmaIntvotcontainment}, every alternative whose interval intersects with $[u+1,v-1]$ is a subordinate of $a_p$.
  Hence, $[i_p,v]$ also satisfies condition~\eqref{good:2}. 
  Since $v\in [\ii,\jj]$, we have that $[i_p,v]\subseteq [\ii,\jj]$, satisfying condition~\eqref{good:6}.
  Interval~$[i_p,v]$ also satisfies condition~\eqref{good:4} since $[i_p,i_q]$ is good and every alternative whose interval intersects $[u+1,v-1]$ is a subordinate of~$i_p$; see \eqref{eq:lemmaIntvotcontainment}.

  It remains to consider the last condition, and let $a_s$ be an arbitrary alternative with $a_s \in \subordinates(a_p) \cup \dominators(a_p)$.
  Since $\mathcal{I}$ is good for~$\{a_s,a_p\}$ and $\{a_s,a_x\}$ and $[i_p,j_p]\cap [i_x,j_x] =\emptyset$,
  it follows that $[i_s,j_s]\subseteq [i_p,j_p]\cup \{i_x,j_x\}$ or $[i_s,j_s]\cap ([i_p,j_p] \cup [i_x,j_x])=\emptyset$.
  
  If $[i_s,j_s]\subseteq [i_p,j_p] \cup [i_x,j_x]$ or $[i_s,j_s] \cap [j_p+1, i_x-1] \neq \emptyset$, then by \eqref{eq: lemmaintintcontainment}, we have $[i_s,j_s]\subseteq [i_p, j_x]$.
  Otherwise, we have that $[i_s,j_s]\cap [i_p,j_x] = \emptyset$.
  Both cases show that $[i_p, v]$ satisfies condition~\eqref{good:5}.
  Since all other intervals have not changed, they satisfy the six conditions as well.
  Altogether, we show that $\mathcal{I}'$ is good for $\solution$, a contradiction to $\mathcal{I}$ be \optimal.
  
  Since $\mathcal{I}$ is good for $\{a_s,a_p\}$, either $[i_s,j_s]\subseteq [i_p,j_p]$ or  $[i_s,j_s]\cap [i_p,j_p] =\emptyset$.
  If $[i_s,j_s]\subseteq [i_p,j_p]$ or $[i_s,j_s] \cap [u+1, v-1] \neq \emptyset$, then  by \eqref{eq:lemmaIntvotcontainment} we have $[i_s,j_s]\subseteq [i_p,v]$ since $[i_p,j_p]\subseteq [i_p, v]$.
  Otherwise, we immediately have that $[i_s,j_s]\cap [i_p,v] = \emptyset$ since $u=i_p$ and no interval covers $v$.
  Altogether, it follows that $[i_p, v]$ satisfies condition~\eqref{good:5}.
  Since all other intervals have not changed, they satisfy the six conditions as well.
  Altogether, we show that $\mathcal{I}'$ is good for $\solution$, a contradiction.
  This concludes the proof.
\end{proof}
}

\toappendix{
The next lemma guarantees that the interval of a dominated alternative is contained in the interval of one of its dominators.

\begin{lemma}\label{LemmaIntintcontainment}
  Let $\solution$ be an \NT\ partial solution wrt.~$[\ii, \jj]$, $\mathcal{I}=([i_1,j_1],\ldots, [i_t,j_t])$ the corresponding \optimal\ collection of intervals, and $a_x\in \committee$ an alternative such that $\dominators(a_x)\cap \committee\neq\emptyset$. If there exists a dominator~$a_y\in \dominators(a_x)\cap \committee$ such that $[i_y,j_y]\neq\emptyset$, then there exists $a_z\in \dominators(a_x)\cap W$ such that $[i_x,j_x]\subseteq[i_z,j_z]$.
\end{lemma}

\begin{proof}
  Let $\solution$, $\mathcal{I}$, and $a_z$ be as defined such that there exists an alternative~$a_y\in \dominators(a_x)\cap \committee$ with $[i_y,j_y]\neq \emptyset$. 
  Suppose, for the sake of contradiction, that for each dominator~$a_y \in \dominators(a_x)\cap \committee$ with $[i_y,j_y]\neq \emptyset$ it holds that $[i_x, j_x] \not\subseteq [i_y, j_y]$.
  The proof works similarly to the one for \cref{LemmaIntvotcontainment}.
  We aim to find an alternative in the committee and extend its interval without violating the goodness, which is a contradiction to $\mathcal{I}$ being \optimal.  
  We start with choosing such an alternative.
  Let $a_{p}\in\dominating(a_x) \cap \committee$ with $[i_p, j_p] \neq \emptyset$ be an alternative that satisfies the following two properties:
  \begin{inparaenum}[(L1)] 
    \item $j_{p} < i_x$ and 
    \item there exists no $a_z\in\dominators(a_{p})\cap \committee$ such that $[i_p,j_p]\subseteq[i_z,j_z]$ and $j_z<i_x$.
  \end{inparaenum}
  Subject to these conditions, we choose~$a_{p}$ that minimizes $i_x-j_{p}$.
  If no such alternative exists, then we choose $a_{p}$ such that: 
  \begin{inparaenum}[(R1)]
    \item $j_x<i_{p}$ and
    \item there exists no $a_z\in\dominators(a_{p})\cap \committee$ such that $[i_p,j_p]\subseteq[i_z,j_z]$ and $j_x<i_z$.
  \end{inparaenum}
  Subject to these conditions, we choose $a_{p}$ that minimizes $i_{p}-j_x$.
  Note that by our assumption about~$a_x$, such an alternative~$a_p$ can be found using the above process. Without loss of generality assume that the found alternative~$a_{p}$ satisfies $j_{p}<i_x$.
  
  Next, due to our choice of $a_{p}$, we observe the following for all $z\in[t]$:
  \begin{align}\label{eq: lemmaintintcontainment}
    \nonumber & \text{if } [i_z, j_z]\cap [j_{p}+1,i_x-1] \neq \emptyset, \\
              &  \text{then } [i_z,j_z]\subseteq[j_{p}+1,i_x-1]    \text{ and }a_{z}\in\subordinates(a_p). 
  \end{align}
  The proof is analogous to the one for \eqref{eq:lemmaIntvotcontainment}.
  We show this for the sake of completeness.
  Consider an arbitrary $z\in[t]$ with $[i_z, j_z]\cap [j_{p}+1,i_x-1] \neq \emptyset$.
  First, we observe that $j_z \le i_x-1$ as otherwise due to $\mathcal{I}$ being good we would have that $[i_x,j_x]\subseteq[i_z,j_z]$ which contradicts our assumption about $a_x$. 
  Next, we observe that $i_z\ge j_p+1$ as otherwise due to the goodness of $\mathcal{I}$
  we would have $[i_p,j_p] \subseteq [i_z,j_z]$, which implies that
  there exits another dominator~$a_{z'}\in\dominators(a_{x})\cap \committee$ such that $[i_{p},j_{p}]\subseteq[i_{z'},j_{z'}]$ and $j_{z'} < i_x$, a contradiction to our choice of $a_p$.
  Altogether, we have that $[i_z,j_z]\subseteq [i_p+1, i_x-1]$, achieving the first part of the implication in~\eqref{eq:lemmaIntvotcontainment}.
  We next show that $a_z\in\subordinates(a_p)$. 
  Suppose, towards a contradiction, that $a_z\notin\subordinates(a_p)$. 
  Then, there are two cases:
  \begin{compactenum}[(a)]
    \item $a_z\in\dominators(a_p)$, or \label{LemmaIntintcontainment_case2}
    \item $a_z\in\incomp(a_p)$.\label{LemmaIntintcontainment_case1}
  \end{compactenum}
  First, consider Case~\eqref{LemmaIntintcontainment_case2}, which implies that $a_z \in \dominators(a_x)$ since $a_p\in \dominators(a_x)$, a contradiction to $a_p$ minimizing $i_x-j_p$.   
  Now, consider Case~\eqref{LemmaIntintcontainment_case1}.
  Recall that $[i_p,j_p]\neq \emptyset$ and $[i_z, j_z]\neq \emptyset$.
  By the goodness of~$\mathcal{I}$, we infer that $a_p$ precedes $a_z$.
  By the neat ordering of the solution, we further infer that $a_p \in \earlier(a_z)$.
  This means that $\rightmost{a_z} > \rightmost{a_p} \ge \rightmost{a_x}$; the last inequality holds since $a_p\in \dominators(a_x)$.
  By our choice of $a_p$, we know that $\leftmost{a_z} > \leftmost{a_x}$ as otherwise $a_z \in \dominators(a_x)\cap \committee$ and $a_p$ would not minimize~$i_x-j_p$.
  This means that $a_x \in \earlier(a_z)$, a contradiction to the neat ordering since $j_z \le i_x-1$.
  Since in both cases we obtain a contradiction, we infer that $a_z \in \subordinates(a_p)$, concluding the proof for \eqref{eq: lemmaintintcontainment}.
  
  Now, let $\mathcal{I}'$ be identical to $\mathcal{I}$ except the interval of $a_p$ is changed from $[i_{p},j_{p}]$ to $[i_{p},j_x]$.
  Clearly, $\mathcal{I}'$ is lexicographically larger than $\mathcal{I}$.
  If we can show that $\mathcal{I}'$ is \good\ for $\solution$,
  then we obtain a contradiction that $\mathcal{I}$ is \optimal.
  The proof of $\mathcal{I}'$ being good for~$\solution$, i.e., it satisfies the six conditions given in \cref{def:good-optimal-intervals}, works analogously to the one for \cref{LemmaIntvotcontainment}. 
  
  Since $\mathcal{I}$ is \good\ and $[i_{p},j_{p}]\neq\emptyset$ it follows that $\happyrepre^{-1}(a_{p})\neq\emptyset$ and therefore condition~\eqref{good:3} is satisfied.
  Since $[i_{p},j_{p}]$ is \good, $[i_x,j_x]\subseteq\VV(a_x)$ and $a_x\in\subordinates(a_{p})$, it follows that $[i_{p},j_x]\subseteq\VV(a_{p})$ and $\happyrepre^{-1}\subseteq[i_{p},j_{p}]\subseteq[i_{p},j_x]\subseteq\VV(a_{p})$.
  Therefore condition \eqref{good:1} is satisfied.
  Let $v \in [i_{p}, j_x]$ be some happy voter. 
  If $v \in [i_{p}, j_{p}]$, then Condition \eqref{good:2} is satisfied because $\mathcal{I}$ is a collection of good intervals.
  If $v \in [j_{p} + 1, j_x]$, then by \eqref{eq: lemmaintintcontainment},
  we have that $[i_{\repre(v)},j_{\repre(v)}]\subseteq[j_{p}+1,i_x-1]$ and, $a_{\repre(v)}\in\subordinates(\repre(i))$. Thus Condition \eqref{good:2} is satisfied as well.
  Due to $[j_{p}+1,j_x]\subseteq[\ii,\jj]$ it follows that condition \eqref{good:6} is satisfied.

  Condition \eqref{good:4} is satisfied by $[i_p,j_x]$ since $[i_p,j_p]$ is good and \eqref{eq: lemmaintintcontainment} implies that every alternative that intersects~$[j_{p}+1, i_x-1]$ is a subordinate of~$a_p$.
  It remains to consider the last condition, and let $a_s$ be an arbitrary alternative with $a_s \in \subordinates(a_p) \cup \dominators(a_p)$.
  Since $\mathcal{I}$ is good for~$\{a_s,a_p\}$ and $\{a_s,a_x\}$ and $[i_p,j_p]\cap [i_x,j_x] =\emptyset$,
  it follows that $[i_s,j_s]\subseteq [i_p,j_p]\cup \{i_x,j_x\}$ or $[i_s,j_s]\cap ([i_p,j_p] \cup [i_x,j_x])=\emptyset$.
  
  If $[i_s,j_s]\subseteq [i_p,j_p] \cup [i_x,j_x]$ or $[i_s,j_s] \cap [j_p+1, i_x-1] \neq \emptyset$, then by \eqref{eq: lemmaintintcontainment}, we have $[i_s,j_s]\subseteq [i_p, j_x]$.
  Otherwise, we have that $[i_s,j_s]\cap [i_p,j_x] = \emptyset$.
  Both cases show that $[i_p, v]$ satisfies condition~\eqref{good:5}.
  Since all other intervals have not changed, they satisfy the six conditions as well.
  Altogether, we show that $\mathcal{I}'$ is good for $\solution$, a contradiction to $\mathcal{I}$ be \optimal.  
  This concludes the proof.
\end{proof}

}
\todoIH{Didn't we talk that the intuitive description here is not quite clea? Done?}
If a dominated alternative is not included in the interval of its dominator, the following lemma tells us that it must have another dominating alternative that contains it. Moreover, this alternative must be assigned happy voters on a specific interval. Contrapositively, we can use this lemma to show that no other dominator can contain the interval of the dominated alternative.
\ifshort\begin{lemma}[\appsymb]\else\begin{lemma}\fi\label{LemmaLongInt}
  Let $(W,\repre)$ be an \NT\ partial solution wrt.\ $[\ii, \jj]$ and $\mathcal{I}=([i_1,j_1],\ldots,[i_t,j_t])$, the corresponding \optimal\ collection of intervals.
  For all pairs of alternatives~$a_x,a_y\in\happyrepre([\ii,\jj])$ such that $a_y\in\dominating(a_x)$ and $[i_x,j_x]\not\subseteq[i_y,j_y]$, there exists an~$a_z\in\dominating(a_x)\setminus\dom(a_y)$ which satisfies $[i_x, j_x] \subseteq [i_z, j_z], [i_z, j_z] \cap [i_y, j_y] = \emptyset$ and the following:
  \begin{compactenum}[(1)]
    \item If $j_y<i_x$, then $\happyrepre^{-1}(a_z)\cap[j_y+1,i_x-1]\neq\emptyset$.
    \item If $j_x<i_y$, then $\happyrepre^{-1}(a_z)\cap[1,i_y-1]\neq\emptyset$.
  \end{compactenum}

\end{lemma}	

\appendixproof{LemmaLongInt}{
\begin{proof}

Let $\solution$ be the solution, $\mathcal{I}$ the collection of intervals, and $a_x$ and $a_y$ the alternatives described in the lemma. Suppose, for the sake of contradiction, no $a_z\in\dominating(a_x)\setminus\dom(a_y)$ exists that satisfies the above properties. We will show that $\mathcal{I}$ is not \optimal\ in this case by providing a \good\ collection of intervals $\mathcal{I}'$ whose signature is lexicographically larger. 
We distinguish between two cases: either (1) $j_y<i_x$ or (2) $j_x<i_y$.
\paragraph*{Case 1: $j_y<i_x$.} Suppose, for the sake of contradiction, that no alternative $a_z$ exists such that $a_z\in\dominating(a_x)\setminus\dom(a_y)$, $[i_x, j_x] \subseteq [i_z, j_z], [i_z, j_z] \cap [i_y, j_y] = \emptyset$ and $\happyrepre^{-1}(a_z)\cap[j_y+1,i_x-1]\neq\emptyset$.
We start by choosing an alternative $a_{\ell}\in\dominators(a_x)\setminus\subordinates(a_y)$ that minimizes $i_x-j_{\ell}$ and satisfies the following properties:
\begin{compactenum}[(i)]
\item $j_{\ell}<i_x$ and \label{LemmaLongInt_case1_cond1}
\item $[i_x,j_x]\not\subseteq[i_{\ell},j_{\ell}]$ and \label{LemmaLongInt_case1_cond2}
\item No alternative in $\dominators(a_{\ell})$ satisfies \eqref{LemmaLongInt_case1_cond1} and \eqref{LemmaLongInt_case1_cond2}. \label{LemmaLongInt_case1_cond3}
\end{compactenum}
Subject to these conditions we choose $a_{\ell}$ that minimizes $i_x-j_{\ell}$.
Note that such an alternative must exist, as $a_y$ satisfies the first two properties.

We now show a useful claim: 
\begin{claim}\label{cl:axtoal}
There exists no $a_z\in\dominating(a_x)\setminus\dom(a_{\ell})$ such that $[i_x, j_x] \subseteq [i_z, j_z], [i_z, j_z] \cap [i_{\ell}, j_{\ell}] = \emptyset$ and $\happyrepre^{-1}(a_z)\cap[j_{\ell}+1,i_x-1]\neq\emptyset$.
  \end{claim}
\begin{proof}[Proof of \cref{cl:axtoal}] \renewcommand{\qedsymbol}{(of \cref{cl:axtoal})~$\diamond$}
Suppose, for the sake of contradiction, there exists an alternative that satisfies the above properties and let that alternative be $a_z$. Since $a_{\ell} \notin \subordinates(a_y)$, there are two possibilities for the relationship of $a_{\ell}$ and $a_y$:
\begin{compactenum}[(a)]
\item $a_{\ell}\in\dominators(a_y)$ or
\item $a_{\ell}\in\incomp(a_y)$.
\end{compactenum}
\paragraph*{Case a: $a_{\ell}\in\dominators(a_y)$.} As in this case $\subordinates(a_x)\subset\subordinates(a_{\ell})$ holds, it follows that $\dominators(a_x)\setminus\subordinates(a_{\ell})\subseteq\dominators(a_x)\setminus\subordinates(a_y)$. This directly implies the statement, as $[j_{\ell}+1,i_x-1]\subseteq[j_y+1,i_x-1]$. 
\paragraph*{Case b: $a_{\ell}\in\incomp(a_y)$.} Due to $\mathcal{I}$ being \good\ it follows that $a_y\in\earlier(a_{\ell})$, as the solution is neatly ordered. Because $[i_z,j_z]\cap[i_{\ell},j_{\ell}]=\emptyset$ and $[i_x,j_x]\subseteq[i_z,j_z]$ it follows that $[i_y,j_y]\cap[i_z,j_z]=\emptyset$, due to $j_y\leq j_z\leq i_x$. Then it follows that $a_z\in(\dominators(a_x)\setminus\subordinates(a_{\ell}))\setminus(\dominators(a_x)\setminus\subordinates(a_y))$, since we assume that the lemma is false. This implies that $a_z\in\subordinates(a_y)\setminus\subordinates(a_{\ell})$. Since $a_z$ is not a subordinate of $a_{\ell}$ and $a_y\in\earlier(a_{\ell})$ it follows that $a_{\ell}\in\earlier(a_z)$, due to the neat ordering and the \good\ intervals. This leads to the following contradiction however:
\begin{align*}
\rightmost{a_y}<\rightmost{a_{\ell}}<\rightmost{a_z}\leq\rightmost{a_y}
\end{align*}
Therefore such an $a_z$ cannot exist and the claim has been shown. 
\end{proof}
Note that all voters in the interval $[j_{\ell}+1,i_x-1]$ are \hassignable\ and therefore part of an interval because of \cref{LemmaIntvotcontainment}.

Let $A=\{a_z\in\dominators(a_x)\setminus\subordinates(a_l)\mid [i_x,j_x]\subseteq[i_z,j_z]\}$ and $B=\{a_z\in\dominators(a_{\ell})\mid [i_{\ell},j_{\ell}]\subseteq[i_z,j_z]\}$. We now partition the alternatives in $W \setminus \{a_{\ell}\}$ into three disjoint subsets:
 \begin{compactenum}
  \item[$C_1$] $\coloneqq A\setminus B$.
  \item[$C_2$] $ \coloneqq \subordinates(a_{\ell})$. Note that this includes $a_x$.
  \item[$C_3$] contains the remaining alternatives.
 \end{compactenum} 
 
We show the following claim:
\begin{claim}\label{lemmalongint_claim_incomal}
For every $a_g \in \incomp(a_{\ell})$, if $[i_g, j_g] \cap [j_{\ell} + 1, i_x] \neq \emptyset$, then $a_g \in C_1$.
\end{claim}

\begin{proof}[Proof of \cref{lemmalongint_claim_incomal}]\renewcommand{\qedsymbol}{(of \cref{lemmalongint_claim_incomal})~$\diamond$}
Let $a_g \in \incomp(a_{\ell})$ such that  $[i_g, j_g] \cap [j_{\ell} + 1, i_x] \neq \emptyset$. Assume, towards a contradiction, that $a_g \notin C_1$. Then either $a_g \notin A$ or $a_g \in B$. If $a_g \in B$, then $a_g \in \dominators(a_{\ell})$, a contradiction to $a_g \in \incomp(a_{\ell})$. Thus $a_g \notin A$.

Since $a_g \notin A$, it must hold that either $a_g \notin \dominating(a_x) \setminus \subordinates(a_{\ell})$ or $[i_x, j_x] \nsubseteq [i_g, j_g]$. Since $a_g \in \incomp(a_{\ell})$, it is trivial that $a_g \notin \subordinates(a_{\ell})$ and the first option reduces to $a_g \notin \dominating(a_x)$. If $a_g\in\subordinates(a_x)$, then it follows that $a_g\in\subordinates(a_{\ell})$ as well, a contradiction. If $a_g\in\incomp(a_x)$, by neat ordering, $a_g$ must be in $\earlier(a_x)$. Then since $a_g\in\incomp(a_{\ell})$, it follows that $\rightmost{a_{\ell}}<\rightmost{a_g}<\rightmost{a_x}\leq\rightmost{a_{\ell}}$, a contradiction.

Now consider the option $[i_x, j_x] \nsubseteq [i_g, j_g]$. Since $\mathcal{I}$ is \good, either $[i_g, j_g] \subset [i_x, j_x]$ or $[i_x, j_x] \cap [i_g, j_g] = \emptyset$. If $[i_g, j_g] \subset [i_x, j_x]$, then $a_g \in \subordinates(a_x) \subseteq \subordinates(a_{\ell})$, a contradiction. Thus $[i_x, j_x] \cap [i_g, j_g] = \emptyset$. If $a_g \in \dominating(a_x)$, our choice of $a_{\ell}$ must have been wrong because $i_x - j_g < i_x - j_{\ell}$. We have already shown that $a_g \in \incomp(a_x)$ leads to a contradiction.
\end{proof} 
 
 Let $j=\max_{\substack{a_u\in\subordinates(a_{\ell})\\ [i_x,j_x]\subseteq[i_u,j_u]\\\min\{\happyrepre^{-1}(a_u)\}\leq \min\{\happyrepre^{-1}(a_x)\}}}\{\happyrepre^{-1}(a_u)\}$. We now create an altered collection of intervals $\mathcal{I}'$ in the following way:
 \begin{compactitem}
 \item For all alternatives $a_b\in C_1$, we replace $[i_b,j_b]$ with $[j+1,j_b]$.
 \item For $a_{\ell}$ we replace $[i_{\ell},j_{\ell}]$ with $[i_{\ell},j]$.
 \item For all alternatives $a_u\in C_2$ we replace $[i_u,j_u]$ with $[\min\{\happyrepre^{-1}(a_u)\},\max\{\happyrepre^{-1}(a_u)\}]$.
 \end{compactitem}

The following claim shows that the interval $[j + 1, j_b]$ is non-empty for every $a_b \in C_1$.
\begin{claim}\label{c1properties}
For every $a_b \in C_1$, $j < \min(\happyrepre^{-1}(a_b))$.
\end{claim}

\begin{proof}[Proof of \cref{c1properties}]\renewcommand{\qedsymbol}{(of \cref{c1properties})~$\diamond$}
Let $a_b \in C_1$ be arbitrary.
By definition of $C_1$, $[i_x, j_x] \subseteq [i_b, j_b]$.
Next we show that $[i_b, j_b] \cap [i_{\ell}, j_{\ell}] = \emptyset$.
If $[i_b, j_b] \subseteq [i_{\ell}, j_{\ell}]$, then $[i_x, j_x] \subseteq [i_{\ell}, j_{\ell}]$, a contradiction.
If $[i_{\ell}, j_{\ell}] \subseteq [i_b, j_b]$, then $b \in B$, a contradiction to $b \in C_1$.
Thus \cref{cl:axtoal} implies that $\happyrepre^{-1}(a_b) \cap [j_{\ell} + 1, i_x - 1] = \emptyset$.

Because $[i_x, j_x] \subseteq [i_b, j_b]$, $j_{\ell} < i_x$ and $[i_b, j_b] \cap [i_{\ell}, j_{\ell}] = \emptyset$, we must have that $j_{\ell} < i_b \leq \min(\happyrepre^{-1}(a_b))$. Since\\ $\happyrepre^{-1}(a_b) \cap [j_{\ell} + 1, i_x - 1] = \emptyset$, we obtain that $i_x \leq \min(\happyrepre^{-1}(a_b))$.

Let $a_s \coloneqq \argmax_{\substack{a_u\in\subordinates(a_{\ell})\\ [i_x,j_x]\subseteq[i_u,j_u]\\\min\{\happyrepre^{-1}(a_u)\}\leq \min\{\happyrepre^{-1}(a_x)\}}}\{\happyrepre^{-1}(a_u)\}$. Let us consider the options for the relationship of $a_s$ and $a_b$.

If $a_b\in \subordinates(a_s)$, then $a_b \in \subordinates(a_{\ell})$, a contradiction to $a_b \in C_1$.

If $a_b$ and $a_s$ are incomparable, then by \good\ interval property their intervals must be disjoint. As $[i_x,j_x]\subseteq [i_b,j_b]\cap[i_s,j_s]$, this is not possible however. Therefore it cannot be the case that $a_b$ and $a_s$ are incomparable.

If $a_s \in \subordinates(a_b)$, then by \NT\ $[\leftmost{a_s}, \max(\happyrepre^{-1}(a_s))] \cap \happyrepre^{-1}(a_b) = \emptyset$. Because $[i_x, j] \subseteq [\leftmost{a_s}, \max(\happyrepre^{-1}(a_s))]$, we get that $j < \min(\happyrepre^{-1}(a_b))$, as required.
\end{proof}
 
The signature of $\mathcal{I}'$ is lexicographically larger than $\mathcal{I}$, because $j_t$ is only decreased for alternatives $a_t$ that are later in the canonical ordering than $a_{\ell}$ and $j_{\ell}$ is increased. Therefore if $\mathcal{I}'$ is \good\ it follows that $\mathcal{I}$ is not \optimal. This contradicts our initial assumption.
 
We now show that $\mathcal{I}'$ is \good. Note that conditions \eqref{good:3} to \eqref{good:6} must hold for all alternatives in $C_3$, as the intervals remain unchanged. Condition \eqref{good:3} is still satisfied, as only non-empty intervals were enlarged and $\mathcal{I}$ is a good collection of intervals.

Condition \eqref{good:1} holds for all alternatives in $C_1$ by \cref{c1properties} and because $\mathcal{I}$ is \good.

Condition \eqref{good:1} holds for $a_{\ell}$, because $\happyrepre^{-1}\{a_{\ell}\}\subseteq[i_{\ell},j_{\ell}]\subseteq[i_{\ell},j]$ and $j\in\VV(a_{\ell})$. 
Condition \eqref{good:1} holds for $C_2$ by definition. 
Condition \eqref{good:2} must be satisfied for $C_1$ and $C_2$, as the intervals became smaller. 
Condition \eqref{good:2} also holds for $a_{\ell}$, as due to \cref{lemmalongint_claim_incomal} each alternative $a_g\in\incomp(a_{\ell})$ that satisfies $[i_g,j_g]\cap[j_{\ell}+1,i_x]\neq\emptyset$ must be part of $C_1$. Those are however not assigned happy voters in this interval. 
Condition \eqref{good:6} is satisfied, as voters outside the interval $[\ii,\jj]$ are not assigned.

For condition \eqref{good:4} and \eqref{good:5}, it suffices to consider only pairs of alternatives where at least one alternative is in $C_1 \cup C_2 \cup \{a_{\ell}\}$, as the intervals for alternatives in $C_3$ were not changed.

We start with condition \eqref{good:4}:
\begin{compactitem}
\item If the alternatives are in $C_1 \cup C_2 \cup C_3$, then the condition holds as the intervals for all alternatives remained the same or got smaller.
\item For $a_u\in C_1$ and $a_{\ell}$ the condition is upheld as the intervals are disjoint.
\item For $a_u\in C_2$ and $a_{\ell}$ the condition is upheld as $a_u\in\subordinates(a_{\ell})$. 
\item Let $a_u\in C_3$ be one alternative and $a_{\ell}$ be the other alternative. Suppose, for the sake of contradiction, that $a_u\in\incomp(a_{\ell})$ and $[i_u,j_u]\cap[i_{\ell},j]\neq\emptyset$. Since $[i_{\ell}, j_{\ell}]$ is a part of $\mathcal{I}$ which is \good, it follows that $[i_{\ell}, j_{\ell}] \cap [i_u,j_u] = \emptyset$. From the contrapositive of \cref{lemmalongint_claim_incomal} we know that $[i_u,j_u]\cap[j_{\ell} + 1,i_x]=\emptyset$. 

It follows that $[i_u,j_u]\cap[i_x + 1, j]\neq\emptyset$. Let $a_s \coloneqq \argmax_{\substack{a_u\in\subordinates(a_{\ell})\\ [i_x,j_x]\subseteq[i_u,j_u]\\\min\{\happyrepre^{-1}(a_u)\}\leq \min\{\happyrepre^{-1}(a_x)\}}}\{\happyrepre^{-1}(a_u)\}$. We will next show that $[i_u, j_u] \cap [i_s, j_s] = \emptyset$.

If $a_s$ and $a_u$ are incomparable, then since the intervals are \good, and $[i_u, j_u] \cap [i_s, j_s] = \emptyset$, as required. If $a_u \in \subordinates(a_s)$, then $a_u \in \subordinates(a_{\ell})$, a contradiction. If $a_s \in \subordinates(a_u)$, then either $[i_u, j_u] \cap [i_s, j_s] = \emptyset$ and we are done, or $[i_s, j_s] \subseteq [i_u, j_u]$. This implies however that either $a_u \in C_1$ or $a_u \in \dominating(a_{\ell})$, both of which are contradictions.

Because $[i_x, j] \subseteq [i_s, j] \subseteq [i_s, j_s]$, this contradicts $[i_u,j_u]\cap[i_x + 1, j]\neq\emptyset$.
\end{compactitem}

\phantom{o}\\
Next, we consider condition \eqref{good:5}:
\begin{compactitem}
\item If both alternatives are in $C_1$, then the condition holds, because if they do not intersect in $\mathcal{I}$, then they do not intersect in $\mathcal{I}'$ either. If they do intersect in $\mathcal{I}$ then their intervals start after $j$, so one must still be contained in the other, as the endpoint did not change. 
\item The pair of alternatives $a_u\in C_1$ and $a_{\ell}$ uphold this condition, because one interval ends with $j$ and the other starts after $j$.
\item Let $a_u\in C_1$ and $a_v\in C_2$ such that $a_v\in\subordinates(a_u)$. Note that $a_u\in\subordinates(a_v)$ cannot be the case by the definition of $C_1$ and $C_2$. If the intervals previously intersected there are two options. If they do not intersect anymore, then the condition is trivially satisfied.

If the new intervals of $a_v$ and $a_u$ intersect, then the intervals had to intersect also in $\mathcal{I}$, meaning that\\ $[j+1,j_u]\cap[\min(\happyrepre^{-1}(a_v)),\max(\happyrepre^{-1}(a_v))]\neq\emptyset$. This implies that $\max(\happyrepre^{-1}(a_v))>j + 1$. Assuming the condition is violated, we must also have that $ \min(\happyrepre^{-1}(a_v))<j+1$, because otherwise the intervals would still be contained. Thus $ \min(\happyrepre^{-1}(a_v))<j+1 < \max(\happyrepre^{-1}(a_v))$. 

Let $a_s \coloneqq \argmax_{\substack{a_u\in\dominators(a_x)\cup\{a_x\}\\ [i_x,j_x]\subseteq[i_u,j_u]\\\min\{\happyrepre^{-1}(a_u)\}\leq \min\{\happyrepre^{-1}(a_x)\}}}\{\happyrepre^{-1}(a_u)\}$. Since $j+1 < \max(\happyrepre^{-1}(a_v))$, it follows that $a_s \neq a_v$ and $ \min(\happyrepre^{-1}(a_v))<  \max(\happyrepre^{-1}(a_s)) < \max(\happyrepre^{-1}(a_v))$. By \NT, this is only possible if $a_v$ dominates $a_s$. Moreover, it is necessary that $[i_s, j_s] \subseteq [i_v, j_v]$. However, since $a_v \in \subordinates(a_{\ell})$, this would imply that $\max(\happyrepre^{-1}(a_v)) > \max_{\substack{a_u\in\dominators(a_x)\cup\{a_x\}\\ [i_x,j_x]\subseteq[i_u,j_u]\\\min\{\happyrepre^{-1}(a_u)\}\leq \min\{\happyrepre^{-1}(a_x)\}}}\{\happyrepre^{-1}(a_u)\}$, a contradiction.
\item For a pair of alternatives $a_u\in C_1$ and $a_v\in C_3$, if they previously did not intersect they still do not intersect. If they intersected previously, it follows that $i_v\geq j$, as else $[i_x,j_x]\subseteq[i_v,j_v]$ would need to hold and then $a_v\notin C_3$.
\item For $a_{\ell}$ and $a_v\in C_2$, the condition must be upheld, because if
$\min\{\happyrepre^{-1}(a_v)\}\leq j$ and $\max\{\happyrepre^{-1}(a_v)\}>j$, it follows that $a_v$ and $\argmax_{\substack{a_u\in\dominators(a_x)\cup\{a_x\}\\ [i_x,j_x]\subseteq[i_u,j_u]\\\min\{\happyrepre^{-1}(a_u)\}\leq i_x}}\{\happyrepre^{-1}(a_u)\}$ would violate the neat ordering.
\item For $a_{\ell}$ and $a_v\in C_3$, if $a_v\in\dominators(a_{\ell})$ and $a_v\in A$, then it follows that the inclusion is still upheld, as $j\in[i_v,j_v]$, because otherwise $\mathcal{I}$ is not a \good\ collection of intervals. If $a_v\in\dominators(a_{\ell})$ and $a_v\notin A$ it follows via our choice of $a_{\ell}$ that $[i_v,j_v]$ and $[i_{\ell},j_{\ell}]$ do not intersect and more precisely $j_v<i_{\ell}$. If $a_{\ell}\in\dominators(a_v)$ then $a_v\in C_2$ must hold and therefore $a_v\notin C_3$.%
\item If both alternatives  $a_u,a_v\in C_2$, then the condition is upheld because of the neat ordering, as either the intervals must be disjoint or $[\min\{\happyrepre^{-1}(a_u)\},\max\{\happyrepre^{-1}(a_u)\}]\subseteq[\min\{\happyrepre^{-1}(a_v)\},\max\{\happyrepre^{-1}(a_v)\}]$ or $[\min\{\happyrepre^{-1}(a_v)\},\max\{\happyrepre^{-1}(a_v)\}]\subseteq[\min\{\happyrepre^{-1}(a_u)\},\max\{\happyrepre^{-1}(a_u)\}]$.
\item For $a_u\in C_2$ and $a_v\in C_3$ if $a_v\in\dominators(a_u)$ the inclusion would still be upheld due to $[\min\{\happyrepre^{-1}(a_u)\},\max\{\happyrepre^{-1}(a_u)\}]\subseteq[i_u,j_u]$. If $a_v\in\subordinates(a_u)$, it follows due to the transitivity of the domination relation that $a_v\in\subordinates(a_{\ell})$ and therefore $a_v\in C_2$.
\end{compactitem}
Since we have found a lexicographically better collection of intervals it follows that $\mathcal{I}$ is no \optimal\ contradicting our initial assumption and concluding the proof for the first case.

\paragraph*{Case 2: $j_x<i_y$.} Once again we suppose that no $a_z$ exists such that $a_z\in\dominators(a_x)\setminus\subordinates(a_y)$, $[i_x,j_x]\subseteq[i_z,j_z]$, $[i_z,j_z]\cap[i_y,j_y]=\emptyset$ and $\happyrepre^{-1}(a_z)\cap[1,i_y-1]\neq\emptyset$. We will lead this to a contradiction by showing that $\mathcal{I}$ is not \optimal\ in this case by providing a \good\ collection of intervals $\mathcal{I}'$ whose signature is lexicographically larger. 

We start by choosing an alternative $a_{\ell}\in\dominators(a_x)$ that minimizes $i_{\ell}-j_x$ and satisfies the following properties:
\begin{compactenum}[(i)]
\item $j_x<i_{\ell}$ and
\item $[i_x,j_x]\nsubseteq [i_{\ell},j_{\ell}]$ and
\item No alternative in $\dominators(a_{\ell})$ satisfies the above properties.
\end{compactenum}
Subject to these conditions we choose $a_{\ell}$ that minimizes $i_{\ell}-j_x$.
Such an alternative must once again exist, as $a_y$ satisfies the first two properties. Before we proceed we show a claim similar to the first case:
\begin{claim}\label{cl:axtoal2}
There exists no $a_z\in\dominating(a_x)\setminus\dom(a_{\ell})$ such that $[i_x, j_x] \subseteq [i_z, j_z], [i_z, j_z] \cap [i_{\ell}, j_{\ell}] = \emptyset$ and $\happyrepre^{-1}(a_z)\cap[1,i_{\ell}-1]\neq\emptyset$.
  \end{claim}
\begin{proof}[Proof of \cref{cl:axtoal2}] \renewcommand{\qedsymbol}{(of \cref{cl:axtoal})~$\diamond$}
Suppose, for the sake of contradiction, there exists an alternative that satisfies the above properties and let that alternative be $a_z$. Since $a_{\ell} \notin \subordinates(a_y)$, there are two possibilities for the relationship of $a_{\ell}$ and $a_y$.
We now distinguish three cases:
\begin{compactenum}[(a)]
\item $a_{\ell}\in\dominators(a_y)$
\item $a_{\ell}\in\incomp(a_y)$
\end{compactenum}
\paragraph*{Case a: $a_{\ell}\in\dominators(a_y)$.} As in this case $\subordinates(a_x)\subset\subordinates(a_{\ell})$ holds, it follows that $\dominators(a_x)\setminus\subordinates(a_{\ell})\subseteq\dominators(a_x)\setminus\subordinates(a_y)$. This directly implies the statement, as $[1,i_{\ell}-1]\subseteq[1,i_y-1]$.
\paragraph*{Case b: $a_{\ell}\in\incomp(a_y)$.} Due to $\mathcal{I}$ being \good\ it follows that $a_{\ell}\in\earlier(a_y)$, as the solution is neatly ordered. Because\\ $[i_z,j_z]\cap[i_{\ell},j_{\ell}]=\emptyset$ and $[i_x,j_x]\subseteq[i_z,j_z]$ it follows that $[i_y,j_y]\cap[i_z,j_z]=\emptyset$, due to $j_y\leq j_z\leq i_x$. Then it follows that $a_z\in(\dominators(a_x)\setminus\subordinates(a_{\ell}))(\dominators(a_x)\setminus\subordinates(a_x))$. This implies that $a_z\in\subordinates(a_y)\setminus\subordinates(a_{\ell})$. Since $a_z$ is not a subordinate of $a_{\ell}$ and $a_{\ell}\in\earlier(a_y)$ it follows that $a_z\in\earlier(a_l)$, due to the neat ordering and the \good\ intervals. This leads to the following contradiction however:
\begin{align*}
\leftmost{a_y}\leq\leftmost{a_z}<\leftmost{a_{\ell}}\leq\leftmost{a_y}
\end{align*}
Therefore such an $a_z$ cannot exist and the claim has been shown. 
\end{proof}
\cref{cl:axtoal} together with \NT\ implies that if any alternative $a_z$ that satisfies $[i_x,j_x]\subseteq[i_z,j_z]$, then it must be a subordinate of $a_{\ell}$ or satisfy $[i_{\ell},j_{\ell}]\subseteq[i_z,j_z]$.
Let $i=\min_{\substack{a_u\in\subordinates(a_{\ell})\\ [i_x,j_x]\subseteq[i_u,j_u]}}\{i_u\}$. 
We now make a new collection of intervals $\mathcal{I}'$ by exchanging $[i_{\ell},j_{\ell}]$ with $[i,j_{\ell}]$ in $\mathcal{I}$. As $\Sig(\mathcal{I}')\succ_{\text{lexico}}\Sig(\mathcal{I})$ we find a contradiction if we prove $\mathcal{I}'$ is \good.

We now show that $\mathcal{I}'$ is \good.

Condition \eqref{good:3} is still satisfied, as only non-empty intervals were manipulated and $\mathcal{I}$ is a good collection of intervals.
Condition \eqref{good:1} holds because $i\in\VV(a_l)$ and therefore $\happyrepre^{-1}(a_{\ell})\subseteq[i_{\ell},j_{\ell}]\subseteq[i,j_{\ell}]\subseteq\VV(a_{\ell})$.
Suppose condition \eqref{good:2} is violated. In this case there would need to be an alternative $a_z$ that is assigned voters in the interval $[j_x+1,i_{\ell}-1]$ that is incomparable to both $a_{\ell}$ and $a_x$, due to \cref{cl:axtoal2}; it cannot be dominated by $a_{\ell}$ or dominate $a_{\ell}$ if it dominates $a_x$ then our choice of $a_{\ell}$ was wrong. This directly leads to the following contradiction however:
\begin{align*}
\leftmost{a_{\ell}}\leq\leftmost{a_x}<\leftmost{a_z}<\leftmost{a_{\ell}}
\end{align*}
Therefore condition \eqref{good:2} must also be satisfied.
Condition \eqref{good:6} is satisfied, as all voters in $[i,j_{\ell}]$ are part of the (partial) solution already.

For conditions \eqref{good:4} and \eqref{good:5} it suffices to look at pairs of alternatives that contain $a_{\ell}$.

Suppose condition \eqref{good:4} were violated by $a_{\ell}$ and $a_u$. We already have shown that $a_u$ cannot be assigned voters in the interval $[j_x+1,i_{\ell}-1]$. As $a_u$ must be assigned voters-otherwise condition \eqref{good:3} would be violated- it follows that $\happyrepre(a_u)\cap[1,j_x]\neq\emptyset$ or $\happyrepre^{-1}(a_u)\cap[i_{\ell},\jj]\neq\emptyset$ must hold. In either case $a_u$ and $a_{\ell}$ or $a_x$ would have violated $\mathcal{I}$ being \good. Therefore condition \eqref{good:4} cannot hold.

Condition \eqref{good:5} is upheld because of our choice of $i$. If an alternative $a_z$ would intersect $[i,j_{\ell}]$ without being completely contained, it follows from $\mathcal{I}$ being \good\ that $i_z<i$. However since $[i_z,j_z]$ would intersect the interval of $\argmin_{\substack{[i_u,j_u]\in\mathcal{I}\\ [i_x,j_x]\subseteq[i_u,j_u]}}\{i_u\}$. Therefore this case cannot happen either.

As $\mathcal{I}'$ is a \good\ interval it follows that $\mathcal{I}$ is not \optimal\ concluding our proof. 
\end{proof}

}
\subsection{The DP table}\label{sec:dynprogtable}
For ease of reasoning, in this section we will aim to find an optimal solution that maximizes the overall satisfaction of the voters rather than minimizes their dissatisfactions.

\mypara{Intuition.} As already mentioned, we will search for a \optimal\ collection of voter intervals that completely characterizes an \MNT\ optimal solution. The idea is to iterate over the possible \good\ voter-intervals for each alternative. The alternatives that can be assigned voters in these intervals are described by the usable sets (see \cref{def: U}). These sets are disjoint for alternatives with disjoint intervals, which allows us to combine them to build bigger partial solutions from the bottom up.

\mypara{The table.}
Table~$T$ has an entry for every tuple~$(\firsttabletuple)$, where $a$ and $b$ are two incomparable alternatives, $[i,j]$ defines a voter interval, $\kUpper,\kLower,\na,\nb,\nd,$ are five non-negative integers with $0\le \kUpper, \kLower \le m$, $0 \le \na,\nb \le \upperB$, and $0 \le \nd \le n$, $B\in \{0,1\}$ being a binary value, and $\promise$ being a promise to be specified shortly.

Informally, the table entries store the number of maximum happy voters on the interval $[i,j]$ of a partial \MNT\ solution, if $a$ is the first and $b$ the last undominated alternative of the partial solution. We can show that in such a solution, we can only assign alternatives that are usable of $a$, $b$ or alternatives between them (see \cref{def: U}). The arguments $c', i',j'$ tell us that we assume there is some so far uncounted alternative $c'$ whose interval is $[i',j']$. This affects the usable sets of $a, b$ and all other alternatives. The value $B$ indicates whether we can use alternatives whose first approving voter is before $i$. The remaining arguments count the assigned and unassigned voters.

Formally, each table entry stores the maximum number of voters from the range~$[i,j]$ that can be satisfied by an \NT\ partial solution $(\committee',\repre)$ wrt. $[i,j]$, where~$\repre\colon [i,j] \to \committee'$ under the following seven conditions,  where %
\todo{H: Use $U'$ to refer to $W_1$? S: I believe this would get easily confused with $U$-set.}
\todo{C: Looks weird. S: Looks fairly fine when the todo is not on the way}
$\committeeR'\coloneqq \committee'\cap \big(\inbetw(a,b) \cup \bigcup\limits_{c\in\{a,b\}\cup\inbetw(a,b)}\usableset(c,\max\{i,\leftmost{c}\},\min\{j,\rightmost{c}\},c',i',j')\big)$:
\begin{enumerate}[(T1)]\label{dyntableprop}
  \item\label{dyntableprop1} Every happy voter from~$[i,j]$ is assigned to an alternative that is either $a$, or $b$,
  or is ``inbetween'' $a$ and $b$,
  or is usable for $a$ or $b$.
  Formally,
  \begin{align*}
    &\happyrepre([i,j]) \subseteq \committee'\subseteq\{a,b\} \cup \inbetw(a,b)\\ &\bigcup_{c\in\{a,b\}\cup\inbetw(a,b)}\usableset(c,\max\{i,\leftmost{c}\},\min\{j,\rightmost{c}\},c',i',j')
  \end{align*}
  \item\label{dyntableprop2} %
  $\committeeR'$ contains $\kUpper$ (resp.\ $\kLower$) many alternatives that are each assigned to $\upperB$ (resp.\ at most $\lowerB$) happy voters from $[i,j]$.

  \item\label{dyntableprop3} Alternative~$a$ (resp.\ $b$) is assigned $\na$ (resp.\ $\nb$) happy voters from $[i,j]$, i.e., $|\repre^{-1}(a)\cap [i,j]|=\na$ (resp.\ $|\repre^{-1}(b) \cap [i,j]|= \nb$).

  \item\label{dyntableprop4} It holds that $|[i,j]\setminus\happyrepre^{-1}(\committee')|\geq\nd$.
  \item\label{dyntableprop5} If $B=0$, then there exists no alternative $c\in \committeeR'\cap\usablesetS(a, i, \min(\rightmost{a}, j), c', i', j')$, such that $\leftmost{c}<i$. %
  \item\label{dyntableprop6} $(\committee',\sigma)$ is monotone with respect to $ \big(\inbetw(a,b)\\ \bigcup\limits_{c\in\{a,b\}\cup\inbetw(a,b)}\usableset(c,\max\{i,\leftmost{c}\},\min\{j,\rightmost{c}\},c',i',j')\big)$.
  \item\label{dyntableprop7} For each voter $v\in[i,j]$ that is unassigned under $\happyrepre$ and each alternative $c\in\happyrepre([i,j])$, it holds that if $v\in\VV(c)$, then $v>\max(\happyrepre^{-1}(c))$. Informally, we want voters that are assignable to an alternative to be only unassigned if they are later than the voters already assigned to that alternative. %
\end{enumerate} 
 
 We say that a table entry is \myemph{correct} if it computes the maximal number of happy voters for an \NT\ partial solution that satisfies (T\ref{dyntableprop1})-(T\ref{dyntableprop7}). Similarly we will say $T$ is \myemph{correct} if every table entry is correct.
 In the following, we describe how to compute~$T(\firsttabletuple)$. We assume that the arguments of the configuration fulfill the following conditions in the configuration.
 \begin{compactenum}[(CT1)]
   \item $[i,j]\subseteq [\leftmost{a}, \rightmost{b}]$.
   \item $a\in \earlier(b)\cup \{b\}$.
   \item If $a \neq b$, then $j-i+1\ge \na+\nb+\nd$; otherwise $\na=\nb$ and $j-i+1\ge \na+\nd$.
   \item If $c'\neq 0$, then $[i,j]\subseteq [i',j']$ and $i' \le \leftmost{a}$; otherwise $i'=j'=0$.
   \item If $c'\neq 0$, $\usableset(c',i',j',c',i',j')\cap\dominators(a)\cap\earlier(b)=\emptyset$.
   \item If $c'=0$, then $\dominators(a)\cap\earlier(b)=\emptyset$.
   \item If $a\neq b$, then $\na\neq0$ and $\nb\neq0$.
 \end{compactenum}
 Intuitively, (CT1)-(CT4) ensure that the chosen arguments make sense, e.g. the interval does not exceed the range of voters that could be assigned to the alternatives.
 Similarly, (CT5) and (CT6) are meant to prevent the combinations of table configurations that could not lead to an \MNT\ solution. 
 (CT7) is meant to ensure monotonicity.
  
 For the case when at least one of the above conditions is violated, the corresponding entry is set to~$-\infty$.
 We start by initializing the table as:
 \begin{align}
   T(a,a,i,j,0,0,\na,\na,\nd,0,c',i',j')\coloneqq \na \label{eq:SPMinit1}
 \end{align}
We update the table, distinguishing between three cases, either ``$a = b$ and $B=0$'' (Case 1), ``$a= b$ and $B=1$'' (Case 2), or ``$a\neq b$'' (Case 3).
 We first define two auxiliary functions
 \begin{align*}
   f_1(c_1,c_2,n_1,n_2) &=
   \begin{cases}
     2 & \, \text{if }c_1\neq c_2\wedge n_1=n_2=\upperB \\
     0 & \, \text{if }n_1\leq\lowerB\wedge n_2\leq\lowerB\\
     1 & \, \text{else}\\
   \end{cases}\\
     f_2(c_1,c_2,n_1,n_2) &=|\{c_1, c_2\} | - f_1(c_1,c_2,n_1,n_2)
   \end{align*}
\begin{figure*}[!t]
\begin{multline}
 \textbf{Case 1:} \quad \label{eq:SPMlower1}
	  T(a,a,i,j,\kUpper,\kLower,\na,\na,\nd,0,c',i',j')\coloneqq\\
	  \max\left\{
	\begin{array}{l}
		\underset{\substack{b,c\in\usablesetS(a,i,j,c',i',j')\\
				\kUpper' = \kUpper - f_1(b,c,\nb,n_c), 
				\kLower' = \kLower-f_2(b,c,\nb,n_c)\\ \leftmost{b}\in[i,j], 0< \nb,n_c\leq\upperB, B\in\{0,1\}}}{\max}
		\left\{ \begin{array}{l}
			T(b,c,i,j,\kUpper',\kLower',\nb,n_c,\na+\nd,B,c',i',j')+\na,\\
			T(b,c,i,j,\kUpper',\kLower',\nb,n_c,\na+\nd,B,a,i,j)+\na \\
		\end{array}\right\},\\
		\underset{\substack{\nd_1+\nd_2=\nd, b_1+b_2=\na\\ i\leq i^*<j^*\leq j, \kUpper_1+\kUpper_2=\kUpper, \kLower_1+\kLower_2=\kLower}}{\max}
		\left\{ \begin{array}{l}
			~~ T(a,a,i,i^*,\kUpper_1,\kLower_1,b_1,b_1,\nd_1,0,c',i',j'), 
			+  T(a,a,j^*,j,\kUpper_2,\kLower_2,b_2,b_2,\nd_2,0,c',i',j')
		\end{array}\right\}
	\end{array}\right\}.
\end{multline}
\begin{multline}
\textbf{Case 2:} \quad \label{eq:SPMlower2}
	 T(a,a,i,j,\kUpper,\kLower,\na,\na,\nd,1,c',i',j')\coloneqq   \\
	\max\left\{
	\begin{array}{l}
		\underset{\substack{b,c\in\usablesetS(a,i,j,c',i',j') \\
				\kUpper=\kUpper'+f_1(b,c,\nb,n_c), 
				\kLower=\kLower'+f_2(b,c,\nb,n_c)\\
				0<\nb,n_c\leq\upperB, B\in\{0,1\}}}{\max}
		\left\{\begin{array}{l}
			T(b,c,i,j,\kUpper',\kLower',\nb,n_c,\na+\nd,B,c',i',j')+\na,\\
			T(b,c,i,j,\kUpper',\kLower',\nb,n_c,\na+\nd,B,a,i,j)+\na                 \\
		\end{array}\right\},\\
		\underset{\substack{\nd_1+\nd_2=\nd,  b_1+b_2=\na\\ i\leq i^*<j^*\leq j, \kUpper_1+\kUpper_2=\kUpper, \kLower_1+\kLower_2=\kLower}}{\max}
		\left\{\begin{array}{l}
			~~T(a,a,i,i^*,\kUpper_1,\kLower_1,b_1,b_1,\nd_1,1,c',i',j')
			+T(a,a,j^*,j,\kUpper_2,\kLower_2,b_2,b_2,\nd_2,0,c',i',j')
		\end{array}\right\}
	\end{array}
	\right\}.
\end{multline}
\begin{multline} 
     \textbf{Case 3:} \quad \label{eq:SPMsum}
	T(a,b,i,j,\kUpper,\kLower,\na,\nb,\nd,B,c',i',j')\coloneqq  \\
	  \max\left\{
	\begin{array}{l}
		\underset{\substack{ c\in\inbetwS(a,b)\\ i\leq i^*<j^*\leq j,  0<n_c\leq\lowerB\\ \kUpper_1+\kUpper_2=\kUpper, \kLower_1+\kLower_2+1=\kLower\\ \nd_1+\nd_2=\nd, B'\in\{0,1\}}}{\max}
		\left\{
		\begin{array}{l}
			~~  T(a,c,i,i^*,\kUpper_1,\kLower_1,\na,n_c,\nd_1,B,c',i',j')
			+ T(b,b,j^*,j,\kUpper_2,\kLower_2,\nb,\nb,\nd_2,B',c',i',j')
		\end{array}\right\},\\
		\underset{\substack{c\in\inbetwS(a,b)\\ i\leq i^*<j^*\leq j,  n_c=\upperB \\ \kUpper_1+\kUpper_2+1=\kUpper, \kLower_1+\kLower_2=\kLower\\\nd_1+\nd_2=\nd, B'\in\{0,1\}}}{\max}
		\left\{
		\begin{array}{l}
			~~T(a,c,i,i^*,\kUpper_1,\kLower_1,\na,n_c,\nd_1,B,c',i',j')
			+T(b,b,j^*,j,\kUpper_2,\kLower_2,\nb,\nb,\nd_2,B',c',i',j')
		\end{array}\right\},\\
		\underset{\substack{i\leq i^*<j^*\leq j\\ \kUpper_1+\kUpper_2=\kUpper, \kLower_1+\kLower_2=\kLower\\\nd_1+\nd_2=\nd, B'\in\{0,1\}}}{\max}
		\left\{
		\begin{array}{l}
			~~T(a,a,i,i^*,\kUpper_1,\kLower_1,\na,\na,\nd_1,B,c',i',j')
			+T(b,b,j^*,j,\kUpper_2,\kLower_2,\nb,\nb,\nd_2,B',c',i',j')
		\end{array}\right\} 
	\end{array}\right\}.
\end{multline}
\end{figure*}
We compute the table in the following order, where the order holds only subject to the previous steps, e.g., $2$ is only relevant subject to $1$ and so on: 
\begin{compactenum}\label{tableorder}
\item Order the entries by $\kUpper + \kLower$ in a non-decreasing order.
\item Order the entries with $a = b$ before the entries where $a \neq b$.
\item Order the entries by $j - i$ in a non-decreasing order.
\item Order the entries by $\na + \nb$ in a non-decreasing order.
\item Order the entries by $\nd$ in an increasing order.
\item Order the entries by $B$ in an increasing order.
\end{compactenum}
For the optimal solution, we return $\max\limits_{\substack{a,b\in\level_\aaa(1), i\leq j\\\kUpper\leq n\mod k -f_1(a,b,\na,\nb)\\\kUpper+\kLower+|\{a,b\}|\leq k\\\na,\nb\leq\upperB,\nd\leq n, B\in\{0,1\}}}T(a,b,i,j,\kUpper,\kLower,\na,\nb,\nd,B,0,0,0)$.

 \todo{I removed the Main theorem subsection because it did not make a lot of sense anymore. A good flow might require new descriptions.}
 \appendixsection{sec:dynprogtable}
\toappendix{ 
We proceed to show the correctness of the algorithm.

  \begin{lemma}\label{lemma:Monroecorr}
  The dynamic programming table is correct. 
   \end{lemma}
   \begin{proof}

   The proof proceeds in four steps. In the first step, we show that the dynamic programming table utilizes only entries that were computed previously to compute the value of a new table entry. The rest of the correctness proof will therefore assume that table entries that are computed earlier in the ordering are already correct. In the second step we show that the initialization~\eqref{eq:SPMinit1} is \correct. In the third step we will show that any table entry corresponds to the number of happy voters in an \NT\ solution that satisfies (T\ref{dyntableprop1})-(T\ref{dyntableprop7}). This step can be understood to be some form of feasibility. In the fourth step we show that given any tuple $(a,b,i,j,\kUpper,\kLower,\na,\nb,\nd,B,c',i',j')$ and a partial solution $(\committee, \repre)$ that is \correct\ (meaning also that the number of happy voters is maximal) for this tuple, the computed dynamic table entry corresponds to either $(\committee, \repre)$ or to another solution with at least as many satisfied voters. This implies that any entry we compute must be \correct.
   \paragraph{Step 1: Computation order.}
   The computation of $T(a, a, i, j, \kUpper, \kLower, \na, \na, \nd, B, c', i', j')$ in Operations \eqref{eq:SPMlower1} and \eqref{eq:SPMlower2} uses the table entries $T(b, c, i, j, \kUpper', \kLower', \nb, n_c$, $\na + \nd$, $B$, $c', i', j')$, $T(b, c, i, j, \kUpper', \kLower', \nb, n_c, \na + \nd$, $B, a$, $i, j)$, $T(a, a, i, i^*, \kUpper_1, \kLower_1, b_1, b_2, \nd_1, B', c', i', j'), T(a, a, j^*, j, \kUpper_2, \kLower_2$, $b_2$, $b_2$, $\nd_2, B', c', i', j')$, where $B' \in \{0,1\}$ and the rest are defined in the operation. The first options satisfy $\kUpper'+\kLower'<\kUpper+\kLower$. Therefore the table entries are computed beforehand due to $1$ in the computation order.
   
The second operation uses table entries where $\kUpper_1,\kUpper_2\leq\kUpper$ and $\kLower_1+\kLower_2\leq\kLower$. Therefore $1$ cannot be used to determine the order. By  As these entries have the same element $a$ in the first two places $2$ cannot be used either. However by $i^*<j^*$ it follows that these table entries are also computed before $T(a, a, i, j, \kUpper, \kLower, \na, \na, \nd, B, c', i', j')$ due to $3$, because $i^*-i<j-i$ and $j-j^*<j-i$. 
   
For Operation \eqref{eq:SPMsum} we can similarly observe that $\kUpper_1,\kUpper_2\leq\kUpper$ and $\kLower_1+\kLower_2\leq\kLower$ and $a\neq b$. Since $i^*<j^*$ we have that $i^*-i<j-i$ and $j-j^*<j-i$ and therefore the necessary entries are computed before due to $3$. 
   \paragraph{Step 2: Initialization.}
   
   We start by showing that the initialization operation~\eqref{eq:SPMinit1} is \correct. Here we want to compute the maximal number of voters on an interval $[i,j]$ that can be satisfied by an alternative $a$, if we assign $\na$ happy voters on this interval to $a$ and leave at least $\nd$ voters unassigned. As we require that $[i,j]\subseteq\VV(a)$ and $j-i+1\geq\na+\nd$ (otherwise the value is $-\infty$ if it is not possible to satisfy), this number is $\na$. This represents a partial solution where $W'=\{a\}$ and $\happyrepre(l)=a$, for all $l\in[i,i+\na-1]$ and the remaining voters are left unassigned. This partial solution satisfies (T\ref{dyntableprop1}), as each happy voter is assigned to $a$. Condition (T\ref{dyntableprop2}) is satisfied, as $W_1=\kUpper=\kLower=0$. As $|[i,i+\na-1]|=\na$ property (T\ref{dyntableprop3}) is fulfilled. Property (T\ref{dyntableprop4}) is satisfied due to $j-i+1\geq\na+\nd$ being required and therefore $j-i+1-\na\geq\nd$. As $\committeeR=\emptyset$, properties (T\ref{dyntableprop5}) and (T\ref{dyntableprop6}) hold. As the first $\na$ voters in the interval $[i,j]$ were assigned to $a$ property (T\ref{dyntableprop7}) also holds. Due to $W=\{a\}$ the neat ordering must hold as well. Maximality of happy voters is also given.
   
   Therefore the \correct ness of the initialization step is shown.
   
   \paragraph{Step 3: Feasibility.}
We now show that a table entry $T(a,b,i,j,\kUpper,\kLower,\na,\nb,\nd,B,c',i',j')$ computed through the recursions~\eqref{eq:SPMlower1}, \eqref{eq:SPMlower2} and \eqref{eq:SPMsum} corresponds to the number of happy voters in an \NT\ solution that satisfies (T\ref{dyntableprop1})-(T\ref{dyntableprop7}). We use proof by induction: our inductive assumption is that all table entries that precede the considered table entry in the computation order are \correct. The base case has been shown in Step 1. Moreover, Step 2 shows that we only use table entries that precede the current table entry when computing the value.

We proceed by distinguishing two types of table entries:
 \begin{compactenum}
   \item $a=b$ (corresponds to the computations \eqref{eq:SPMlower1} and \eqref{eq:SPMlower2})
   \item $a\neq b$ (corresponds to the computation \eqref{eq:SPMsum})
   \end{compactenum}

   For the table entries of type 1, we start by showing that any of the operations in \eqref{eq:SPMlower1} or \eqref{eq:SPMlower2} lead to a partial solution that satisfies properties (T\ref{dyntableprop1})-(T\ref{dyntableprop7}). We will first show the following useful claim:
       \begin{claim}\label{cl:same}
       Let $b,d\in\aaa$ be alternatives such that $b,d\in\subordinates(a)$, $b\in\earlier(d)$ and $\leftmost{d}\leq j$ hold. Then if $b\in\usableset(a,i,j,c',i',j')$ it holds that $d\in\usableset(a,i,j,c',i',j')$.
          \end{claim}
          \begin{proof}[Proof of \cref{cl:same}] \renewcommand{\qedsymbol}{(of \cref{cl:same})~$\diamond$}
          Since $\leftmost{d} < j$, we must either have that $\VV(d) \cap [i, j] \neq \emptyset$ or $\rightmost{d} < i$. However, because $b\in\usableset(a,i,j,c',i',j')$, $i< \rightmost{b} < \rightmost{d}$. This leads to a contradiction if $\rightmost{d} < i$. Moreover, if $\leftmost{b} \in [i, j]$, then $\leftmost{d} \in [i, j]$.

            Suppose for the sake of contradiction, $d\notin\usableset(a,i,j,c',i',j')$. Therefore we know that $\leftmost{d}\notin[i,j]$. Then there exists $z\in\earlier(a)\cap\dominating(d)$ that blocks $d$ from being usable. If $z\in\earlier(a)\cap\dominating(d)$ it would however follow that $z \in \dominating(b)$, due to $\leftmost{z}<\leftmost{a}\leq\leftmost{b}$ and $\rightmost{z}\geq\rightmost{d}>\rightmost{b}$. Then $z$ also blocks $b$ from being usable. This however contradicts $b\in\usableset(a,i,j,c',i',j')$.
          \end{proof}
   
We will show that all of the operands of the $\max$-operation in \eqref{eq:SPMlower1} and \eqref{eq:SPMlower2} correspond to a partial solution that satisfies (T\ref{dyntableprop1})-(T\ref{dyntableprop7}). Then the result of the $\max$-operation must also correspond to a partial solution that satisfies (T\ref{dyntableprop1})-(T\ref{dyntableprop7}).
   
    We start by looking at the first operand of the $\max$-operation in \eqref{eq:SPMlower1} and \eqref{eq:SPMlower2}. Consider the partial solution $(\committee',\repre')$ corresponding to the configuration of the used table entry. Note that it must correspond to one of the operands of the inner $\max$-operation.
    
     By inductive hypothesis, as all table entries that could be used in the computation precede the considered table entry $(\committee',\repre')$ satisfies (T\ref{dyntableprop1})-(T\ref{dyntableprop7}). We create a partial solution $(\committee,\repre)$ for the new configuration. 
    As all voters in the interval $[i,j]$ approve of $a$ and there are $\na+\nd$ unassigned voters (satisfying Property (T\ref{dyntableprop7})), it follows that by assigning the first $\na$ unassigned voters to $a$ we get an extended partial solution that satisfies property (T\ref{dyntableprop3}).

To Property (T\ref{dyntableprop1}) we must show that $\committee'\subseteq\usableset(a,i,j,c',i',j')$. Since we know that $\committee'$ satisfies Property (T\ref{dyntableprop1}), it is enough to show that for every $b, c \in \usablesetS(a,i,j,c',i',j')$, it holds that $\{b, c\} \cup \inbetw(b, c) \cup \bigcup_{b' \in \{b, c\} \cup \inbetw(b, c)} \usableset(b',\max\{i+\na,\leftmost{b'}\},\min\{j,\rightmost{b'}\},c',i',j') \subseteq \{a\} \cup  \usableset(a,\max\{i,\leftmost{a}\},\min\{j,\rightmost{a}\},c',i',j')$ and $\bigcup_{b' \in \{b, c\} \cup \inbetw(b, c)} \usableset(b',\max\{i,\leftmost{b'}\},\min\{j,\rightmost{b'}\},a,i,j) \subseteq \{a\} \cup  \usableset(a,\max\{i,\leftmost{a}\},\min\{j,\rightmost{a}\},c',i',j')$.
Since $b, c \in \usablesetS(a,i,j,c',i',j')$ by definition, this follows trivially. \cref{cl:same} shows that $\inbetw(b, c) \subseteq \usableset(a,i,j,c',i',j')$, as each alternative $d\in\inbetw(b,c)$ satisfies $\leftmost(d)\leq j$ and $b\in\earlier(d)$.

Let $d \in \bigcup_{b' \in \{b, c\} \cup \inbetw(b, c)} \usableset(b',\max\{i,\leftmost{b'}\}$, $\min\{j,\rightmost{b'}\}$, $c',i',j')$ be arbitrary. Assume, towards a contradiction, that $d \notin \usableset(a,\max\{i,\leftmost{a}\},\min\{j,\rightmost{a}\},c',i',j')$. If $\leftmost{d} \geq  \max\{i,\leftmost{b'}\}$, then $\leftmost{d} \geq \max\{i,\leftmost{a}\}$ and  $d \in \usableset(a,\max\{i,\leftmost{a}\},\min\{j,\rightmost{a}\},c',i',j')$, a contradiction.
If $\leftmost{d} < i'$ there must be some alternative $\hat{b} \in \dominating(d) \cap \incomp(a) \cap \earlier(c)$ that blocks $d$ from being usable. If $\hat{b} \in \incomp(b')$ as well, it blocks $b'$ form being usable, a contradiction. If $\hat{b} \in \dominating(b')$, then it blocks $b'$ from being usable to $a$, a contradiction.
If $i' \leq \leftmost{d} < \max\{i,\leftmost{a}\}$ there must be some alternative $\hat{b} \in \dominating(a) \cap \dom(c') \cap \earlier(c)$ such that $\leftmost{\hat{b}} \geq i'$ or $\earlier(c') \cap \dominating(\hat{b}) = \emptyset$. However, because $a \dominate d$, this alternative also blocks $d$ from being in $\usableset(b',\max\{i,\leftmost{b'}\},\min\{j,\rightmost{b'}\},c',i',j')$.

The case for showing that $\bigcup_{b' \in \{b, c\} \cup \inbetw(b, c)} \usableset(b',\max\{i,\leftmost{b'}\}$, $\min\{j,\rightmost{b'}\}$, $a,i,j) \subseteq \{a\} \cup  \usableset(a,\max\{i,\leftmost{a}\},\min\{j,\rightmost{a}\},c',i',j')$ is near-identical.

For Property (T\ref{dyntableprop2}) we observe that $b$ and $c$ are the only voters who belong to $\committeeR$ of $\committee$ but not to $\committeeR'$ of $\committee'$. Because $f_1$ and $f_2$ count the number of unique alternatives among $b$ and $c$ that are assigned $\upperB$ and fewer voters, Property (T\ref{dyntableprop2}) is satisfied.
    Property (T\ref{dyntableprop4}) holds as before assigning $\na$ voters there were at least $\na+\nd$ voters unassigned.
    Property (T\ref{dyntableprop5}) holds because $\leftmost{b}\in[i,j]$ and each other alternative used in the table configuration starts after $b$. Therefore no used alternative in $\committeeR$ could be assigned any voter before $i$. For the operation \eqref{eq:SPMlower2} we have that $B=1$ and thus the condition is satisfied trivially.
    Since $(\committee', \repre')$ satisfies (T\ref{dyntableprop6}),  $b, c$ are undominated in $\committeeR \cap \{b, c\}$ and $\nb>0$ and $n_c>0$ it follows that the generated solution is monotone with respect to $\usableset(a,i,j,c',i',j')$. From this it follows that this solution satisfies (T\ref{dyntableprop6}) as well.
    Property (T\ref{dyntableprop7}) is satisfied, because the original solution satisfies (T\ref{dyntableprop7}) and the first $\na$ voters are assigned to $a$. 
   
   We now move on to the second operand of the $\max$-operation in \eqref{eq:SPMlower1} and \eqref{eq:SPMlower2} (the sum). 
    Let $(\committee',\repre_1)$ and $(\committee'',\repre_2)$ be the partial solutions corresponding to the first and second of the table configurations in the sum. 
    
    We will now compute $(\committee,\repre)$ for the table entry of the sum. We will set $\committee=\committee'\cup \committee''$ and \begin{align*}
    \repre(v)\coloneqq\begin{cases}
    \repre_1(v)\text{, if }v\in[i,i^*]\\
    \repre_2(v)\text{, if }v\in[j^*,j]
    \end{cases}
    \end{align*}
Note that, due to $i^*<j^*$ the second partial solution is assigning voters on a later interval than the first. 
    As $\usableset(a,i,i^*,c',i',j')\subseteq\usableset(a,i,j,c',i',j')$ and $\usableset(a,j^*,j,c',i',j')\subseteq\usableset(a,i,j,c',i',j')$, condition (T\ref{dyntableprop1}) is satisfied.
    Due to $B=0$ for the right hand side of the equation it follows that no alternative other than $a$ is used on both sides of the table. No alternative $d$ with $\leftmost{d} > i^*$ satisfies voters on the interval $[i, i^*]$. Because $B = 0$, the second operation of the sum cannot contain any alternative $d$ such that $\leftmost{d} < j^*$. Therefore by summing up $\kUpper$ and $\kLower$ it follows property (T\ref{dyntableprop2}) must also be satisfied for $\committee,\repre)$.
    Property (T\ref{dyntableprop3}) is satisfied, as $a$ is assigned $b_1$ voters in the first partial solution and $b_2$ in the second, because we require $b_1+b_2=\na$. As we require that at least $\nd_1$ voters in the first partial solution are unassigned and at least $\nd_2$ voters in the second partial solution are left unassigned, it follows that at least $\nd_1+\nd_2=\nd$ voters are left unassigned on the whole interval (due to the intervals of the partial solutions being disjoint), thereby ensuring property (T\ref{dyntableprop4}).
    Finally property (T\ref{dyntableprop5}) is satisfied in operation \eqref{eq:SPMlower1}, because no voter in $\committeeR''$ could be assigned voters before $i<j^*$ and no voter in $\committee_1'$ could be assigned voters before $i$, because the first entry of the sum satisfies (T\ref{dyntableprop5}).
    Property (T\ref{dyntableprop5}) is satisfied in \eqref{eq:SPMlower2}, because $B=1$ and therefore the condition is satisfied either way.
    
    We can find a solution that satisfies the same number of voters and satisfies (T\ref{dyntableprop6}) by \cref{lemmaMNO}. By creating an instance where all unassigned voters are assigned to some alternative that satisfies all alternatives and then applying \cref{lemmaMNO} we then get a solution that satisfies (T\ref{dyntableprop7}) with the same number of satisfied voters.

   For table entries computed in the operation \eqref{eq:SPMsum}, we can once again assume that all table entries that may be used to compute $T(a,b,i,j,\kUpper,\kLower,\na,\nb,\nd,B,c',i',j')$ are already correct and satisfy (T\ref{dyntableprop1})-(T\ref{dyntableprop7}) by inductive assumption. We start with showing a useful claim.
    	\begin{claim}\label{cl:samec}
    	Let $a_x,a_y,a_z$ be three alternatives and $[i_x,j_x]$, $[i_y,j_y]$, $[i_z,j_z]$ three corresponding intervals (not necessarily good), such that for all $\ell\in\{x,y,z\}$, $V(a_{\ell})\supseteq [i_{\ell},j_{\ell}]$, $[i_x,j_x]\cap[i_y,j_y]=\emptyset$, $[i_x,j_x]\cup[i_y,j_y]\subseteq[i_z,j_z]$, $i_z\leq\leftmost{a_x}<\leftmost{a_y}$ and $a_x\in\incomp(a_y)$. Then $\usableset(a_x,i_x,j_x,a_z,i_z,j_z)\cap\usableset(a_y,i_y,j_y,a_z,i_z,j_z)=\emptyset$.
    			\end{claim}
    			\begin{proof}[Proof of \cref{cl:samec}] \renewcommand{\qedsymbol}{(of \cref{cl:samec})~$\diamond$}
    			Suppose there exists $b\in\usableset(a_x,i_x,j_x,a_z,i_z,j_z)\cap\usableset(a_y,i_y,j_y,a_z,i_z,j_z)$. We can deduce that $\leftmost{b}\leq j_x$, as else $b\notin\usableset(a_x,i_x,j_x,a_z,i_z,j_z)$. As we require $\leftmost{a_x}\geq i_z$ it follows however that $b\notin\usableset(a_y,i_y,j_y,a_z,i_z,j_z)$, as $a_x\in\earlier(a_y)\cap\dom(a_z)$. This contradicts our initial assumption and concludes the proof.  
    			\end{proof}
    
We first show that assuming the previous entries satisfy (T\ref{dyntableprop1})--(T\ref{dyntableprop7}), any of the operands of the $\max$-operation in the operation \eqref{eq:SPMsum} satisfies (T\ref{dyntableprop1})--(T\ref{dyntableprop7}). Then we show that a maximal solution that satisfies (T\ref{dyntableprop1})--(T\ref{dyntableprop7}) can be found by using the $\max$-operation. 
    Unless stated otherwise, we consider all of the operands of the outer $\max$-operation at once, as they only differ in how $\kUpper$ respectively $\kLower$ relate to $\kUpper_1$ and $\kUpper_2$ respectively $\kLower_1$ and $\kLower_2$. 
    Let $(W',\repre')$ and $(\committee'',\repre'')$ be the partial solutions corresponding to the table configurations of the first respectively second part of the sum.
    We will make a solution $(\committee,\repre)$ that represents the generated table configuration in the following way: Let $\committee=\committee'\cup \committee''$ and $\repre(v)=\begin{cases}
    \repre'(v)\text{, if }v\in[i,i^*]\\
    \repre''(v)¸\text{, if }v\in[j^*,j].
    \end{cases}$
    
   Since the two partial solutions satisfy (T\ref{dyntableprop1})--(T\ref{dyntableprop7}) no alternative that is used in the first table configuration can be used in the second table configuration due to \cref{cl:samec}. Therefore, as $c\in\inbetw(a,b)$, $\inbetw(a,c)\subset\inbetw(a,b)$, $\usableset(a,i,\min\{\rightmost{a},i^*\},c',i',j')\subseteq\usableset(b,i,\min\{\rightmost{a},j\},c',i',j')$ and $\usableset(b,\max\{j^*,\leftmost{b}\},j,c',i',j')\subseteq\usableset(b,\max\{i,\leftmost{b}\},j,c',i',j')$, it follows that (T\ref{dyntableprop1}) is satisfied.
    
    Note that $\committeeR=\committeeR'\cup \committeeR''\cup\{c\}$, if $c\neq a$ and $\committeeR=\committeeR'\cup \committeeR''$, if $c=a$. If $c \neq a$, $c$ may have been assigned either $\upperB$ voters or fewer. In the first case $c$ will need to be counted as a part of $\kUpper$ and in the second case as a part of $\kLower$. In the first operand of the outer $\max$-operation $c$ is assigned at most $\lowerB$ voters and all the other alternatives in $\committee_1'$ and $\committee_2'$ are assigned the same number of voters by construction. Therefore  (T\ref{dyntableprop2}) is satisfied due to $\kUpper_1+\kUpper_2=\kUpper$ and $\kLower_1+\kLower_2 + 1=\kLower$. In the second operand of the outer $\max$-operation $c$ is assigned $\upperB$ voters and all the other alternatives in $\committeeR'$ and $\committeeR''$ are assigned the same number of voters by construction. Therefore  (T\ref{dyntableprop2}) is satisfied due to $\kUpper_1+\kUpper_2 + 1=\kUpper$ and $\kLower_1+\kLower_2=\kLower$. In the third operand $c=a$ and thus $c$ is not added to $\committee_1$. We have that $\kUpper_1+\kUpper_2=\kUpper$ and $\kLower_1+\kLower_2=\kLower$.
    
    Because $a$ is assigned voters only in $[i,i^*]$, $b$ is only assigned voters on in $[j^*,j]$ and we assume the previous entries also satisfy (T\ref{dyntableprop3}), the alternatives $a$ and $b$ are assigned $\na$ and $\nb$ voters respectively, satisfying (T\ref{dyntableprop3}).
    
    Condition (T\ref{dyntableprop4}) is satisfied as the number of unassigned voters on each side sum up to $\nd$ and therefore the total number of unassigned voters on the interval must be at least $\nd$. 
    
    As we know that no alternatives are used in $\committee'$ and $\committee''$, property (T\ref{dyntableprop5}) is satisfied, as the table configuration corresponding to $(\committee',\repre')$ satisfied the condition.
    
    Using the same steps as in Case $1$ we can then by applying \cref{lemmaMNO} once again get a solution that satisfies (T\ref{dyntableprop6}) and (T\ref{dyntableprop7}).
    
As each of the operands of the $\max$-operation satisfies (T\ref{dyntableprop1})--(T\ref{dyntableprop7}) for all of the operations \eqref{eq:SPMlower1}--\eqref{eq:SPMsum}, the results of the operations must satisfy them as well. This concludes Step 2.
   \paragraph{Step 4: Optimality.}
   
	We now show that the dynamic table finds an optimal solution. We show this by showing that given an optimal partial solution $(\committee', \repre')$ corresponding to some tuple $(a, b, i, j, \kUpper, \kLower, \na, \nb, \nd,B,c',i',j')$ that satisfies (T\ref{dyntableprop1})--(T\ref{dyntableprop7}), the algorithm will find $(\committee', \repre')$ or a different partial solution satisfying  (T\ref{dyntableprop1})--(T\ref{dyntableprop7}) that satisfies at least as many voters as this solution. %
Throughout this proof, we assume that the previous table entries give optimal solutions. The base case has been shown in Step 1.	
	
	We again distinguish between two types of table entries: 
	
   \begin{compactenum}
      \item $a=b$ (corresponds to the operations \eqref{eq:SPMlower1} and \eqref{eq:SPMlower2})
      \item $a\neq b$ (corresponds to the operation \eqref{eq:SPMsum})
      \end{compactenum}
   
   In the first case, suppose we are given a table configuration $T(a,a,i,j,\kUpper,\kLower,\na,\na,\nd,B,c',i',j')$. Let $(W',\repre')$ be a maximal \NT\ partial solution with respect to $[i,j]$ that fulfills (T\ref{dyntableprop1})-(T\ref{dyntableprop7}) for this configuration and let $\mathcal{I}'$ be the corresponding \optimal\ intervals. Let $j^*\in[i,j]$ be the last voter in this interval that is happy with the alternative it is assigned to.
   
If $\repre'(j^*)=a$, then we can compute the table entry with the following formula:
    \begin{align}
    &T(a,a,i,j,\kUpper,\kLower,\na,\na,\nd,B,c',i',j')=\\&T(a,a,i,j^*-1,\kUpper,\kLower,\na-1,\na-1,0,B,c',i',j')+\\&T(a,a,j^*,j,0,0,1,1,\nd,0,c',i',j')
    \end{align}
This follows as each alternative $a_t\in\usableset(a,i,j,c',i',j')\setminus\usableset(a,i,j^*-1,c',i',j')$ must satisfy $\leftmost{a_t}\geq j^*$ and can therefore not be assigned happy voters in the interval $[i,j^*-1]$.
    Note that due to (T\ref{dyntableprop7}) it follows that all the unassigned voters must be after $j'$.
    
One can verify that there are values for the $\max$-operation in \eqref{eq:SPMlower1} and \eqref{eq:SPMlower2} that corresponds to these calculations. Therefore the result of the corresponding computation of the entry $T(a,a,i,j,\kUpper,\kLower,\na,\na,\nd,B,c',i',j')$ must be at least $T(a,a,i,j^*-1,\kUpper,\kLower,\na-1,\na-1,0,B,c',i',j')+ T(a,a,j^*,j,0,0,1,1,\nd,0,c',i',j')$. If we can show that the optimal solution can be correctly computed using this computation, then the table entry must correspond to a solution with at least as high number of satisfied voters.

Assume $a_s=\repre'(j')\neq a$.
We proceed by showing the following claim:
 
  \begin{claim}\label{lemma:inUinU3}
If $a_s\in\usableset(a,i,j,c',i',j')\setminus\usablesetS(a,i,j,c',i',j')$, then there exists an alternative $a_t\in\usablesetS(a,i,j,c',i',j')$ such that $[i_s, j_s] \subseteq [i_t, j_t]$ and $a_s\in\usableset(a_t,i_t,j_t,c',i',j')$ or $a_s\in\usableset(a_t,i_t,j_t,a,i,j)$.
  \end{claim}
  \begin{proof} [Proof of \cref{lemma:inUinU3}] \renewcommand{\qedsymbol}{(of \cref{lemma:inUinU3})~$\diamond$}
We know $\usablesetS(a,i,j,c',i',j')\cap\dominators(a_s)\neq\emptyset$, due to $a_s \in\usableset(a,i,j,c',i',j')\setminus\usablesetS(a,i,j,c',i',j')$. Thus  we can look at the non-empty set $D\coloneqq\usablesetS(a,i,j,c',i',j')\cap\dominators(a_s)$. 
  
  Since $a_s\in\usableset(a,i,j,c',i',j')\setminus\usablesetS(a,i,j,c',i',j')$, we know by \cref{LemmaIntintcontainment} that for one alternative $a_t\in \committee' \setminus \{a\}$ it must hold that $[i_s,j_s]\subseteq[i_t,j_t]$. Because $[i_s, j_s] \subseteq [i, j]$, we also have that $[i_t, j_t] \subseteq [i, j]$ or $[i, j] \subseteq [i_t, j_t]$ by the definition of \good\ intervals. Because $\committee' \subseteq \usableset(a,i,j,c',i',j')\cup\{a\}$, we must have that $a_t \in  \usableset(a,i,j,c',i',j') \subseteq \subordinates(a)$.
  We need to distinguish three cases:
  \begin{compactenum}[(i)]
  \item $\leftmost{a_s}\in[i_t,j_t]$
  \item $\leftmost{a_s}\in[i,i_t-1]$
  \item $\leftmost{a_s}\in[i',i-1]$
  \end{compactenum}
  Note that the case $\leftmost{a_s}<i'$ can be ignored as we know $i'\leq\leftmost{a}\leq\leftmost{a_s}$.
  
  \noindent\textbf{Case 1: $\leftmost{a_s}\in[i_t,j_t]$.} In this case trivially $a_s\in\usableset(a_t,i_t,j_t,c',i',j')$ and $a_s\in\usableset(a_t,i_t,j_t,a, i, j)$ holds.
  
  \noindent\textbf{Case 2: $\leftmost{a_s}\in[i,i_t-1]$.} Suppose, for the sake of contradiction that $a_s\notin\usableset(a,i,j,c',i',j')$ holds. Let $b\in\dominators(a_s)\cap\subordinates(a)\cap\earlier(a_t)$ be a blocking alternative. Due to $b$ being a blocking alternative it must satisfy either $\leftmost{b}>i$, or $\leftmost{b}<i$ and $\earlier(a)\cap\dominators(b)=\emptyset$. In either case $b\in\usableset(a,i,j,c',i',j')$ would hold. Due to the neat ordering $a_s$ would need to be assigned happy voters before $a_t$. This however conflicts with $\mathcal{I}$ being \optimal, because if we choose a blocking alternative $b$ such that $b$ is undominated amongst blocking alternatives and $\rightmost{b}$ is maximal, then it follows via \cref{LemmaLongInt} that the interval of $b$ would need to contain $[i_s,j_s]$ a contradiction to our choice of $a_r$. Therefore $a_s\in\usableset(a,i,j,c',i',j')$ must hold.
  
  \noindent\textbf{Case 3: $\leftmost{a_s}\in[i',i-1]$.} Due to $a_s\in\usableset(a,i,j,c',i',j')$, it follows that there does not exist an alternative $\earlier(a)\cap\dominators(a_s)\cap\subordinates(c')$ that would lead to $a_s$ not being in $\usableset(a,i,j,c',i',j')$. Therefore it follows that $\earlier(a_t)\cap\dominators(a_s)\cap\subordinates(c')\subseteq\usableset(a,i,j,c',i',j')$, because any alternative that would dominate an alternative in $\dominators(a_s)\cap\subordinates(c')$ (leading to it not being in $\usableset(a,i,j,c',i',j')$ ) would also dominate $a_s$. Then one can proceed as in the previous case.
  \end{proof}

  Note that because of \cref{lemma:inUinU3} we know that either $a_s\in\usablesetS(a,i,j,c',i',j')$ or there exists an alternative $a_t\in\usablesetS(a,i,j,c',i',j')$ such that $a_s\in\usableset(a_t,i_t,j_t,a,i,j)$ or $a_s\in\usableset(a_t,i_t,j_t,c',i',j')$ and $[i_s,j_s]\subseteq[i_t,j_t]$.

    We can therefore look at the \optimal\ intervals of alternatives in $\usablesetS(a,i,j,c',i',j')\cap W'$. Let $b_1,\ldots,b_x$ be those alternatives ordered based on the first voter that approves of them. (Note that due to them not being dominated in this set, it follows that they cannot have the same voter approving them first.) Let $[i_{\ell},j_{\ell}]$ be the \optimal\ interval for each $b_{\ell}$. Let $j_0=i-1$ and $i_{x+1}=j$. We distinguish two cases:
    \begin{compactenum}[(a)]
    \item There exists $\ell\in[0,x]$, such that $j_{\ell}+1\neq i_{\ell+1}$.
    \item There does not exist such $\ell$.\label{caseb}
    \end{compactenum}
    
    We distinguish between these cases because $j_{\ell}+1\neq i_{\ell+1}$ implies that the voter $j_{\ell}+1$ does not approve of any alternative except $a$ that is used in the partial solution corresponding to the configuration in the table - otherwise the collection of intervals is not \optimal\ as we could lengthen one of the larger intervals. In the first case we can split the table along $j_{\ell+1}$ in the following way:
    \begin{align*}
    &T(a,a,i,j,\kUpper,\kLower,\na,\na,\nd,B,c',i',j')=\\&T(a,a,i,j_{\ell},\kUpper_1,\kLower_1,\na',\na',\nd_1,B,c',i',j')+\\&T(a,a,j_{\ell}+1,j,\kUpper_1,\kLower_1,\na'',\na'',\nd_2,0,c',i',j')
    \end{align*}
    
Again can verify that there are values for the $\max$-operation in \eqref{eq:SPMlower1} and \eqref{eq:SPMlower2} that corresponds to these calculations. Therefore the result of the corresponding computation of the entry $T(a,a,i,j,\kUpper,\kLower,\na,\na,\nd,B,c',i',j')$ must be at least the result of the computation.
    Due to $j_{\ell}+1\neq i_{\ell+1}$ we know that splitting the interval has no alternative on the right side, except $a$ that could be assigned voters before $j_{\ell}+1$. We show a useful claim to show the correctness of this split:
    
    \begin{claim}\label{cl:Uinclusion}
         For any $i^*\in[i,j-1]$, it holds that $\usableset(a,i,j,c',i',j')=\usableset(a,i,i^*,c',i',j')\cup\usableset(a,i^*+1,j,c',i',j')$.
       \end{claim}
       \begin{proof}[Proof of \cref{cl:Uinclusion}] \renewcommand{\qedsymbol}{(of \cref{cl:Uinclusion})~$\diamond$}
         It is clear that $\usableset(a,i,j,c',i',j')\supseteq\usableset(a,i,i^*,c',i',j')\cup\usableset(a,i^*+1,j,c',i',j')$, since both sets on the right side are subsets of $\usableset(a,i,j,c',i',j')$. We now consider an alternative $e\in\usableset(a,i,j,c',i',j')\setminus\usableset(a,i,i^*,c',i',j')$. For this alternative it must holds that $\VV(e)\cap[i,i^*]=\emptyset$ but $\VV(e)\cap[i,j]\neq\emptyset$. Therefore it follows that $\leftmost{e}\in[i^*+1,j]$ and from that $e\in\usableset(a,i^*+1,j,c',i',j')$.
       \end{proof}

    We know from \cref{cl:Uinclusion} that the alternatives usable on each side of the interval correspond to the alternatives that are used on the whole interval. Now if we choose the configuration on each side to correspond with the value of the partial solutions on that interval the value must be the expected value due to both intervals being correctly computed per precondition.

    In case \eqref{caseb}, it follows that all voters in $[i,j]$ approve of some alternative in $\usablesetS(a,i,j,c',i',j')\cap \committee$.
    
    We can now consider two further cases:
    \begin{compactenum}[(\ref{caseb}.i)]
   \item $i=\leftmost{b_1}$\label{caseb1}
   \item $i>\leftmost{b_1}$\label{caseb2}
    \end{compactenum}
    
    In case (\ref{caseb}.\ref{caseb1}) we show that $T(b_1,b_x,i,j,\kUpper',\kLower',n_1,n_x,\nd+\na,B,a,i,j)$ can be used to create $T(a,a,i,j,\kUpper,\kLower,\na,\na,\nd,B,c',i',j')$. One can again verify that this is one of the operands of the $\max$-operation for \eqref{eq:SPMlower1} or \eqref{eq:SPMlower2} and thus gives a lower bound for the result. We now show that the alternatives that are be used in $(\committee,\repre)$ with the exception of $a$ can also be used in this table entry. Suppose there exists an alternative $a'\in (\committee\setminus\{a\})\setminus(\{b_1,b_x\} \cup \inbetw(b_1,b_x)\\ \bigcup_{c\in\{a,b\}\cup\inbetw(a,b)}\usableset(c,\max\{i,\leftmost{c}\},\min\{j,\rightmost{c}\},c',i',j')$. Suppose it is not dominated by $b_1$ or $b_x$. In this case it follows that $a' \in \inbetw(b_1,b_x)$ and must therefore be in the set. Suppose $a'$ is not dominated by $b_1$ but dominated by $b_x$. Then it must be in $\usableset(b_x, \max(i,\leftmost{b_x}), j, a, i, j)$, as $i\leq\leftmost{b_x}$ and all alternatives $a_p$ that are dominated by $b_x$ must satisfy $\leftmost{b_x}\leq\leftmost{a_p}$. Similarly, if $a'$ is dominated by $b_1$ it must be in $\usableset(b_1, i, \min(\rightmost{b_1}, j), c', i', j')$. Therefore all cases have been considered and the same alternatives must be usable. Let $(\committee',\repre')$ be the corresponding solution that satisfies (T\ref{dyntableprop1})-(T\ref{dyntableprop7}). Suppose that $T(b_1,b_x,i,j,\kUpper',\kLower',n_1,n_x,\nd+\na,B,c,i,j)$ is not equal to the number of happy voters in $(W,\repre)$ minus $\na$. If it is larger, then the number of happy voters in $(\committee,\repre)$ cannot be optimal, as by assigning the first $\na$ unassigned voters to $a$ we get a better solution. It cannot be lower either, as in this case by leaving the voters that were previously assigned to \na\ unassigned we would find a better solution for the table entry contradicting our inductive hypothesis.

    In case (\ref{caseb}.\ref{caseb2}) we show that $T(b_1,b_x,i,j,\kUpper',\kLower',n_1,n_x,\nd+\na,B,c',i',j')$ can be used to create $T(a,a,i,j,\kUpper,\kLower,\na,\na,\nd,B,c',i',j')$. One can again verify that this is one of the operands of the $\max$-operation for \eqref{eq:SPMlower1} or \eqref{eq:SPMlower2} and thus gives a lower bound for the result. We now show that the alternatives that are be used in $(\committee,\repre)$ with the exception of $a$ can also be used in this table entry.
    Suppose there exists an alternative $a' \in (\committee\setminus\{a\})\setminus(\{b_1,b_x\} \cup \inbetw(b_1,b_x)\\ \bigcup_{c\in\{a,b\}\cup\inbetw(a,b)}\usableset(c,\max\{i,\leftmost{c}\},\min\{j,\rightmost{c}\},c',i',j')$. Per our choice of $b_1$ and $b_x$ it follows that this alternative cannot be undominated in $\usableset(a,i,j,c',i',j')$. Suppose it is not dominated by $b_1$ or $b_x$. In this case it follows that $a'\in\inbetw(b_1,b_x)$. Suppose $a'\in\dominators(b_1)\cup\dominators(b_x)$. Without loss of generality we will assume that $a'\in\dominators(b_1)$. Note that $i>\leftmost{a'}$ must hold, as else $a'\in\usableset(b_1, i, \min(\rightmost{b_1}, j), c', i', j')$.
    
    Let $b'$ be blocking such that $a'$ is not usable in $\usableset(b_1, i, \min(\rightmost{b_1}, j), c', i', j')$. Because we require (T\ref{dyntableprop6}) of $(\committee,\repre)$ it follows that $b'\notin\usableset(a,i,j,c',i',j')$. Therefore there must exist an alternative $b''$ that blocks $b'$ from being usable. Since $i>\leftmost{a'}$ it follows that $b''$ also blocks $a'$ from being usable in $\usableset(a,i,j,c',i',j')$. This directly contradicts  $a'\in (W\setminus\{a\})$ and completes this part of the proof.\\

We now show the optimality of the table entries of type 2. Let $T(a,b,i,j,\kUpper,\kLower,\na,\nb,\nd,B,c',i',j')$ be a tuple and let $(\committee,\repre)$ be a partial solution for $[i,j]$ that is optimal, neatly ordered and satisfies (T\ref{dyntableprop1})-(T\ref{dyntableprop7}) for the given table configuration and let $\mathcal{I}$ be the corresponding \optimal\ collection of intervals.
 
  Let $[i_b,j]$ be the \optimal\ interval of $b$ in this partial solution. We know from \cref{LemmaLongInt} that the right endpoint of the interval must be $j$ because of optimality. We show that $\usableset(b,i_b,j,c',i',j')\cup\{b\}$ must contain all alternatives that are used on the interval $[i_b,j]$. Suppose, to contradiction, that there exists an alternative $d\in \committee\setminus\usableset(b,i_b,j,c',i',j')\cup\{b\}$ that is assigned voters in the interval. Then $\leftmost{d}<(i_b)$ must hold. Otherwise either $d$ would have to be in $\usableset(b, i_b, j, c', i', j')$ or it could not be assigned voters in the interval. As we require $\usableset(c',i',j',c',i',j')\cap\dominators(a)\cap\earlier(b)=\emptyset$ we can lead this to a contradiction in the following way: Let $b'$ be an alternative that blocks $d$ from being usable. It cannot satisfy $b'\in\earlier(a)$, as then $
 b'\in\dominators(a)$ which would contradict $\usableset(c',i',j',c',i',j')\cap\dominators(a)\cap\earlier(b)=\emptyset$. If $b'\in\subordinates(a)$, then $a$ would also be a blocking alternative and then it follows via \cref{LemmaLongInt} that $d$ cannot be used in the interval $[i_b,j]$. If $a\in\earlier(b')$, $d$ could not be used in the interval with the same reasoning, as $f$ would be in the solution. This leads to a contradiction. Therefore such an alternative $d$ can not exist.
 
 Let $c$ be the last alternative in $\inbetwS(a,b)\cup\{a\}$ (ordered by $\leftmost{c}$) that is used in the partial solution.
 Then $T(a,b,i,j,\kUpper,\kLower,\na,\nb,\nd,B,c',i',j')=T(a,c,i,i^*,\kUpper_1,\kLower_1,\na,n_c,\nd_1,B,c',i',j')+T(b,b,j^*,j,\kUpper_2,\kLower_2,\nb,\nb,\nd_2,B',c',i',j')$, where $i'$ is the first voter in the interval of $b$, $j'$ is the last voter in the interval of $c$, $\kUpper_1,\kUpper_2,\kLower_1,\kLower_2$ are the number of alternatives assigned $\lceil\frac{n}{k}\rceil$ used in the intervals, excluding $a,b$ and $c$, $n_c$ is the number of voters allocated to $c$, $\nd_1,\nd_2$ are lower bounds on the number of unassigned voters on either half such that $\nd_1+\nd_2=\nd$, and $B'$ is $0$ or $1$.

 Now we show that if an alternative is used in $[i,i_b-1]$ it is part of \begin{align*}
 &L=\{a,c\} \cup \inbetw(a,c) \\&\bigcup_{d\in\{a,c\}\cup\inbetw(a,c)}\usableset(d,\max\{i,\leftmost{d}\},\min\{i_b-1,\rightmost{d}\},c',i',j')
 \end{align*}
 Towards a contradiction, let $a'\in W\setminus L$ be an alternative that is used but not in $L$.
 Note that, as $c$ is the last alternative in $\inbetwS(a,b)\cup\{a\}$ (ordered by $\leftmost{c}$) that is used in $(\committee,\repre)$, it follows that if $a'$ is assigned happy voters in $[i,i_b-1]$ and it is not in $L$ it must be comparable to either $a$ or $c$, because otherwise it would be in $\inbetw(a,c)$. Note that $a'$ cannot be incomparable to $c$ and later than $c$, as in this case our choice of $c$ would be incorrect.
 
 If $a'\in\subordinates(a)$ it follows from $a'\in\usableset(a,i,$ $\min\{j,\rightmost{a}\},$ $c',i',j')\setminus\usableset(a,i,\min\{i_b-1,$  $\rightmost{a}\},$ $c',i',j')$ that no voters in $[i,i_b-1]$ approve of $a'$ thereby contradicting that any voter in the interval is happy with $a'$. If $a'\in\subordinates(c)\setminus\subordinates(a)$, then it follows that there must be an alternative in $\inbetw(a,c)$ that blocks it from being usable. However since this alternative is also considered in $\bigcup_{d\in\{a,c\}\cup\inbetw(a,c)}\usableset(d,\max\{i,\leftmost{d}\},\min\{i_b-1,\rightmost{d}\},c',i',j')$ it follows that $a'$ must be in the set as well, a contradiction.

 We have shown that the table configuration in this case can use all alternatives in $\committee$. Once again if the sum of the two table entries exceeds the number of happy voters in $(\committee,\repre)$, we contradict its optimality, whereas if sum is less than the number of happy voters the previous table entries were incorrect. Combining the two table entries therefore leads to the correct table entry.
 
 This concludes the proof.

   \end{proof}
}   
   
\toappendix{
   \begin{lemma}\label{lemma:Monroetime}
   The dynamic programming table can be computed in $O(n^{11} m^5 \csize)$.
   \end{lemma}
   \begin{proof}
   First consider the table $T(a,b,i,j,\kUpper,\kLower,\na,\nb,\nd,B,c',i',$ $j')$. We know that $a, b, c' \in C$ and thus they can take $m$ different values. Similarly, $i, j, i', j' \in V$ so they can take $n$ different values. The variables $\kUpper,\kLower$ describe how many alternatives can represent some specific number of voters. Their values must thus be bounded above by $\csize$. The variables $\na$ and $\na$ describe the number of voters $a$ and $b$ represent. Thus their values are bounded above by $\lceil \frac{n}{\csize} \rceil$. The variable $\nd$ represents a number of voters and thus its value is bounded by $n$. Boolean $B$ can take at most two different values. Thus the size of the table is in $O(n^5 m^3 (\frac{n}{\csize})^2 \csize^2) = O(n^7 m^3)$.\\
    
   Next we show that each cell can be computed in polynomial time.
   
   We note that $\usableset(a,i,j,c',i',j')$ can be computed in $O(m^2)$ time, by checking for each subordinate of $a$ whether there exists an alternative that blocks it from being usable. 
   
   The initialization step \eqref{eq:SPMinit1} can be done in constant time.
   
   In Case $1$ \eqref{eq:SPMlower1} and Case $2$ \eqref{eq:SPMlower2} we take the maximum over two components. The first component iterates over two alternatives and the possibilities of number of voters assigned to them and two possible values of $B$ and thus takes at most $O(m^2\frac{n^2}{k^2})$ time. The second component iterates over two voters and different ways to represent $\na,\nd, \kUpper$ and $\kLower$ as a sum of two values. Because $\nd$ is bounded by $n$,  $\na$ by $\lceil \frac{n}{\csize} \rceil$ and $\kUpper$ and $\kLower$ by $\csize$, the second component takes at most $O(n^4 \csize)$. Thus these steps takes at most $O(n^4 \csize + m^2\frac{n^2}{k^2})$ time to compute.
   
   Case $3$ \eqref{eq:SPMsum} consists of three components. In them we iterate over two voters, up to one alternative, a value in the range $[0, \lceil \frac{n}{\csize} \rceil]$ and different ways to represent values $\nd, \kUpper$ and $\kLower$ as sums of two values. Like before, $\nd$ is bounded above by $n$ and $\kUpper$ and $\kLower$ by $\csize$. Thus the step takes at most $O(n^3 m \frac{n}{\csize}  \csize ^2) = O(n^4 m \csize )$ time.
   
   By taking the maximum over the times to compute table entries, we obtain that the first table can be computed in time $O(n^7 m^3 \max(n^4 \csize + m^2, n^4 m \csize))$ which is bounded above by $O(n^{11} m^5 \csize)$.
   
   \end{proof}
}   

   \ifshort\begin{thm}[\appsymb]\else\begin{thm}\fi\label{thm:MonroeP}
       For SC profiles, \app-\Monroe-\summw and \app-\Monroe-\maxmw can be solved in polynomial time.
   \end{thm}

\appendixproofwithsketch{thm:MonroeP}{
\todoIH{The following proof idea is a bit too disconnected from the rest. Could you refer to the lemmas and definitions whenever possible? The tables, the order, etc. Also please try to be a bit more precise.}
\begin{proof}[Proof idea]
  We briefly discuss why the DP given in \eqref{eq:SPMlower1}--\eqref{eq:SPMsum} is correct.
  Each DP entry corresponds to an optimal partial solution on interval $[i,j]$ when only alternatives that are usable for $a, b$, alternatives between them and alternatives usable for them may be assigned happy voters. By \cref{lemmaMNO} we can always find an optimal solution which is \MNT. In such a solution, we can construct a collection of \good\ intervals (\cref{def:good-optimal-intervals}). If the interval of a voter $a \in W$ is $[i,j]$, we can show that apart from $a$, the only alternatives that are usable for~$a$ (\cref{def: U}) can be assigned happy voters on the interval $[i,j]$. Thus each partial solution considers all the alternatives that could be used for it and ignores the rest.

In Operations~\eqref{eq:SPMlower1}--\eqref{eq:SPMlower2} we use partial solutions from lower level alternatives (\cref{def:levels}) to build a solution for the dominating alternative. In the first $\max$-operation we assume $b$ and $c$ are in the usable set of $a$, and we build a new committee that also assigns some voters to $a$. Because we keep track of unassigned voters with $\nd$, we know that there are enough unassigned voters to assign them to $a$. We can show that all the alternatives that are usable for $b, c$ and the alternatives inbetween must also be usable for $a$, and thus the table requirements are preserved.
In the second $\max$-operation we combine two committees that use alternatives usable for $a$ but assign voters on disjoint intervals. With the parameters counting the number of assigned voters, we can be sure that we do not violate proportionality. The value $B=0$ in the second entry enforces that the alternatives we use do not overlap.
In Operation \eqref{eq:SPMsum} we combine two disjoint table entries on the same level (\cref{def:levels}) to obtain a ``wider" committee. By \cref{LemmaLongInt}, we can show that the usable sets must be disjoint; the proof is in the full version of the paper. Therefore, as long as we take care to not assign too many voters, we will obtain a valid partial solution.

The final calculation goes over all intervals of $\level$-1 alternative and all possible voter intervals they could cover. As \cref{LemmaIntvotcontainment} states that \good\ intervals must cover all the \hassignable\ voters, we do not ignore any voters when we compute the final optimal solution.

It is easy to see that the algorithm for \summw\ can be adapted to solve \maxmw.
\end{proof}

}{   
   \begin{proof}
     Using \cref{lemma:Monroecorr,lemma:Monroetime} we obtain that all table entries are correct and computed in polynomial time. It remains to show that $\max\limits_{\substack{a,b\in\{c\in C\mid \dominators(c)=\emptyset\}\\ i\leq j\\\kUpper\leq n\mod k -f_1(a,b,\na,\nb)\\\kUpper+\kLower+|\{a,b\}|\leq k\\\na,\nb\leq\upperB}}T(a,b,i,j,\kUpper,\kLower,\na,\nb,0,1,0,0,0)$ is the maximum number of satisfiable voters under any assignment. From the correctness of the table entries it follows that the value is the number of happy voters in a valid solution. 
     
     We proceed to show this corresponds to the number of happy voters in an optimal solution. Let $\solution$ be an optimal \MNT\ solution that satisfies (T\ref{dyntableprop7}). By applying the same method as in the proof of \cref{cl:NT-2}, one can derive an optimal solution with this property from an optimal solution that is \MNT. Let $a\in W$ be an undominated alternative that minimizes $\leftmost{a}$ and $b\in W$ an undominated alternative that maximizes $\leftmost{b}$. Let $\na$ and $\nb$ be the number of happy voters assigned to $a$ and $b$, respectively. Let $i$ be the first voter that is happily assigned and $j$ the last voter that is happily assigned. Let $\kUpper$ and $\kLower$ be the number of alternatives excluding $a$ and $b$ that are assigned $\lceil\frac{n}{k}\rceil$ respectively $\lfloor\frac{n}{k}\rfloor$ happy voters. We now show that $T(a,b,i,j,\kUpper,\kLower,\na,\nb,0,1,0,0,0)$ is the number of happy voters in $\solution$. To achieve this, we show that $(W,\repre)$ must satisfy (T\ref{dyntableprop1})-(T\ref{dyntableprop7}).  Due to our choice of $a$ and $b$ it follows that all undominated alternatives in $W$ must be in $\inbetw(a,b)$. As $\solution$ is \MNT, for each alternative $c\in W$ it holds that $\dominators(c)\subseteq W$. Without loss of generality, let $c\in W\cap\happyrepre^{-1}([i,j])$ be a dominated alternative. We choose $d\in\dominators(c)$ such that it is undominated and it minimizes $\leftmost{d}$. Then $\earlier(d)\cap\dominators(c)=\emptyset$ and $d\in W$ must hold. Due to our choice of $a$ and $b$ it follows that $d\in\inbetwS(a,b)\cup\{a,b\}$. Then it follows that $c\in\usableset(d,\max\{\leftmost{d},i\},\min\{\rightmost{d},j\},0,0,0)$, since $c$ is assigned happy voters. Therefore it follows that all alternatives that are in $W$ are part of $\committeeR\cup\{a,b\}$ and the solution satisfies (T\ref{dyntableprop1}). (T\ref{dyntableprop2}) and (T\ref{dyntableprop3}) are satisfied by choice of the variables. As $\nd=0$, property (T\ref{dyntableprop4}) is satisfied as well. Similarly, as $B=1$, property (T\ref{dyntableprop5}) is satisfied for $\solution$. Because the solution is \MNT\, it is also monotone for any subset of $\aaa$ and therefore (T\ref{dyntableprop6}) must be satisfied for $\solution$ as well. As we have chosen $\solution$ such that it satisfies (T\ref{dyntableprop7}) and shown that there exists an optimal solution that satisfies (T\ref{dyntableprop7}), $\solution$ satisfies all properties we required of it. This concludes the proof.

   \end{proof}
 }

 \section{Parameterized algorithms}\label{sec:FPTSPCC}
 \appendixsection{sec:FPTSPCC}
 In this section, we show parameterized results for \CC-\mw, \appMsum\ and \Monroe-\maxmw\ for nearly \SP\ and \SC\ profiles wrt. the number of alternatives or voters to delete. We achieve all the results in this section by extending DP approaches from the literature and \cref{sec:monroe-SC}. Note that one can find a set of $\delv$ voters (resp. alternatives) deleting which yields an SP (resp. SC) profile in FPT time wrt.\ $\delv$. Hence we assume in this section we are given such a set as part of the input. 

 \todoIC{Removed for linear preferences. I think the sentence is fine because the result for approval preferences solve the max-problem through the observation. Max-problem was also an open question. S: Changed the sentence a bit.}

\toappendix{
Throughout this section, let $I=(\profiletuple, \misr, \csize)$ be an instance and let \myemph{$\delV$} denote the set of \myemph{$\delv$} voters that need to be deleted in order to make the profile \SP or \SC.
For notational convenience, let $\aaa=[m]$ and \myemph{$\reV = \vvv\setminus \delV$} such that $\reV=[n]$.
For nearly SP preferences, let \myemph{$1\rhd 2 \rhd \ldots \rhd m$} denote an SP order, and for nearly SC preferences, let \myemph{$1 \rhd 2 \ldots \rhd n$} denote an SC order after $\delV$ is deleted.
Given a voter subset~$V'\subseteq \vvv$ and two alternatives~$j, j'\in \aaa$, let \myemph{$\misr(V',j)$} denote the sum of misrepresentations of all voters towards~$j$  and \myemph{$\costdiff(V', j, j')$} the misrepresentation difference of voters in~$V'$ that are more satisfied with $j$ than with $j'$,
i.e., $\costdiff(V',j,j') = \sum_{v\in V'} \max\big(\misr(v, j') - \misr(v, j), 0\big)$.
Note that for approval preferences, this is exactly the number of voters that approve~$j$ but not~$j'$.
Finally, for a subset of alternatives~$C'\subseteq \aaa$ and a voter subset~$V'\subseteq \vvv$, let \myemph{$\misr(V', C')$} denote the minimum sum of misrepresentations of $V'$ towards~$C'$, i.e., $\misr(V',C') = \sum_{v\in V'}\min_{c\in C'}\misr(v,c)$.}

\ifshort\begin{thm}[\appsymb]\else\begin{thm}\fi\label{thm:nearSPvoter}
  \CCsum\ is \FPT wrt. the number~$\delv$ of voters to delete to obtain an SP profile.
\end{thm}
\appendixproofwithsketch{thm:nearSPvoter}{

\begin{proof}[Proof idea]
 For approval based voter deletion, the basic idea is to group the alternatives together that are approved by the same set of deleted voters, and observe that for each group we need at most one alternative of each group to represent the deleted voters. Hence, we can extend the DP algorithm by Betzler et al.~\cite{BetzlerMW2013} by trying all possible subsets of groups.
 
 The idea for linear voter deletion is a bit more involved. We instead guess in \FPT\ time an ordered partition of the deleted voters which corresponds to the ordering of the alternatives according to the SP ordering. This allows us to extend the DP algorithm by Betzler et al.~\cite{BetzlerMW2013} by trying all possible ordered partitions.
\end{proof}

}{ 
\begin{proof}
  We first consider approval preferences, and then linear preferences.

  \noindent \textbf{Approval preferences.}
  Before we show the proof, we introduce several necessary notion. 
  We say that two alternatives have  the same \myemph{type} if they are approved by exactly the same set of ``deleted'' voters, and let \myemph{$\type(a)$} $ = \VV(a)\cap \delV$, and \myemph{$\type(C')$} $= \{\type(a) \mid a \in C'\}$ for all $C'\subseteq \aaa$.
  Let \myemph{$\Alltypes$} denote the set of all types of alternatives, i.e, $\Alltypes = \{\type(a) \mid a \in \aaa\}$, and for each subset~$T\subseteq \Alltypes$ of types, let \myemph{$\cost_{\delV}(T)$} denote the sum of misrepresentations of the ``deleted'' voters towards the alternatives with types~$T'$, i.e.,
  $\cost_{\delV}(T) = |\delV \setminus (\bigcup\limits_{V'\in T}V')|$.

  The algorithm is based on the observation that to represent the voters from~$\delV$ we need at most one alternative for each type.
  Hence, we can compute for each subset of types an optimal partial solution which additionally covers all these types by modifying the DP algorithm of Betzler et al.~\cite{BetzlerMW2013}.
  Since there are $\delv$~voters in $\delV$, there are at most $2^{\delv}$ many types of alternatives, i.e., $|\Alltypes| \le 2^{\delv}$,
  and we show later that the new algorithm runs in \FPT\ time wrt.~$\delv$.

  More precisely, we compute a DP table~$A_1$ such that for each $\csize' \in [\csize]$, $j\in \aaa$, and $T\subseteq \Alltypes$ 
  it holds that
  \begin{align}\nonumber
    A_1(j, \csize', T)  =  \min \{\misr(\vvv, \committee) \mid & \committee\subseteq [j] \wedge |\committee| = \csize' \wedge j \in W \\
                                                         & \wedge T = \type(\committee) \}.\label{DP-ApprovalCC-SP-t}
  \end{align}
For this reason we require that $\type(j)\in T$ for any table entry $ A_1(j, \csize', T)$.
  We initialize the table as follows.
  \begin{align*}
    A_1(j,\csize',T)=
    \begin{cases}
      \misr(\vvv,j),& \text{if } \csize' = 1 \text{ and }T=\{\type(j)\},\\
      \infty, & \, \text{if } \csize' = 1 \text{ and }T\neq\{\type(j)\},\\
      \infty, & \, \text{if } j < \csize'.
    \end{cases}
  \end{align*}
  
  We compute the other values as follows:
  \begin{align}
    A_1(j, \csize', T) =\!
    \!\min\limits_{\substack{j' \in [j-1]\\T'\in \left\{T,T\setminus \{\type(j)\}\right\}}}\!\! & \big\{ A_1(j', \csize'-1, T') - \costdiff(\reV, j, j') +\nonumber \\
    &  \cost_{\delV}(T) - \cost_{\delV}(T') \big\}.\label{A_1-rec}
  \end{align}
  Finally, to compute the misrepresentation of an optimal solution, we compute and return
  $\min_{j\in \aaa, T \subseteq \Alltypes} A_1(j, \csize , T)$.

  It is straightforward to verify that the DP runs in $O(m^2\cdot k\cdot n\cdot 2^{\delv})$ time as the number of entries is $O(m\cdot k\cdot 2^{\delv})$ and computing each entry needs $O(m\cdot n)$ time.\todoIC{C: Check, please.}
 
  The correctness proof works similarly to the one given by Betzler et al.~\cite{BetzlerMW2013}. 
  We show this for the sake of completeness.
  Let $\committee^*$ be a size-$\csize$ committee that minimizes $\misr(\vvv, \committee^*)$.
  If we can show that the entries in~$A_1$ comply with~\eqref{DP-ApprovalCC-SP-t}, then we can show that $A_1(\max(\committee^*), \csize, \type(\committee^*))$ is equal to $\misr(\repre^*)$.
  Thus, it remains to show that each entry in~$A_1$ equals the RHS of~\eqref{DP-ApprovalCC-SP-t}.
  We prove this by induction on~$\csize'$.

  \smallskip
  \noindent \emph{Base case:} $\csize'=1$. In this case, $\{j\}$ is the only committee that satisfies the first three conditions given on the RHS of \eqref{DP-ApprovalCC-SP-t}.
  All voters need to be represented by~$j$.
  If the type of $j$ also covers the alternative types in~$T$, then the last condition on the RHS is also satisfied, and we compute misrepresentations of all voters towards $j$ (see the first case in the initialization step), which complies with  \eqref{DP-ApprovalCC-SP-t}.
  Otherwise, no committee satisfies all conditions on RHS, and the misrepresentation is set to infinity which complies with  \eqref{DP-ApprovalCC-SP-t}.
  
  \smallskip
  \noindent \emph{Induction step.} Assume that $A_1(j, \csize', T)$ complies with \eqref{DP-ApprovalCC-SP-t} for all $j\in \aaa$ and $T\subseteq \Alltypes$.
  Let $\committee$ denote a committee with  
  $\committee \subseteq [j]$, $j\in \committee$,  $|\committee|=\csize'+1$,
  and $T = \type(\committee)$ such that $\misr(\vvv, \committee)$ is minimum among all such committees.
  We aim to show that $A_1(j, \csize'+1, T) = \misr(\vvv, \committee)$.
  First, define $\committee'=\committee \setminus \{j\}$, $j'=\max(\committee')$, and $T' = \type(\committee')$.
  Since $\committee' \subset \committee$, 
  no voter in $\reV$ can be misrepresented more in $\committee$ than in $\committee'$.
  The only voters that improve are those that approve $j$ and no one in $\committee'$. 
  Due to the SP property, we cannot have a voter that approves $j$, $p$ and not $j'$, where $j' < p < j$. 
  Thus all voters~$v$ that improve in $\committee$ have $\misr(v,j') > \misr(v,j)=0$.
  The difference in the misrepresentation of $\reV$ is thus precisely $\costdiff(\reV, j, j')$. 
  For the voters in $\delV$, the difference in misrepresentation between $\committee$ and $\committee'$ is $\cost_{\delV}(T) - \cost_{\delV}(T')$ by the definition of $\cost$.
  Consequently, since $\committee$ is optimal for the configuration~$(j,\csize'+1, T)$, it must hold that $\committee$ is optimal for the configuration~$(j', \csize', T')$. 
  By induction assumption, we have that $A_1(j', \csize', T') = \misr(\vvv, \committee')$.
  Since in \eqref{A_1-rec},  we take the minimum over all the possible $j'$ and $T'$, the correctness of the calculation follows. 
  Altogether, we show that $A_1(j, \csize'+1, T) = \misr(\vvv, \committee)$, as desired. 
  \todoIH{Add running time for the above.}
  
  \medskip  
  \noindent \textbf{Linear preferences.} 
  In this case, we need to guess more structure about the solution since we do not have the types that were useful for the approval preferences.
  We first guess a partition of $\delV$ into $\tilde{t}\leq \min\{\delv,\csize\}$ subsets such that each subset shall be represented by the same alternative, and we also guess in which order their representatives are (according to the SP order).
  Let $V_1,\ldots V_{\tilde{\delv}}$ denote an ordered partition of~$\delV$ such that for each~$z\in [\tilde{\delv}]$ all voters from $V_z$ are represented by the same alternative and for each two indices~$z,z'\in [\tilde{\delv}]$ with $z< z'$ the representative of $V_z$ precedes the representative of $V_{z'}$ in the SP order.
  
  Next, we modify the DP approach from Betzler et al.~\cite{BetzlerMW2013} by computing the misrepresentation when choosing an optimal committee of smaller size from a subset of alternatives that also represents the first $j$ voter subsets.
  We compute $A_2(a,b,\csize',j)$, where $a \in \aaa$, $b\in \aaa \cup \{0\}$,
 $\csize' \in [\csize]$, and $j\in [\tilde{\delv}]\cup \{0\}$.
 We later claim that it equals the misrepresentation of a committee~$\committee$ and an assignment~$\repre$ satisfying  
\begin{compactenum}[(1)]
  \item\label{lin-CC-t-SP:1}  $|\committee|=\csize'$, $\committee\subseteq [a]$, $a \in \committee$,
  \item\label{lin-CC-t-SP:2} for every~$z \in [j]$ it holds that $|\repre(V_z)| = 1$,
\item\label{lin-CC-t-SP:3} for every pair $z,z' \in [j]$ with $z < z'$ it holds that
$c \rhd c'$ where $\repre(V_z) = \{c\}$ and $\repre(V_{z'}) = \{c'\}$, 
\item $b \in \committee$ and $\repre(V_{j})=\{b\}$ if $j > 0$.
\end{compactenum}
that \emph{minimizes} the sum of misrepresentations over $\reV \cup \bigcup_{z \in [j]} V_{z}$.
We initialize the table as follows:
\begin{align*}
  A_2(a,b,1,j)=
\begin{cases}
  \misr(\reV, a), & \text{ if } j=0,\\
  \misr(\reV\cup V_1, a), &\text{ if } a=b \text{ and } j=1,\\
  \infty, & \text{ else}.
\end{cases}
\end{align*}
For all configurations~$(a, b, \csize', j)$ with $a < \csize'$ or $\csize' < j$ or $b < j$ we set $A_2(a,b,\csize, j) = \infty$ as no committee can satisfy the conditions~\eqref{lin-CC-t-SP:1}--\eqref{lin-CC-t-SP:3} under such configuration.
For the other configurations, we compute as follows, distinguishing between $j=0$ and $j > 0$:
\begin{align*}
   A_2(a,b,\csize',0)  =
    \min\limits_{a'\in [a-1]}\big\{A_2(a', 0, \csize'-1, 0) -  \costdiff(\reV, a, a')\big\}. 
\end{align*}
\begin{align*}
  & A_2(a,b,\csize',j)  =\\
  & \; \begin{cases}
    \min\limits_{\substack{a'\in [a-1]\\ b' \in [a']}}  \big\{A_2(a', b', \csize'-1, j-1) -  \costdiff(\reV, a, a')+ \misr(V_j, b)\big\}, \\
     ~\hfill ~~\text{ if } a=b,\\
     \min\limits_{a'\in \{b, \cdots, a-1\}}A_2(a', b, \csize'-1, j) -  \costdiff(\reV, a, a'), 
    ~\hfill ~~\text{ if } a > b.
   \end{cases}
\end{align*}
Finally, to determine the misrepresentation of an optimal solution we compute and return $\min\limits_{(V_1,\ldots, V_{\tilde{\delv}})\colon \text{ordered partition of } \delV}\min\limits_{a, b\in \aaa} A_2(a,b,\csize, \tilde{\delv})$.

  The correctness proof works similarly to the one for the approval preferences.
  Let $(\committee^*, \repre^*)$ be a solution that minimizes~$\misr(\repre^*)$. 
  Accordingly, let $V_1, \ldots, V_{\tilde{\delv}}$ denote an ordered partition of the voters from~$\delV$ such that for all~$z\in [\tilde{\delv}]$ it holds that $|\repre^*(V_z)| = 1$, 
  and for all voters~$v\in V_z$ and $v'\in V_{z'}$ such that $z < z'$ alternative~$\repre^*(v)$ precedes $\repre^*(v')$ along the SP order, i.e., $\repre^*(v) < \repre^*(v')$.
  Then, if we can show that $A_2(a, b, \csize', j)$ computes the minimum sum of misrepresentations of all committees that satisfy conditions~\eqref{lin-CC-t-SP:1}--\eqref{lin-CC-t-SP:3} for the partition~$(V_1,\cdots, V_{\tilde{\delv}})$,
  then we show that  $A_2(\max(\committee^*), b, \csize, {\tilde{\delv}}) = \misr(\repre^*)$ where $\{b\} = \repre^*(V_{\tilde{\delv}})$, as desired.
  
  We prove this by induction on $\csize'$.
  \smallskip
  
  \noindent \emph{Base case:} $\csize'=1$.
  In this case, $\{a\}$ is the only committee that satisfies condition~\eqref{lin-CC-t-SP:1}. 
  All voters from $\reV\cup \bigcup_{z\in [j]}V_z$ need to be represented by~$a$.
  Hence, if $j = 0$, the misrepresentation of any optimal solution for the voter set~$\reV$ is exactly $\misr(\reV, a)$  (see the first operation in the initialization step), while $j = 1$, we need to make sure that $a=b$ and compute the misrepresentation of $\reV\cup V_1$ towards $\{a\}$ (see the second operation in the initialization step).
  Otherwise, no committee satisfies all conditions from~\eqref{lin-CC-t-SP:1}-\eqref{lin-CC-t-SP:3}, hence the misrepresentation is set to infinity. 

  \smallskip
  \noindent \emph{Induction step.} Assume that $A_2(a, b, \csize', j)$ is the minimum sum of misrepresentations of all solutions~$(\committee', \repre')$ satisfying conditions~\eqref{lin-CC-t-SP:1}-\eqref{lin-CC-t-SP:3}.
  Let $(\committee, \repre)$ denote a solution that satisfies \eqref{lin-CC-t-SP:1}-\eqref{lin-CC-t-SP:3} and minimizes the misrepresentation for voters from $\reV\cup \bigcup_{z\in [j]}V_z$ for the configuration~$(a, b, \csize'+1, j)$.
  If $j=0$ (see the first case in the recurrence step), then $(\committee, \repre)$ is an optimal solution for the voter set~$\reV$. 
  Correctness follows as the algorithm corresponds to the one in \cite{BetzlerMW2013}.

  Otherwise, let $b=\max(\repre(\cup_{z\in [j]}V_z))$ and $a = \max(\committee)$.
  There are two cases, either $a=b$ or $a > b$.
  If $a = b$, then we have that $\repre(V_j) = \{b\} = \{a\}$.
  Define $\committee'= \committee\setminus \{a\}$, $a'=\max(\committee')$, and $b' = \max(\cup_{z\in [j-1]}\repre(V_z))$. 
  Then, by induction assumption and by the optimality of $(\committee, \repre)$ for the configuration~$(a, b, \csize'+1, j)$, we have that $A_2(a', b', \csize', j-1) = \misr(\reV\cup \bigcup_{z\in [j-1]}V_z, \committee')$.
  Similarly to the arguments for approval preferences, by the SP property, the voters from $\reV$ that can improve wrt.~$\committee$ are those~$v$ where $\misr(v, a) < \misr(v, a')$.
  This corresponds to $\costdiff(\vvv, a, a')$.
  The voters in $V_j$ are represented by $b$, and hence the misrepresentation is the last summand in the second operation of the recurrence. 
  This concludes the compliance of the second operation in the recurrence.
  
  If $a > b$, then by definition, alternative~$a$ does not represent any voter from $\cup_{z\in [j]}V_z$.
  Define $\committee'= \committee\setminus \{a\}$ and $a'=\max(\committee')$. 
  Then, by induction assumption, we have that $A_2(a', b, \csize', j) = \misr(\reV\cup \bigcup_{z\in [j]}V_{z}, \committee')$.
  Similarly to the case of $a=b$, one can verify that the third operation in the recurrence computes $\misr(\repre)$. 

  As for the running time, we guess by brute-force searching for all ordered voter partition in $O(\delv^\delv)$ time.
  For each guessed partition, we compute a DP table with $O(m(m+1)\csize \delv)$ entries in $O(m^4n\csize\delv)$ time as the computation of each entry requires $O(m^2n)$ time. \todoH{Please complete the running time analysis.}
  Altogether, our algorithm runs in \FPT\ time wrt.~$\delv$.
  \todoIC{Check, please.}
\end{proof}
}
Before we continue with our next FPT result, we observe that a similar continuous block property utilized by Skowron et al.~\cite[Lemma 5]{SYFE2015} also holds for SC approval preferences.
This property is crucial for designing FPT result for the parameter number of voters (resp.\ alternatives) to delete to obtain SC preferences,
and hence may be of independent interest.
It is worth of noting that Elkind et al.~\cite{EL15} give an algorithm for \appCCsum\ that does not explicitly use continuous block property. This property is however the foundation of our \FPT\ algorithm.

\todoIH{We need a stronger result...}
\ifshort\begin{lemma}[\appsymb]\else\begin{lemma}\fi\label{lem:Approval-CC-SC}
  Let $I=(\profiletuple, \misr, \csize)$ be an instance of \appCCsum\ with SC preferences such that $\aaa=\{c_1,\ldots,c_m\}$ and $\vvv=[n]$, and $1 \rhd \ldots \rhd n$ is an SC order of the voters.
  Further, let $c_1 \blacktriangleright \ldots \blacktriangleright c_m$ be an order of the alternatives that orders the alternatives non-decreasingly according to their first approving voter~$\leftmost{}$, subject to that non-increasingly according to their last approving voter~$\rightmost{}$ in~$\rhd$.
  Then, for each solution~$(\committee, \repre^*)$, there exists another one~$\solution$ for the same committee~$\committee$ with $\misr(\repre) \le \misr(\repre^*)$ such that the following holds:
  \begin{compactenum}[(i)]
    \item\label{lem:Approval-CC-SC-voter} For each alternative~$c\in \committee$ the assigned voters~$\repre^{-1}(c)$ defines an interval in $\rhd$.
    \item\label{lem:Approval-CC-SC-alt} For each two voters~$u,v\in \vvv$ with $u \rhd v$, $\repre(u)=c_i$ and $\repre(v)=c_j$ it holds that $i < j$.
  \end{compactenum}
\end{lemma}

\appendixproof{lem:Approval-CC-SC}{
\begin{proof}
	\todoIC{Added proof}
Let $(\committee^,\repre^*)$ be an arbitrary solution. We will create a new assignment $\repre$ such that $(\committee,\repre)$ satisfies our desired properties.

We now order the alternatives in $\committee=\{a_1,\ldots,a_k\}$ such that $a_i\blacktriangleright a_j$ if $i<j$. In other words, $a_1,\ldots a_k$ is the ordering of the alternatives restricted to $\committee$. We will now iteratively describe the interval of voters that is assigned to each alternative. In each iteration let $p$ be the last voter that was assigned. We furthermore define $\leftmost{a_{k+1}}=n+1$. We initialize as $\repre^{-1}(a_1)=[1,\max\{\rightmost{a_1},\leftmost{a_2}-1\}]$. Then for each $a_i$ in increasing order we set $\repre^{-1}(a_i)=[p+1,\max\{\rightmost{a_i},\leftmost{a_{i+1}}-1\}]$. As the interval of $a_1$ starts with $1$, the interval of $a_k$ ends with $n$, and each interval starts directly after the previous non-empty interval ended, due to our definition of $p$, it follows that $\repre$ is an assignment. Per construction $\repre$ satisfies \cref{lem:Approval-CC-SC}\eqref{lem:Approval-CC-SC-voter}. As voters are assigned to the alternatives in increasing order, it follows for all alternatives $a_i,a_j\in W$ with $i<j$ that, if $\repre^{-1}(a_i)\neq\emptyset$ and $\repre^{-1}(a_j)\neq\emptyset$, then $\max\{\repre^{-1}(a_i)\}<\min\{\repre^{-1}(a_j)\}$. From this it follows that \cref{lem:Approval-CC-SC}\eqref{lem:Approval-CC-SC-alt} is satisfied as well. 

It remains to show that $\repre$ satisfies $\misr(\repre)\leq\misr(\repre^*)$. Suppose for the sake of contradiction that there exists a voter $v$ such that $\misr(v,\repre(v))=1$ but $\misr(v,\repre^*(v))=0$. Let $\repre(v)=a_j$ and $\repre^*(v)=a_i$. We distinguish two possibilities:
\begin{compactenum}[(i)]
\item\label{lem:Approval-CC-SC-c1} $i<j$
\item\label{lem:Approval-CC-SC-c2} $j<i$
\end{compactenum}
We start by considering Case \eqref{lem:Approval-CC-SC-c1}. We know that $\repre^{-1}(a_i)=[p+1,\max\{\rightmost{a_i},\leftmost{a_{i+1}}-1\}]$, where $p$ is the last voter that was assigned to an alternative before $a_i$ was assigned voters. Due to $j>i$ it follows that $p<v$ and therefore $p+1\leq v$. Furthermore it follows from $\leftmost{a_j}\leq p$ that $\leftmost{a_{i+1}}\leq p$. This however implies that $v\in[p+1,v]\subseteq[p+1,\rightmost{a_i}]\subseteq\repre^{-1}(a_i)$, a contradiction to our assumption that $\repre(v)=a_j$.

Now consider Case \eqref{lem:Approval-CC-SC-c2}. Let $a_x$ be the first alternative in the ordering that satisfies $v\in\VV(a_x)$. It must hold that $j<x$, as otherwise we would get a contradiction same as in \ref{lem:Approval-CC-SC-c1}. We can now observe the following property that holds due to our ordering of the alternatives: For all $i'<x$ it must hold that $\leftmost{i'}\leq\leftmost{x}\leq v$. If this property did not hold for some $i'$, then $i'$ and $x$ would violate the ordering. This directly contradicts our assumption that $v\in\repre^{-1}{a_j}$, as $\repre^{-1}(a_j)=[p+1,\max\{\rightmost{a_j},\leftmost{a_{j+1}}-1\}]\subseteq[1,\max\{\rightmost{a_j},\leftmost{a_{j+1}}-1\}]\subseteq[1,\max\{v-1,\leftmost{a_{j+1}}-1\}]\subseteq[1,v-1]$. Therefore $v\notin\repre^{-1}(a_j)$ in this case, contradicting our assumption.

As it is not possible for a voter $v\in\aaa$ to exist such that $\misr(v,\repre(v))=1$ but $\misr(v,\repre^*(v))=0$, it follows that $\misr(\repre)\leq\misr(\repre^*)$.
\end{proof}
}

\ifshort\begin{thm}[\appsymb]\else\begin{thm}\fi\label{thm:Approval-CC-SC-t-voters:FPT}
    For $\delv$-voters nearly SP profiles, \CCsum\ is \FPT wrt.~$\delv$.
\end{thm}

\appendixproof{thm:Approval-CC-SC-t-voters:FPT}{
\begin{proof}
  The proof idea is the same as the one for \cref{thm:nearSPvoter}.
  Hence, we use the same notion introduced there. 
  Similarly to that proof, we first consider approval and then linear preferences.
  
  \noindent \textbf{Approval preferences.} We will modify the DP approach by \cite{SYFE2015} instead.
  Recall that the SC order of the remaining voters~$\reV$ is $1\rhd \ldots \rhd n$.
  Without loss of generality, let $1\blacktriangleright \ldots \blacktriangleright m$ be an order that orders the alternatives non-decreasingly according to their first approving voters in $\reV$, and subject to that according to their last approving voters in $\reV$.
  The idea is to iterate through each voter~$n'$ in~$\rhd$ and each alternative~$j$ in~$\blacktriangleright$ and find an optimal partial solution for the voters~$[n']\cup \delV$, which uses only alternatives from $[j]$ and covers a given subset of types. 
  Formally, we compute a DP table~$B_1$ such that for each $\csize' \in [\csize]$, $j\in \aaa$, and $T\subseteq \Alltypes$, and $n'\in \vvv \cup \{0\}$  
  such that
  $B_1(j,\csize', T, n')$ stores the minimum misrepresentation of all solutions~$(\committee, \repre)$ for the voters~$\delV\cup [n']$ that satisfies the following conditions:
   \begin{compactenum}[(B1)]
     \item \label{CCneaSCvoter-1} $\committee \subseteq [j]$ and $|\committee| = \csize'$.
     \item \label{CCneaSCvoter-2} $T=\type(\committee)$ (i.e., $T$ is the set of types in the committee.)
   \end{compactenum}
   For the sake of readability, we call a solution that satisfies the above two conditions an \myemph{optimal} solution for $(j,\csize', T, n')$, and
   we say that a table entry is \myemph{correct} for  $(j,\csize', T, n')$ if it equals the misrepresentation of the corresponding optimal solution. 
Note that for $n'=0$, we optimize the misrepresentation over~$\delV$.
We initialize the table as follows.
\begin{align*}
  B_1(j, 1, T, n') = \min_{\substack{j' \in [j]\\ T=\type(j')}} \misr(\delV\cup [n'], j').
\end{align*}
Then, for $n'=0$ we continue with the following, where we require that :
\begin{align}
  B_1(j, \csize', T, 0) \\ & = 
  \min_{\substack{j'\in [j]\\\type(j')\in T}}  \{
                 B_1(j'-1, \csize'-1, T\setminus \{\type(j')\}, 0) \nonumber \\
                 & \quad -
                   \cost_{\delV}(T\setminus \{\type(j')\}) + \cost_{\delV}(T),\nonumber\\ &\quad B_1(j'-1, \csize'-1, T, 0)\big\}.\label{SC-t-voters-2}
\end{align}
We compute the other entries as follows where we assume that $j\ge \csize' > 1$, $n' \ge 1$, and $|\csize'|\ge |T|$; the other entries are set to $\infty$: 
  \begin{align}
 &   B_1(j, \csize', T, n') = \nonumber \\
    &      \min\limits_{\mathclap{\substack{j'\in [j]\\\type(j')\in T\\ x \in [n'+1]\\ V'=\{x,\ldots,n'\}\\T'\in \left\{T,T\setminus \{\type(j')\}\right\}}}}\quad \left\{
    \begin{array}{l}
      B_1(j'-1, \csize'-1, T', x-1) + \misr(V', j') \\
      +  \cost_{\delV}(T) - \cost_{\delV}(T').\\
      \end{array}\right\}
    \label{SC-t-voters-3}
  \end{align}
  The minimum misrepresentation is given by $\min\limits_{T\subseteq \mathcal{T}}B_1(m, k, T, n)$.
  To prove that this indeed solves our problem, we show that each table entry is correct.
  
  \smallskip
  \noindent \emph{Base case:} $\csize'=1$. It is straightforward that the table entry is correct as every optimal solution has committee of size one which is a subset of~$[j]$, contains all the types in~$T$ and minimizes the misrepresentation over $\delV\cup [n']$.

  \noindent \emph{Inductive hypothesis:} Assume that for all $j\in \aaa$, $T\subseteq \Alltypes$, and $n' \in [n]\cup \{0\}$, table entry~$B_1(j, \csize', T, n')$ is correct. %

  \noindent \emph{Inductive step:} $\csize' > 1$. 
  Let $(\committee, \repre)$ denote an optimal solution for $(j, \csize'+1, T, n')$,
  i.e., $\committee \subseteq [j]$,   $|\committee|=\csize'+1$,
  and $T = \type(\committee)$ such that $\misr(\delV\cup [n'], \committee)$ is minimum among all such committees.
  We aim to show that $A_1(j, \csize'+1, T, n') = \misr(\delV\cup [n'], \committee)=\misr(\repre)$.
  We distinguish between two cases, either $n' = 0$ or $n' > 0$.
  First, consider the case when $n'=0$.
  Then, we need to optimize the overall misrepresentation for all voters from $\delV$.
  Let $j^*=\max(\committee)$. 
  Hence, the misrepresentation of $\solution$  for $\delV$ can be decomposed into two parts, one induced by $\committee\setminus \{j^*\}$, and another induced by $j^*$.
  Since we iterate through all possible alternative~$j'$ for $j^*$ in \eqref{SC-t-voters-2}, we computes the desired value~$\misr(\delV, \committee)$. 
  
  Now, consider the case when $n' > 0$.
  Again let $j^*=\max(\committee)$, i.e., the last ranked alternative by voter~$1$ restricted to~$\committee$,
  $\committee' = \committee\setminus \{j^*\}$, $T' = \type(\committee')$,
  and let $x^*=\min(\repre^{-1}(j^*)\cap [n'])$ if $\repre^{-1}(j^*)\cap [n']\neq \emptyset$; $x^*=n'+1$ otherwise.
  Since $\csize > 1$, we know that $j^* \ge 2$. 
  By \cref{lem:Approval-CC-SC}\eqref{lem:Approval-CC-SC-voter}, we can assume that the voters that are assigned to~$j^*$ by $\repre$ form an interval in $\vvv=[n]$.
  By \cref{lem:Approval-CC-SC}\eqref{lem:Approval-CC-SC-alt}, every voter from $\reV$ that is assigned to an alternative other than $j^*$ is ordered ahead of $\repre^{-1}(j^*)$ in $\rhd$.  
  This means that $(\repre', \committee')$ is a valid solution for the voters from~$[x^*-1] \cup (\delV \setminus \repre^{-1}(j^*))$, where $\repre'$ is derived from $\repre$ by ignoring the assignment of the voters from $\repre^{-1}(j^*)$ and reassign each voter~$v$ from $\repre^{-1}(j^*)\cap \delV$ with $\repre'(v) = a$ whenever there exists an alternative~$a\in \committee$ with $\misr(v, a) = 0$.
  By the optimality of $(\repre, \committee)$ for $(j, \csize'+1, T, n')$, we infer that $(\repre', \committee')$ is optimal for $(j^*-1, \csize', T', x^*-1)$.

  Now, let us consider the misrepresentation of $\solution$.
  It can be divided into three parts, one induced by $(\committee', \repre')$, one induced by $j^*$ for $\repre^{-1}(j^*) = \{x+1, \ldots, n'\}$, and the remaining one by $j^*$ for the voters in $\delV\cap \repre^{-1}(j^*)$.
  Since we iterate through all possible alternatives~$j'$ for $j^*$ and all possible voters~$x\in [n'+1]$ for $x^*$, we can find one table entry with $B_1(j'-1, \csize', T', x-1) = \misr([x-1]\cup\committee')$, which implies that
  the table entry is correctly computed according to~\eqref{SC-t-voters-3}.
  This concludes the proof for the correctness of the DP for approval preferences.

  It is straightforward to verify that the DP runs in $O(2^tm^2n^2k)$ time as the number of entries is $O(2^tnkm)$ and computing each entry needs $O(mn)$ and the necessary precomputation needs at most $O(2^t)$ time. 

  \noindent \textbf{Linear preferences.} The idea for linear preferences is to combine the brute-forcing part of \cref{thm:nearSPvoter} with our previous DP approach.
  As before, we use the same notion introduced in the proof for \cref{thm:nearSPvoter}.
  We additionally assume that the first voter~$1$ in the SC order~$\rhd$ has preference order~$1\succ 2 \ldots \succ m$.
  Now, in the first step, we guess a partition of $\delV$ into $\tilde{t}\leq \min\{\delv,\csize\}$ subsets such that each subset shall be represented by the same alternative, and we also guess in which order their representatives are according to the preference order of voter~$1$.
  Let $V_1,\ldots V_{\tilde{\delv}}$ denote an ordered partition of~$\delV$ such that for each~$z\in [\tilde{\delv}]$ all voters from $V_z$ are represented by the same alternative and for each two indices~$z,z'\in [\tilde{\delv}]$ with $z< z'$ the representative~$a$ of $V_z$ is preferred to the representative~$b$ of~$V_{z'}$ by voter~$1$ in $\reV$, i.e., $a < b$. 
  After that, we compute a DP table~$B_2$ such that for each alternative~$j\in \aaa$, $\csize' \in [\csize]$, $g\in [\tilde{t}]\cup \{0\}$, and $n'\in \vvv\cup \{0\}$,
table entry~$B_2(j, \csize', g, n')$ equals the misrepresentation of a solution~$\solution$ satisfying for voters~$[n']\cup \bigcup_{z \in [g]} V_{z}$
\begin{compactenum}[(1)]
  \item\label{lin-CC-t-SC:1}  $|\committee|=\csize'$, and $\committee\subseteq [j]$,
  \item\label{lin-CC-t-SC:2} for every~$z \in [g]$ it holds that $|\repre(V_z)| = 1$,
\item\label{lin-CC-t-SC:3} for every pair $z,z' \in [g]$ with $z < z'$ it holds that
$a < b$ where $\repre(V_z) = \{a\}$ and $\repre(V_{z'}) = \{b\}$.
\end{compactenum}
that \emph{minimizes} the sum of misrepresentations over voters~$[n'] \cup \bigcup_{z \in [g]} V_{z}$.
Similarly, we call $\solution$ that optimizes the above an optimal solution for $(j, \csize', g, n')$,
and we say that a table entry~$B_2(j, \csize', g, n')$ if correct if $B_2(j, \csize', g, n')$ equals the misrepresentation of the optimal solution.

\smallskip
\noindent \emph{Initialization of $B_2$.}
To initialize we set all~$B_2(j, \csize', g, n')$ to~$\infty$ if $j < \csize'$ or $k' < g$. 
For $\csize' = 1$, we initialize as follows; note that in this case $g \in \{0,1\}$:
\begin{align*}
   B_2(j,1,g,n') =  \min_{j'\in [j]} \misr([n']\cup \bigcup_{z\in [g]}V_z, j').
\end{align*}
For $n'=0$, we compute as follows:
\begin{align*}
 & B_2(j,\csize',g,0) = \\
&  \min 
\begin{cases}
  0, & \text{ if } g = 0,\\
 ~~~\min\limits_{\mathclap{j'\in \{g, \ldots, j\}}} B_2(j'-1, \csize'-1, g-1, 0) \\+ \misr(V_g(j')), & \text{ if } j \ge \csize' \ge g, \\
\infty, & \text{ else.}
\end{cases}
\end{align*}
\noindent \emph{Recurrence.} We compute the remaining entries as follows, where we assume that $j \ge \csize' > 1$ and $n' > 0$:
\begin{align*}
 & B_2(j,\csize',g,n') = \min\{w_1,w_2\},\text{ where}\\ 
  &w_1 = \min_{\substack{j'\in [j]\\0 \le x < n'}}B_2(j'-1,\csize'-1, g, x) + \misr(\{x+1,\ldots,n'\}, j'), \\ %
  & w_2 =  \min_{
   \substack{j'\in [j]\\0 \le x \le n'}}(B_2(j'-1,\csize'-1, g-1, x)\\&+\misr(\{x+1,\ldots, n'\}\cup V_g,j). %
\end{align*}
To determine the minimum misrepresentation, we return $\min\limits_{(V_1,\ldots, V_{\tilde{\delv}})\colon \text{ordered partition of } \delV}B_2(m, \csize, \tilde{t}, n)$.

To prove the correctness, we show that each table entry is \myemph{correct} for $(j, \csize', g, n')$, i.e., it equals $\misr(\repre)$ where $\solution$ is an optimal solution for $[n']\cup \bigcup_{z\in [g]}V_z$ that satisfies the conditions \eqref{lin-CC-t-SC:1}--\eqref{lin-CC-t-SC:2}.
Clearly, the initialization for $\csize'=1$ is correct since in this case we look at all possible committee of size one that is a subset of $[j]$.
For $n'=0$, the table entries are also correct since we iterate through all possible cases of assigning alternatives to $V_g, V_{g-1}, \ldots, V_{1}$ such that the assigned alternatives correspond to the preference order of the first voter~$1$. 
For $\csize' \ge 1$, we assume that all table entries for $(j, \csize'-1, g, n')$ are correct.
Let $\solution$ be an optimal solution for the voters~$[n']\cup \delV$ regarding the configuration~$(j, \csize', g, n')$.
We aim to show that $B_2(j, \csize', g, n') = \misr(\repre)$.
Let $j^*=\max(\committee)$ and let $\committee'=\committee\setminus \{j^*\}$.
Then, by a similar reasoning for the lemma by Skowron et al.~\cite[Lemma 5]{SYFE2015},
we can assume that the voters~$\repre^{-1}(j^*)\cap [n']$ that are assigned to $j^*$ form an interval in $[n']$
such that every voter from $\reV$ that is assigned to an alternative other than~$j^*$ is ordered ahead of all~$\repre^{-1}(j^*)$.
We distinguish between two cases, $j^*$ is assigned to $V_g$, or else.
Consider first that $j^*$ is not assigned to~$V_g$ by $\repre$.
Then, by our requirement on the order $V_1,\ldots, V_{\tilde{t}}$,
we know that~$j^*$ is not assigned any voter in $\delV$.
Since $n' > 0$, alternative~$j^*$ is assigned to some voter in $[n']$, and let $x^*=\min(\repre^{-1}(j^*)\cap [n'])$.
Observe that we can divide the misrepresentation of $\solution$ into two parts, one part is equal to the misrepresentation of $(\committee', \repre')$ where $\repre'$ is derived from $\repre$ by ignoring the assignments of the voters to $j^*$, and the other part $\misr({x,\ldots, n'}, j^*)$.
By the optimality of $\solution$, it follows that $(\committee', \repre')$ is optimal for $(j^*-1, \csize'-1, g, x-1)$.
This means that $B_2(j^*-1, \csize'-1, g, x^*) = \misr(\repre')$, the correctness of table entry $B_2(j^*, \csize, g, x^*)$ follows as we will have $w_1=B_2(j^*-1, \csize'-1, g, x^*) + \misr(\{x^*+1,\ldots, n'\}, j^*)$.
The second case that $j^*$ is assigned to $V_g$ works similarly.
The only difference is that there may be no voters from $[n']$ that is assigned to $j^*$.
Hence, we iterate also on $x=n'$ in which case $\{x+1, \ldots, n'\}=\emptyset$.

\todoIH{Add running time analysis.}
For the running time, we guess by brute-force searching for all ordered voter partitions in $O(t^t)$ time. For each guessed partition we compute a DP table with $O(mk\tilde{t}n)$ entries in $O(m^2n^2k\tilde{t})$, as each table entry takes $O(mn)$ time.
\end{proof}
}
Together with \cref{obs:reduction}, we obtain the following.
\begin{corollary}\label{corr:max}
   For $\delv$-voters nearly SP (resp.\ SC) profiles, \CCmax\ is \FPT wrt.~$\delv$. %
\end{corollary}

Using analogous methods to \cref{thm:nearSPvoter,thm:Approval-CC-SC-t-voters:FPT}, we extend the DP approaches for \appMsum\ for SP (resp. SC) profiles to account for the deleted voters in $\delV$.

\ifshort
\begin{thm}[\appsymb]\else\begin{thm}\fi
    \label{thm:MonroenearSCvot}
   For $\delv$-voters nearly SP profiles, \appMsum\ and \Mmax\ are \FPT\ wrt.~$\delv$. %
   For $\delv$-voters nearly SC profiles, \appMsum\ and \emph{\app}-\Mmax\ are \FPT\ wrt.~$\delv$. %
  \end{thm}
  
  \appendixproof{thm:MonroenearSCvot}{
\begin{proof}[Proof sketch]
  We consider the SP and SC cases separately.
  To this end, let $\delV$ be the set of voters that needs to be deleted in order to make the profile SP respectively SC. The term deleted voters will refer to this set of voters. $V\setminus\delV$ will be the set of voters considered in the general DPs for the orderings and such.

\subsection*{The $\delv$-voters nearly SP case}
The approach is similar to the one for \cref{thm:nearSPvoter}: We adapt the algorithm by Betzler et al.~\cite{BetzlerMW2013} for the case with SP preferences under the Monroe rule. For $t$ voters, there are at most $\min\{t,\csize\}$ many alternatives they can be assigned to. We first guess for each deleted voter whether he will be assigned happily. Then we guess an ordered partition of the voters that will be happily assigned. This construction is the same as in the proof of Theorem 2\todo{Change to cref}.

We start by partitioning the deleted voters into two parts: the part of the voters that will be happily assigned and the part that will not be happily assigned. There are $2^t$ such partitions.
We then partition the voters of the first part into groups. We then check all possible orderings of these groups as the number of groups is upper-bounded by $\min\{t,\csize\}$. There are at most $t^t$ such divisions into groups and their orderings. We will refer to these groups as $V_1,\ldots,V_{\tilde{t}}$.

Each group of voters will be assigned to a separate alternative. We additionally define the set $\mathcal{U}(0,x_1,x_2)=\{0\}$. Moreover, $\mathcal{U}(u,x_1,x_2),\mathcal{U}_l(u,x,x_1,x_2)$ and $\mathcal{U}_r(u,x,x_1,x_2)$, where $u$ is a voter and $x,x_1,x_2$ are alternatives, all contain $0$ in addition to their definition by Betzler et al.~\cite{BetzlerMW2013}.\\

We compute $\dyn(u_i,x_1,x_2,\csize_c',\csize_f',b,t_1,t_2)$. Intuitively this is the maximum number of voters from $\mathcal{U}(u_i,x_1,x_2)$ and the groups $V_{t_1},\ldots,V_{t_2}$ that can be covered by $\csize_c'+\csize_f'+1$ alternatives from 
$[x_1,x_2]$, where:
\begin{compactitem}
\item $x_1$ is one of them and covers at most $b$ voters,
\item  $u_i$ is covered by an alternative from $[x_1,x_2]$,
\item at most $\csize_c'$ alternatives different from $x_1$ are assigned $\lceil\frac{n}{\csize}\rceil$ voters and
\item at most $\csize_f'$ alternatives different from $x_1$ are assigned less than $\lceil\frac{n}{\csize}\rceil$ voters and
\item $\repre(V_i)<\repre(V_j)$ iff.\ $i<j$.
\end{compactitem}

The parameters $t_1$ and $t_2$ are added to keep track of the groups, but the rest is identical to the original DP.

We start by initializing the table:
The first initialization step is the same as in the original:\begin{align*}
\dyn(u_i,&x_1,x_2,0,0,b,0,0)\\&= \min(b,|\{u\in\mathcal{U}(u_i,x_1,x_2):\rank_{x_1}(u)=0\}|)
\end{align*}
The second is due to the newly added groups:\begin{align*}
\dyn(0,&x_1,x_1,0,0,b,T) \\ &= \begin{cases}
b & \text{if all voters in }V_{t_1} \text{ approve of }x_1,\text{ and }b=|V_{t_1}|, \\
-\infty & \text{else.}
\end{cases}
\end{align*}
The idea is to use left-most interval property, which allows us to split the voters into two parts based on an interval of alternatives and the first voter covered by alternatives in this interval. Our algorithm extends the DP by splitting the partitions of deleted voters in this manner (due to us guessing the ordering as well) as well. We show the update table operation, which is similar to how it was done in the paper by Betzler et al., in Algorithm~\ref{alg:VoteDel}.
\begin{algorithm*}[t!]
	\caption{Computing $\dyn(u_i,x_1,x_2,k_c',k_f',b,t_1,t_2)$ for the proof of \cref{thm:MonroenearSCvot}.}\label{alg:VoteDel}	
	$M,M_l,M_r\coloneqq0$
	
	\lIf{$|V_{t_1}|<b$ and all voters in $V_{t_1}$ approve of $x_1$}{
		$M\coloneqq \max\{M,\dyn(u_i,x_1,x_2,k_c',k_f',b-|V_{t_1}|,t_1+1,t_2)+|V_{t_1}|\}$}
	\If{$|V_{t_1}|=b$ and all voters in $V_{t_1}$ approve of $x_1$}{
		\ForAll{$x'\in[x_1+1,x_2]$}{
			$M\coloneqq \max\{M,\dyn(u_i,x',x_2,k_c'-1,k_f',\upperB,t_1+1,t_2)+|V_{t_1}|)\}$
			
			\lForAll{$0<j<\upperB$}{
				$M\coloneqq \max\{M,\dyn(u_i,x',x_2,k_c',k_f'-1,j,t_1+1,t_2)+|V_{t_1}|\}$}}}
	\If{$b>1$}{
		\lForAll{$u_j\in\mathcal{U}(u_i,x_1,x_2)\setminus\{u_i\}$ with $u_j$ approving of $x_1$}{$M\coloneqq\max\{M,\dyn(u_j,x_1,x_2,k_c',k_f',b-1,t_1,t_2)+1\}$}}
	\Else{	\ForAll{$x'\in[x_1+1,x_2]$ with $|[x,x_2]|\geq k_c+k_f$}{
			\ForAll{$u_j\in\mathcal{U}(u_i,x_1,x_2)\setminus\{u_i\}$ with $u_j$ approving of $x'$}{
				\lIf{$k_c'>0$}{$M\coloneqq\max\{M,\dyn(u_j,x',x_2,k_c'-1,k_f',\upperB,t_1,t_2)+1\}$}
				\If{$k_f'>0$}{
					\lForAll{$0<j<\upperB$}{%
						$M\coloneqq\max\{M,\dyn(u_j,x',x_2,k_c',k_f'-1,j,t_1,t_2)+1\}$}}}}}
	\ForAll{$x'\in[x+k_c'+k_f'+1,x_2]$}{
		\lIf{$V_{t_2}$ approves of $x'$ and $|V_{t_2}|=\upperB$}{ $M\coloneqq\max\{M,\dyn(u_i,x_1,x'-1,k_c'-1,k_f',b,t_1,t_2-1)+|V_{t_2}|\}$}
		\lIf{$V_{t_2}$ approves of $x'$ and $|V_{t_2}|<\upperB$}{$M\coloneqq\max\{M,\dyn(u_i,x_1,x'-1,k_c',k_f'-1,b,t_1,t_2-1)+|V_{t_2}|\}$}}
	\ForAll{$x\in[x_1+1,r(u_i)]$ with $u_i$ approving of $x'$}{
		\ForAll{$t'\in \{t_1-1,t_2\}$}{
			\ForAll{$k_c^l\geq0$ and $k^r_c\geq0$ with $k_c^l+k_c^r=k_c'$}{
				\ForAll{$k_f^l\geq0$ and $k^r_f\geq0$ with $k_f^l+k_f^r=k_f'$}{
					\If{$|[x_1,x-1]|\geq k_c^l+k_f^l+1$ and $|[x,x_2]|\geq k_c^r+k_f^r$}{
						\ForAll{$u_j\in\mathcal{U}_l(u_i,x,x_1,x_2)$ with $u_j$ approving of $x_1$}{
						\lIf{$t'=t-1-1$}{$M_l\coloneqq\max\{M_l,\dyn(u_j,x_1,x-1,k_c^l,k_f^l,b,0,0)\}$}
						\lElse{$M_l\coloneqq\max\{M_l,\dyn(u_j,x_1,x-1,k_c^l,k_f^l,b,t_1,t')\}$}}
						\ForAll{$u_j\in\mathcal{U}_r(u_i,x,x_1,x_2)$ with $u_j$ approving of $x$}{
							\If{$t'=t_2$}{
							\If{$k_c^r>0$}{$M_r\coloneqq\max\{M_r,\dyn(u_j,x,x_2,k_c^r-1,k_f^r,\upperB-1,0,0)+1\}$}
							\If{$k_f^r>0$}{
							\lForAll{$0<j<\upperB$}{$M_r\coloneqq\max\{M_r,\dyn(u_j,x,x_2,k_c^r,k_f^r-1,j-1,0,0)+1\}$}}}
							\Else{
							\lIf{$k_c^r>0$}{$M_r\coloneqq\max\{M_r,\dyn(u_j,x,x_2,k_c^r-1,k_f^r,\upperB-1,t'+1,t_2)+1\}$}
							\If{$k_f^r>0$}{
								\ForAll{$0<j<\upperB$}{$M_r\coloneqq\max\{M_r,\dyn(u_j,x,x_2,k_c^r,k_f^r-1,j-1,t'+1,t_2)+1\}$}}}}
						
						$M\coloneqq\max\{M,M_r+M_l\}$
	}}}}}
	\Return{$M$}
\end{algorithm*}
The correctness of this FPT algorithm follows from the correctness of the DP algorithm by Betzler et al.~\cite{BetzlerMW2013}, as this DP is an extension of Betzler et al. For $t=0$ the two DPs coincide. Let $V_1,\ldots V_{\tilde{t}}$ be an ordered partition of the voters which corresponds to an optimal assignment. As the alternatives are split by their induced ordering the correctness of the DP can be argued via induction over $t'$. The induction beginning is given by the case where $t=0$. The induction step can be verified by verifying that all possibilities of a group of deleted voters being assigned is captured and counted correctly. 

First we note that we iterate over $O((t+1)^t)$ different partitions of deleted voters in order to at the end find the correct grouping. For each of these partitions we need to compute the described dynamic programming table. The table has size $O(mn^3kt^2)$ and the computation of a single table entry takes at most $O(m\frac{n^2}{k}t)$ time. Therefore the total running time is bounded by $O(m^2n^5(t+1^{t+3}))$ which is FPT with respect to $t$.

\subsection*{The $\delv$-voters nearly SC case}\label{sec:monroe_sc_vdel_fpt}
The idea behind this FPT algorithm is the same as the idea of the corresponding algorithm for $\delv$-voters nearly SP preferences. We extend our previously presented DP approach to also keep track of the voters which were deleted. We add $t$ table entries to the table which represent each deleted voter and whether the voter has been happily assigned. Each of these table entries can take the value of either $0$ (meaning the voter has not been happily assigned) or $1$ (meaning the voter has been happily assigned). We will represent these table entries using $\delVect$, where $\delVect$ is the vector of the table entries. We will use $\size(\delVect)$ to describe the number of table entries that are $1$.
\begin{figure*}[!t]
\begin{multline}
 \textbf{Case 1:} \quad \label{eq:SPMlower1FPT}
	  T(a,a,i,j,\kUpper,\kLower,\na,\na,\nd,0,c',i',j',\delVect)\coloneqq\\
	  \max\left\{
	\begin{array}{l}
		\underset{\substack{b,c\in\usablesetS(a,i,j,c',i',j')\\
				\kUpper' = \kUpper - f_1(b,c,\nb,n_c), 
				\kLower' = \kLower-f_2(b,c,\nb,n_c)\\ \leftmost{b}\in[i,j], 0< \nb,n_c\leq\upperB, B\in\{0,1\}}}{\max}
		\left\{ \begin{array}{l}
			T(b,c,i,j,\kUpper',\kLower',\nb,n_c,\na+\nd,B,c',i',j',\delVect)+\na,\\
			T(b,c,i,j,\kUpper',\kLower',\nb,n_c,\na+\nd,B,a,i,j,\delVect)+\na \\
		\end{array}\right\},\\
		\underset{\substack{\nd_1+\nd_2=\nd, b_1+b_2=\na\\ i\leq i^*<j^*\leq j, \kUpper_1+\kUpper_2=\kUpper, \kLower_1+\kLower_2=\kLower\\\delVect_1+\delVect_2=\delVect}}{\max}
		\left\{ \begin{array}{l}
			~~ T(a,a,i,i^*,\kUpper_1,\kLower_1,b_1,b_1,\nd_1,0,c',i',j',\delVect_1),\\ 
			+  T(a,a,j^*,j,\kUpper_2,\kLower_2,b_2,b_2,\nd_2,0,c',i',j',\delVect_2)
		\end{array}\right\},\\
		\underset{\substack{b_1+b_2=\na\\\delVect_1+\delVect_2=\delVect}}{\max}
				\left\{ \begin{array}{l}
					~~ T(a,a,i,j,\kUpper,\kLower,b_1,b_1,\nd,0,c',i',j',\delVect_1),\\ 
					+  T(a,a,0,0,0,0,b_2,b_2,0,0,c',i',j',\delVect_2)
				\end{array}\right\}
	\end{array}\right\}.
\end{multline}
\begin{multline}
\textbf{Case 2:} \quad \label{eq:SPMlower2FPT}
	 T(a,a,i,j,\kUpper,\kLower,\na,\na,\nd,1,c',i',j')\coloneqq   \\
	\max\left\{
	\begin{array}{l}
		\underset{\substack{b,c\in\usablesetS(a,i,j,c',i',j') \\
				\kUpper=\kUpper'+f_1(b,c,\nb,n_c), 
				\kLower=\kLower'+f_2(b,c,\nb,n_c)\\
				0<\nb,n_c\leq\upperB, B\in\{0,1\}}}{\max}
		\left\{\begin{array}{l}
			T(b,c,i,j,\kUpper',\kLower',\nb,n_c,\na+\nd,B,c',i',j',\delVect)+\na,\\
			T(b,c,i,j,\kUpper',\kLower',\nb,n_c,\na+\nd,B,a,i,j,\delVect)+\na                 \\
		\end{array}\right\},\\
		\underset{\substack{\nd_1+\nd_2=\nd,  b_1+b_2=\na\\ i\leq i^*<j^*\leq j, \kUpper_1+\kUpper_2=\kUpper, \kLower_1+\kLower_2=\kLower\\\delVect_1+\delVect_2=\delVect}}{\max}
		\left\{\begin{array}{l}
			~~T(a,a,i,i^*,\kUpper_1,\kLower_1,b_1,b_1,\nd_1,1,c',i',j',\delVect_1)\\
			+T(a,a,j^*,j,\kUpper_2,\kLower_2,b_2,b_2,\nd_2,0,c',i',j',\delVect_2)
		\end{array}\right\}\\
		\underset{\substack{b_1+b_2=\na\\\delVect_1+\delVect_2=\delVect}}{\max}
				\left\{\begin{array}{l}
					~~T(a,a,i,j,\kUpper,\kLower,b_1,b_1,\nd,1,c',i',j',\delVect_1)\\
					+T(a,a,0,0,0,0,b_2,b_2,0,0,c',i',j',\delVect_2)
				\end{array}\right\}
	\end{array}
	\right\}.
\end{multline}
\begin{multline} 
     \textbf{Case 3:} \quad \label{eq:SPMsumFPT}
	T(a,b,i,j,\kUpper,\kLower,\na,\nb,\nd,B,c',i',j')\coloneqq  \\
	  \max\left\{
	\begin{array}{l}
		\underset{\substack{ c\in\inbetwS(a,b)\\ i\leq i^*<j^*\leq j,  0<n_c\leq\lowerB\\ \kUpper_1+\kUpper_2=\kUpper, \kLower_1+\kLower_2+1=\kLower\\ \nd_1+\nd_2=\nd, B'\in\{0,1\}\\\delVect_1+\delVect_2=\delVect}}{\max}
		\left\{
		\begin{array}{l}
			~~  T(a,c,i,i^*,\kUpper_1,\kLower_1,\na,n_c,\nd_1,B,c',i',j',\delVect_1)\\
			+ T(b,b,j^*,j,\kUpper_2,\kLower_2,\nb,\nb,\nd_2,B',c',i',j',\delVect_2)
		\end{array}\right\},\\
		\underset{\substack{c\in\inbetwS(a,b)\\ i\leq i^*<j^*\leq j,  n_c=\upperB \\ \kUpper_1+\kUpper_2+1=\kUpper, \kLower_1+\kLower_2=\kLower\\\nd_1+\nd_2=\nd, B'\in\{0,1\}\\\delVect_1+\delVect_2=\delVect}}{\max}
		\left\{
		\begin{array}{l}
			~~T(a,c,i,i^*,\kUpper_1,\kLower_1,\na,n_c,\nd_1,B,c',i',j',\delVect_1)\\
			+T(b,b,j^*,j,\kUpper_2,\kLower_2,\nb,\nb,\nd_2,B',c',i',j',\delVect_2)
		\end{array}\right\},\\
		\underset{\substack{ c\in\inbetwS(a,b)\\0<n_c\leq\lowerB\\ B'\in\{0,1\}\\\delVect_1+\delVect_2=\delVect}}{\max}
				\left\{
				\begin{array}{l}
					~~  T(a,c,i,j,\kUpper,\kLower-1,\na,n_c,\nd,B,c',i',j',\delVect_1)\\
					+ T(b,b,0,0,0,0,\nb,\nb,0,B',c',i',j',\delVect_2)
				\end{array}\right\},\\
				\underset{\substack{c\in\inbetwS(a,b)\\  n_c=\upperB \\ B'\in\{0,1\}\\\delVect_1+\delVect_2=\delVect}}{\max}
				\left\{
				\begin{array}{l}
					~~T(a,c,i,j,\kUpper-1,\kLower_,\na,n_c,\nd,B,c',i',j',\delVect_1)\\
					+T(b,b,0,0,0,0,\nb,\nb,0,B',c',i',j',\delVect_2)
				\end{array}\right\},\\
		\underset{\substack{i\leq i^*<j^*\leq j\\ \kUpper_1+\kUpper_2=\kUpper, \kLower_1+\kLower_2=\kLower\\\nd_1+\nd_2=\nd, B'\in\{0,1\}\\\delVect_1+\delVect_2=\delVect}}{\max}
		\left\{
		\begin{array}{l}
			~~T(a,a,i,i^*,\kUpper_1,\kLower_1,\na,\na,\nd_1,B,c',i',j',\delVect_1)\\
			+T(b,b,j^*,j,\kUpper_2,\kLower_2,\nb,\nb,\nd_2,B',c',i',j',\delVect_2)
		\end{array}\right\} 
	\end{array}\right\}.
    \end{multline}
    \caption{DP descriptions for the $\delv$-voters nearly SC case in the proof of \cref{thm:MonroenearSCvot}.}
\end{figure*}

\mypara{The table.}
Table~$T$ has an entry for every tuple~$(\firsttabletuple,\delVect)$, where the table entries are the same as in the original table with the addition of allowing $i=j=0$ for the case where an alternative may only be assigned deleted voters and $\delVect$ is the binary vector of size $t$ that we described earlier. We keep track of the same information as in the original table with the addition of keeping track of which of the deleted voters we assigned happily.

Formally, each table entry stores the maximum number of voters from the range~$[i,j]$ and the voters in $\delVect$ whose entry is $1$ that can be satisfied by an \NT\ partial solution $(\committee',\repre)$ wrt. $[i,j]$, where~$\repre\colon [i,j] \to \committee'$ under the initial seven conditions and the following three additional conditions. The three conditions add an additional requirement that the deleted voters whose entry is $1$ must be happily assigned.%

 In the following, we describe how to compute~$T(\firsttabletuple,\delVect)$. We assume that the arguments of the configuration fulfill the same conditions in the configuration, as for the original table and 
 \begin{compactitem}
 \item[(CT8)] Either $i=j=0$ or $i,j\neq0$ and
\item[(CT9)] $j-i+1+\size(\delVect)\geq\na$ if $i,j\neq0$ and
\item[(CT10)] $\size(\delVect)\geq\na$ if $i,j=0$.
 \end{compactitem}
Condition (CT8) is meant to ensure that $0$ is only used when an alternative is at that point not assigned any non-deleted voter and Condition (CT9) and Condition (CT10) just clarify the connection between the number of deleted voters that are satisfied and the number of voters assigned to an alternative.
  
 For the case when at least one of the conditions is violated, once again the corresponding entry is set to~$-\infty$.
 We start by initializing the table as:
 \begin{align}
   T(a,&a,i,j,0,0,\na,\na,\nd,0,c',i',j',\delVect) \coloneqq \label{eq:SPMinitFPT} \\ &\begin{cases} \nonumber
\na \text{, if all deleted voters whose entry is $1$ approve of }a,\\
-\infty\text{, else.}
   \end{cases} 
 \end{align}
We update the table, distinguishing between three cases, either ``$a = b$ and $B=0$'' (presented in Equation \eqref{eq:SPMlower1FPT}), ``$a= b$ and $B=1$'' (presented in Equation \eqref{eq:SPMlower2FPT}), or ``$a\neq b$'' (presented in Equation \eqref{eq:SPMsumFPT}). A small change from the original DP is necessary, as we allow $i,j$ to take the value of $0$ in such cases we will handle the alternative as though it covered an interval of length zero.
 We again use the two auxiliary functions
 \begin{align*}
   f_1(c_1,c_2,n_1,n_2)=
   \begin{cases}
     2 & \, \text{if }c_1\neq c_2\wedge n_1=n_2=\upperB, \\
     0 & \, \text{if }n_1\leq\lowerB\wedge n_2\leq\lowerB,\\
     1 & \, \text{else.}\\
   \end{cases}\\
     f_2(c_1,c_2,n_1,n_2)=|\{c_1, c_2\} | - f_1(c_1,c_2,n_1,n_2)
   \end{align*}

We compute the table in the following order, where the order holds only subject to the previous steps, e.g., $2$ is only relevant subject to $1$ and so on: 
\begin{compactenum}\label{tableorderFPT}
\item Order the entries by $\kUpper + \kLower$ in a non-decreasing order.
\item Order the entries with $a = b$ before the entries where $a \neq b$.
\item Order the entries by $j - i$ in a non-decreasing order.
\item Order the entries by $\na + \nb$ in a non-decreasing order.
\item Order the entries by $\nd$ in an increasing order.
\item Order the entries by $B$ in an increasing order.
\item Order the entries by $\delVect$ in increasing lexicographical order.
\end{compactenum}
For the optimal solution, we return $\max\limits_{\substack{a,b\in\level_\aaa(1)\\ i\leq j\\\kUpper\leq n\mod k -f_1(a,b,\na,\nb)\\\kUpper+\kLower+|\{a,b\}|\leq k\\\na,\nb\leq\upperB\\\nd\leq n\\ B\in\{0,1\}\\\delVect\in\{0,1\}^t}}T(a,b,i,j,\kUpper,\kLower,\na,\nb,\nd,B,0,0,0,\delVect)$.

The correctness of this algorithm follows similarly as the correctness for the original DP, as the vectors must add up to a binary vector again, thereby assuring that each deleted voter is assigned at most once. As each table entry is built using the maximum over all valid table entries the maximality is preserved for the same reasons as in the original DP.

The size of the DP is increased by a factor of $2^t$, due to $\delVect\in\{0,1\}^t$. Furthermore, the time to compute a table entry is also increased by at most $2^t$, due to the additional need to split $\delVect$ into $\delVect_1$ and $\delVect_2$. This leaves us with a total FPT running time of $O(2^{2t}n^{11} m^5 \csize)$.

\end{proof}
}
While the parameter ``number of voters to delete to obtain an SP (resp. SC) profile'' turned out to be very useful for \app-\Monroe-\mw, similar to \CC-\mw, the parameter ``number of alternatives to delete'', for which \CC-\mw\ was \FPT, does not immediately yield \FPT-result for \app-\Monroe.
By guessing how many voters each deleted alternative will be assigned, we can extend the existing DP approaches and obtain the following XP result:

\ifshort\begin{thm}[\appsymb]\else\begin{thm}\fi\label{thm:MonroenearSCalt}
For $\delv$-alternatives nearly SP profiles, \appMsum\ and \Mmax\ are in XP wrt.~$\dela$. 
For $\delv$-alternatives nearly SC profiles, \appMsum\ and \emph{\app}-\Mmax\ are in XP wrt.~$\dela$. 
\end{thm}

\appendixproof{thm:MonroenearSCalt}{

\begin{proof}[Proof sketch]
Let $\tilde{A}$ be the set of alternatives that need to be deleted in order to make the preference profile single-peaked resp. single-crossing. Deleted alternatives will refer to the alternatives in $\tilde{A}$. The intuition behind this result is to ``guess'' a subset $\hat{A}$ of alternatives in $\tilde{A}$ that will be used in an optimal solution and for each alternative in the subset, how many voters will be assigned it, and then run an extension of the DPs described in \Cref{sec:monroe-SC} respectively~\cite{BetzlerMW2013}. We will split the proof into two parts: One for the SC case and one for the SP case.

\subsection*{The $\dela$-alternatives nearly SP case}
We divide voters in types based on which of the $t$ deleted alternatives they approve. If two voters approve the  same deleted alternatives, they are in the same type. There are at most $2^t$ such types.
We can now guess how many voters of each type will be assigned to deleted alternatives and how many deleted alternatives will be used. There are at most $n^{2^t}\cdot 2^t$ possibilities. As Monroe voting is FPT wrt. the number of alternatives we can then compute the maximal number of happy voters when the guessed number of voters of each type are assigned to the guessed number of deleted alternatives. Therefore our dynamic programming will try to compute the maximal number of happy voters that can be assigned to alternatives in the SP part without assigning the guessed number of voters for each type.
We fix an arbitrary ordering of the different types. Let $\typenr$ be the number of types. We furthermore assume that all the types of voters can be happily assigned to the guessed number of deleted alternatives - if not, our guess must have been wrong. %

We compute the dynamic table $\dyn(u_i,x_1,x_2,k_c',k_f',b,T)$, where $T\subset \{0,\ldots,n\}^\typenr$ is a vector whose length is equal to the number of types. We will use this vector to keep track of how many voters of each type we have assigned to deleted alternatives in each step. Each entry is limited by the number of voters of the corresponding type that are assigned to the deleted alternatives.

The table can be intuitively read as the maximum number of voters from the set $\mathcal{U}(u_i,x_1,x_2)$, as described in the paper by Betzler et al.~\cite{BetzlerMW2013} that can be covered by $k_c'+k_f'+1$ alternatives from $[x_1,x_2]$, where:
\begin{compactitem}
\item $x_1$ is one of them and covers at most $b$ voters and
\item $u_i$ is covered by an alternative from $[x_1,x_2]$ or a deleted alternative and
\item at most $k_c'$ alternatives different from $x_1$ are assigned $\lceil\frac{n}{k}\rceil$ voters
\item at most $k_f'$ alternatives different from $x_1$ are assigned less than $\lceil\frac{n}{k}\rceil$ voters and
\item each entry of $T$ is equal to the number of voters of the corresponding type that will be assigned to deleted alternatives.
\end{compactitem}

 The computation generally works the same as the DP by Betzler et al.~\cite{BetzlerMW2013} except that we allow voters to be reserved for deleted alternatives. We allow $x_1$ and $x_2$ to take the value $0$ in order to accommodate the possibility for voters to be assigned to deleted alternatives. Note that if one of the two takes the value $0$ both need to take the value $0$, and $b$ must be $0$ as well. This corresponds to the case where voters are assigned to a deleted alternative that are before or after all voters that are assigned to non-deleted alternatives.

We start by initializing the table:

The first initialization step is the same as in the original:\begin{align*}
	\dyn(u_i,&x_1,x_2,0,0,b,(0,\ldots,0))\\ &=\min(b,|\{u\in\mathcal{U}(u_i,x_1,x_2)\colon \rank_{x_1}(u)=0\}|)
\end{align*}	
The second is due to the newly added groups:\begin{align*}
		\dyn(u_i,&0,0,0,0,0,T)=\\ &\begin{cases}
			  1\text{, if }|T|=1 \text{ and }u_i\text{ is of the type that has the entry }1. \\
			-\infty \text{, else.}
		\end{cases}
	\end{align*}
The last initialization step is meant to allow us to assign multiple alternative towards the end of the voter ordering to be assigned to deleted alternatives. The order of computation start with the smallest $|T|$ and increases the size in each step. In the following $T'$ will be a vector that is identical to $T$ except for one entry, where $T'$ has a value that is smaller by one. It will be symbolically written down as $T'=T-1$.

\begin{align*}
	\dyn(u_i,0,0,0,0,0,T)= \begin{cases}
		&\dyn(u_j,0,0,0,0,0,T') \text{, where }j>i,\\&T'=T-1 \text{ and }u_i\text{ is of the type}\\ &\text{that is smaller in }T'. \\
		&-\infty \text{, else}
	\end{cases}
\end{align*}
In Algorithm $\ref{alg:AltDel}$ we show the update table similarly to how it was done in the original algorithm~\cite{BetzlerMW2013}, as described in the previous section, except we keep track of voters assigned to deleted alternatives as well.

\begin{algorithm*}[t!]
	\caption{Computing $\dyn(u_i,x_1,x_2,k_c',k_f',b,T)$ for the proof of \cref{thm:MonroenearSCalt}.}\label{alg:AltDel}
	
	$M,M_l,M_r\coloneqq0$\\
	\lIf{$size(t_1)<b$ and all voters in $V_{t_1}$ approve of $x_1$}{%
		$M\coloneqq \max\{M,\dyn(u_i,x_1,x_2,k_c',k_f',b-size(t_1),t_1+1,t_2)+size(t_1)\}$}
	\If{$size(t_1)=b$ and  all voters in $V_{t_1}$ approve of $x_1$}{
		\ForAll{$x'\in[x_1+1,x_2]$}{
			$M\coloneqq \max\{M,\dyn(u_i,x',x_2,k_c'-1,k_f',\upperB,t_1+1,t_2)+\size(t_1)\}$
			
			\lForAll{$0<j<\upperB$}{%
				$M\coloneqq \max\{M,\dyn(u_i,x',x_2,k_c',k_f'-1,j,t_1+1,t_2)+\size(t_1)\}$}}}
	\If{$b>1$}{
		\lForAll{$u_j\in\mathcal{U}(u_i,x_1,x_2)\setminus\{u_i\}$ with $u_j$ approving of $x_1$}{$M\coloneqq\max\{M,\dyn(u_j,x_1,x_2,k_c',k_f',b-1,T)+1\}$}
	}
	\Else{	\ForAll{$x\in[x_1+1,x_2]$ with $|[x,x_2]|\geq k_c+k_f$}{
			\ForAll{$u_j\in\mathcal{U}(u_i,x_1,x_2)\setminus\{u_i\}$ with $u_j$ approving of $x$}{
				\lIf{$k_c'>0$}{$M\coloneqq\max\{M,\dyn(u_j,x,x_2,k_c'-1,k_f',\upperB,T)+1\}$}}
			\If{$k_f'>0$}{
				\lForAll{$0<j<\upperB$}{$M\coloneqq\max\{M,\dyn(u_j,x,x_2,k_c',k_f'-1,j,T)+1\}$}}
			
	}}
	\ForAll{$x'\in[x+k_c'+k_f'+1,x_2]$}{
		\lIf{$t_2$ approves of $x'$ and $\size(t_2)=\upperB$}{$M\coloneqq\max\{M,\dyn(u_i,x_1,x'-1,k_c'-1,k_f',b,t_1,t_2-1)+\size(t_2)\}$}
		\lIf{$t_2$ approves of $x'$ and $\size(t_2)<\upperB$}{$M\coloneqq\max\{M,\dyn(u_i,x_1,x'-1,k_c',k_f'-1,b,t_1,t_2-1)+\size(t_2)\}$}}
	\ForAll{$x\in[x_1+1,r(u_i)]$ with $u_i$ approving of $x'$}{
		\ForAll{$T',T'':$ $T'+T''=T$}{
			\ForAll{$k_c^l\geq0$ and $k^r_c\geq0$ with $k_c^l+k_c^r=k_c'$}{
				\ForAll{$k_f^l\geq0$ and $k^r_f\geq0$ with $k_f^l+k_f^r=k_f'$}{
					\If{$|[x_1,x-1]|\geq k_c^l+k_f^l+1$ and $|[x,x_2]|\geq k_c^r+k_f^r$}{
						\lForAll{$u_j\in\mathcal{U}_l(u_i,x,x_1,x_2)$ with $u_j$ approving of $x_1$}{$M_l\coloneqq\max\{M_l,\dyn(u_j,x_1,x-1,k_c^l,k_f^l,b,T')\}$}
						\ForAll{$u_j\in\mathcal{U}_r(u_i,x,x_1,x_2)$ with $u_j$ approving of $x$}{
							\lIf{$k_c^r>0$}{$M_r\coloneqq\max\{M_r,\dyn(u_j,x,x_2,k_c^r-1,k_f^r,\upperB-1,T'')+1\}$}
							\If{$k_f^r>0$}{
								\ForAll{$0<j<\upperB$}{$M_r\coloneqq\max\{M_r,\dyn(u_j,x,x_2,k_c^r,k_f^r-1,j-1,T'')+1\}$}}}
						
						$M\coloneqq\max\{M,M_r+M_l\}$}}}}}
	\Return{$M$}
\end{algorithm*}
Correctness again follows from the correctness of the algorithm by Betzler et al.~\cite{BetzlerMW2013}. As we guess all possibilities for alternatives to be used in an optimal solution, and then compute for all number of assigned happy voters the maximal satisfaction, the correctness follows. 

The guessing which type of alternative will be assigned how many voters requires $n^{2^t}\cdot 2^t$ different iterations at most. We then compute a table of size $O(m^2kn^{t+2})$. The time to compute a single table entry is upper bounded by $O(m+\frac{n^2}{k}+2^tnk)$. By multiplying out the runtimes we can see that the problem is in XP wrt. $t$.
\subsection*{The $\dela$-alternatives nearly SC case}
The general idea behind this XP algorithm is similar to the $\dela$-alternatives nearly SP case in this proof, as it is also an extension of our previous DP approach. We first guess a subset of alternatives that is in the committee. %
Then we will partition the subset into alternatives that will be assigned at most $\lfloor\frac{n}{\csize}\rfloor$ voters and those that may be assigned at most $\lceil\frac{n}{k}\rceil$ voters. Since there are $3^t$ possible ways to partition voters in such a way, we get an XP algorithm as long as we solve the instance for one guess in XP time. We denote the vector of alternatives that have been guessed to be in the solution as $\mathcal{A}$. It has one entry for each deleted alternative that will be used and that entry can take integer values between $0$ and $\lfloor\frac{n}{k}\rfloor$ for alternatives that will be assigned at most $\lfloor\frac{n}{k}\rfloor$ and between $0$ and $\lceil\frac{n}{k}\rceil$ for alternatives that may be assigned up to $\lceil\frac{n}{k}\rceil$ voters. We will use $\size(\mathcal{A})$ to refer to the sum of all table entries.

\mypara{The table.}
Table~$T$ has an entry for every tuple~$(\firsttabletuple,\mathcal{A})$, where the table entries as described previously are the same as in the original table with the addition of allowing $a=b=0$ for the case where a voter is assigned to a deleted alternative and $\mathcal{A}$ is the vector of size at most $t$, that we described earlier.

Informally, we keep track of the same information as in the original table with the addition of keeping track of how many happy voters are assigned to each deleted alternative that was guessed to be in the solution. Note that the upper bound for $\kUpper$ and $\kLower$ has to be updated with the information of the deleted alternatives, i.e. if the number of deleted alternatives guessed to be assigned up to $\lceil\frac{n}{k}\rceil$ voters is $r$, then the number of non-deleted alternatives that may be assigned $\lceil\frac{n}{k}\rceil$ voters is reduced by $r$ in order to keep proportionality. Furthermore the deleted alternatives will not be counted in $\kUpper$ and $\kLower$ so the value of $\kUpper+\kLower$ has to be adjusted as well.

Formally, each table entry stores the maximum number of voters from the range~$[i,j]$ under the same conditions as the original DP with the addition of voters possibly being assigned to the previously designated deleted alternatives. Note that the \NT\ respectively \MNT\ property only makes sense with regards to alternatives which adhere to the SC property and therefore an \NT\ respectively \MNT\ solution will be a solution that is \NT\ or \MNT\ when restricted to only non-deleted alternatives.

 In the following, we describe how to compute~$T(\firsttabletuple,\mathcal{A})$. We assume that the arguments of the configuration fulfill the same conditions in the configuration, as for the original table and 
 \begin{compactitem}
 \item[(CT8)] Either $a=b=0$ or $i,j\neq0$ and
\item[(CT9)] If $a=b=0$, then $\na=\nb=0$ and
\item[(CT10)] If $j-i+1\geq\na+\nb+\size(\mathcal{A})$.
 \end{compactitem}
Condition (CT8) is meant to ensure that $0$ is only used when voters are only assigned to deleted alternatives. Condition (CT9) clarifies that alternative $0$ is only used as a placeholder and therefore cannot be assigned any voters. Condition (CT10) is used to make sure that only the available number of voters are assigned.
  
 For the case where at least one of the conditions is violated, the corresponding entry is set to~$-\infty$.
 We start by initializing the table as:
 \begin{align}
   &T(a,a,i,j,0,0,\na,\na,\nd,0,c',i',j',0,\ldots,0)\coloneqq\na\label{eq:SPMinit1XP}\\
   &T(0,0,i,i,0,0,0,0,0,c',i',j',\mathcal{A})\coloneqq\begin{cases}
   1\text{ if }\size(\mathcal{A})=1\text{ and }i\\\text{ approves of the alternative}\\\text{that is assigned one voter.}\\
   -\infty\text{, else.}\nonumber
   \end{cases}
 \end{align}
We update the table, distinguishing between three cases, either ``$a = b$ and $B=0$'' (presented in Equation \eqref{eq:SPMlower1XP}), ``$a= b$ and $B=1$'' (presented in Equation \eqref{eq:SPMlower2XP}), or ``$a\neq b$'' (presented in Equation \eqref{eq:SPMsumXP}). A small change from the original DP is necessary, as we allow $i,j$ to take the value of $0$ in such cases we will handle the alternative as though it covered an interval of length zero.
 We again use the two auxiliary functions
 \begin{align*}
   f_1(c_1,c_2,n_1,n_2)=
   \begin{cases}
     2 & \, \text{if }c_1\neq c_2\wedge n_1=n_2=\upperB, \\
     0 & \, \text{if }n_1\leq\lowerB\wedge n_2\leq\lowerB,\\
     1 & \, \text{else.}\\
   \end{cases}\\
     f_2(c_1,c_2,n_1,n_2)=|\{c_1, c_2\} | - f_1(c_1,c_2,n_1,n_2)
   \end{align*}
\begin{figure*}[!t]
\begin{multline}
 \textbf{Case 1:} \quad \label{eq:SPMlower1XP}
	  T(a,a,i,j,\kUpper,\kLower,\na,\na,\nd,0,c',i',j',\mathcal{A})\coloneqq\\
	  \max\left\{
	\begin{array}{l}
		\underset{\substack{b,c\in\usablesetS(a,i,j,c',i',j')\\
				\kUpper' = \kUpper - f_1(b,c,\nb,n_c), 
				\kLower' = \kLower-f_2(b,c,\nb,n_c)\\ \leftmost{b}\in[i,j], 0< \nb,n_c\leq\upperB, B\in\{0,1\}}}{\max}
		\left\{ \begin{array}{l}
			T(b,c,i,j,\kUpper',\kLower',\nb,n_c,\na+\nd,B,c',i',j',\mathcal{A})+\na,\\
			T(b,c,i,j,\kUpper',\kLower',\nb,n_c,\na+\nd,B,a,i,j,\mathcal{A})+\na \\
		\end{array}\right\},\\
		\underset{\substack{\nd_1+\nd_2=\nd, b_1+b_2=\na\\ i\leq i^*<j^*\leq j, \kUpper_1+\kUpper_2=\kUpper, \kLower_1+\kLower_2=\kLower\\\mathcal{A}_1+\mathcal{A}_2=\mathcal{A}}}{\max}
		\left\{ \begin{array}{l}
			~~ T(a,a,i,i^*,\kUpper_1,\kLower_1,b_1,b_1,\nd_1,0,c',i',j',\mathcal{A}_1),\\ 
			+  T(a,a,j^*,j,\kUpper_2,\kLower_2,b_2,b_2,\nd_2,0,c',i',j',\mathcal{A}_2)
		\end{array}\right\},\\
		\underset{\substack{\nd_1+\nd_2=\nd, i\leq i^*<j^*\leq j\\\mathcal{A}_1+\mathcal{A}_2=\mathcal{A}}}{\max}
				\left\{ \begin{array}{l}
					~~ T(a,a,i,i^*,\kUpper,\kLower,\na,\na,\nd_1,0,c',i',j',\mathcal{A}_1),\\ 
					+  T(0,0,j^*,j,0,0,0,0,\nd_2,0,c',i',j',\mathcal{A}_2)
				\end{array}\right\}
	\end{array}\right\}.
\end{multline}
\begin{multline}
\textbf{Case 2:} \quad \label{eq:SPMlower2XP}
	 T(a,a,i,j,\kUpper,\kLower,\na,\na,\nd,1,c',i',j')\coloneqq   \\
	\max\left\{
	\begin{array}{l}
		\underset{\substack{b,c\in\usablesetS(a,i,j,c',i',j') \\
				\kUpper=\kUpper'+f_1(b,c,\nb,n_c), 
				\kLower=\kLower'+f_2(b,c,\nb,n_c)\\
				0<\nb,n_c\leq\upperB, B\in\{0,1\}}}{\max}
		\left\{\begin{array}{l}
			T(b,c,i,j,\kUpper',\kLower',\nb,n_c,\na+\nd,B,c',i',j',\mathcal{A})+\na,\\
			T(b,c,i,j,\kUpper',\kLower',\nb,n_c,\na+\nd,B,a,i,j,\mathcal{A})+\na                 \\
		\end{array}\right\},\\
		\underset{\substack{\nd_1+\nd_2=\nd,  b_1+b_2=\na\\ i\leq i^*<j^*\leq j, \kUpper_1+\kUpper_2=\kUpper, \kLower_1+\kLower_2=\kLower\\\mathcal{V}_1+\mathcal{V}_2=\mathcal{V}}}{\max}
		\left\{\begin{array}{l}
			~~T(a,a,i,i^*,\kUpper_1,\kLower_1,b_1,b_1,\nd_1,1,c',i',j',\mathcal{A}_1)\\
			+T(a,a,j^*,j,\kUpper_2,\kLower_2,b_2,b_2,\nd_2,0,c',i',j',\mathcal{A}_2)
		\end{array}\right\}\\
		\underset{\substack{\nd_1+\nd_2=\nd, i\leq i^*<j^*\leq j\\\mathcal{A}_1+\mathcal{A}_2=\mathcal{A}}}{\max}
				\left\{ \begin{array}{l}
					~~ T(a,a,i,i^*,\kUpper,\kLower,\na,\na,\nd_1,1,c',i',j',\mathcal{A}_1),\\ 
					+  T(0,0,j^*,j,0,0,0,0,\nd_2,0,c',i',j',\mathcal{A}_2)
				\end{array}\right\}
	\end{array}
	\right\}.
\end{multline}
\begin{multline} 
     \textbf{Case 3:} \quad \label{eq:SPMsumXP}
	T(a,b,i,j,\kUpper,\kLower,\na,\nb,\nd,B,c',i',j')\coloneqq  \\
	  \max\left\{
	\begin{array}{l}
		\underset{\substack{ c\in\inbetwS(a,b)\\ i\leq i^*<j^*\leq j,  0<n_c\leq\lowerB\\ \kUpper_1+\kUpper_2=\kUpper, \kLower_1+\kLower_2+1=\kLower\\ \nd_1+\nd_2=\nd, B'\in\{0,1\}\\\mathcal{A}_1+\mathcal{A}_2=\mathcal{A}}}{\max}
		\left\{
		\begin{array}{l}
			~~  T(a,c,i,i^*,\kUpper_1,\kLower_1,\na,n_c,\nd_1,B,c',i',j',\mathcal{A}_1)\\
			+ T(b,b,j^*,j,\kUpper_2,\kLower_2,\nb,\nb,\nd_2,B',c',i',j',\mathcal{A}_2)
		\end{array}\right\},\\
		\underset{\substack{c\in\inbetwS(a,b)\\ i\leq i^*<j^*\leq j,  n_c=\upperB \\ \kUpper_1+\kUpper_2+1=\kUpper, \kLower_1+\kLower_2=\kLower\\\nd_1+\nd_2=\nd, B'\in\{0,1\}\\\mathcal{A}_1+\mathcal{A}_2=\mathcal{A}}}{\max}
		\left\{
		\begin{array}{l}
			~~T(a,c,i,i^*,\kUpper_1,\kLower_1,\na,n_c,\nd_1,B,c',i',j',\mathcal{A}_1)\\
			+T(b,b,j^*,j,\kUpper_2,\kLower_2,\nb,\nb,\nd_2,B',c',i',j',\mathcal{A}_2)
		\end{array}\right\},\\
		\underset{\substack{i\leq i^*<j^*\leq j\\ \kUpper_1+\kUpper_2=\kUpper, \kLower_1+\kLower_2=\kLower\\\nd_1+\nd_2=\nd, B'\in\{0,1\}\\\mathcal{A}_1+\mathcal{A}_2=\mathcal{A}}}{\max}
		\left\{
		\begin{array}{l}
			~~T(a,a,i,i^*,\kUpper_1,\kLower_1,\na,\na,\nd_1,B,c',i',j',\mathcal{A}_1)\\
			+T(b,b,j^*,j,\kUpper_2,\kLower_2,\nb,\nb,\nd_2,B',c',i',j',\mathcal{A}_2)
		\end{array}\right\}\\
		\underset{\substack{i\leq i^*<j^*\leq j\\ \kUpper_1+\kUpper_2=\kUpper, \kLower_1+\kLower_2=\kLower\\\nd_1+\nd_2=\nd, B'\in\{0,1\}\\\mathcal{A}_1+\mathcal{A}_2=\mathcal{A}}}{\max}
				\left\{
				\begin{array}{l}
					~~T(a,b,i,i^*,\kUpper,\kLower,\na,\nb,\nd_1,B,c',i',j',\mathcal{A}_1)\\
					+T(0,0,j^*,j,0,0,0,0,\nd_2,B',c',i',j',\mathcal{A}_2)
				\end{array}\right\}
	\end{array}\right\}.
    \end{multline}
    \caption{DP descriptions for the $\dela$-alternatives nearly SC case in the proof of \cref{thm:MonroenearSCalt}.}
\end{figure*}

We compute the table in the following order, where the order holds only subject to the previous steps, e.g., $2$ is only relevant subject to $1$ and so on: 
\begin{compactenum}\label{tableorderXP}
\item Order the entries by $\kUpper + \kLower$ in a non-decreasing order.
\item Order the entries with $a = b$ before the entries where $a \neq b$.
\item Order the entries by $j - i$ in a non-decreasing order.
\item Order the entries by $\na + \nb$ in a non-decreasing order.
\item Order the entries by $\nd$ in an increasing order.
\item Order the entries by $B$ in an increasing order.
\item Order the entries by $\mathcal{A}$ in increasing lexicographical order.
\end{compactenum}
For the optimal solution, we return $\max\limits_{\substack{a,b\in\level_\aaa(1)\\ i\leq j\\\kUpper\leq n\mod k -f_1(a,b,\na,\nb)\\\kUpper+\kLower+|\{a,b\}|\leq k\\\na,\nb\leq\upperB\\\nd\leq n\\ B\in\{0,1\}}}T(a,b,i,j,\kUpper,\kLower,\na,\nb,\nd,B,0,0,0,\mathcal{A})$, where $\mathcal{A}$ has to fulfill the constraints of the vector we specified earlier, i.e. only the designated alternatives being assigned $\lceil\frac{n}{k}\rceil$ alternatives.

The correctness of this algorithm follows similarly as the correctness for the original DP, as the vectors must add up to an allowed vector again, i.e. the deleted alternatives are allowed at most the number of voters we guessed beforehand. As the algorithm finds an optimal solution for each possible number of happy voters assigned to each deleted alternatives it follows that final solution is optimal for each subset of alternatives and number of voters they will be assigned that is checked.

In total our approach makes $3^t$ DP tables. The size of the DP table is increased by a factor of $(\lceil\frac{n}{k}\rceil+1)^t$, due to $\mathcal{A}\in[0,\lceil\frac{n}{k}\rceil]^t$. Furthermore, the time to compute a table entry is also increased by at most $(\lceil\frac{n}{k}\rceil+1)^t$, due to the additional need to split $\mathcal{A}$ into $\mathcal{A}_1$ and $\mathcal{A}_2$. This leaves us with a total XP running time of $O((\lceil\frac{n}{k}\rceil+1)^{2t}n^{11} m^5 \csize)$.

Since both dynamic programming approaches run in polynomial time the problem is therefore XP wrt. $\dela$.

\end{proof}
}

\section{Conclusion and open questions}
We provide several efficient algorithms for \Mmw\ and \CCmw\ under (nearly) SP and SC preferences.
Our work leads to some immediate open questions.
First, it remains open whether our XP results from \cref{thm:MonroenearSCalt} could be improved to \FPT\ algorithms.
Perhaps the flow network approach from Betzler et al.~\cite{BetzlerMW2013} could be useful here.
Second, our polynomial-time algorithm for \cref{thm:MonroeP} is rather complicated and has a high running time. It would be interesting to know whether it can be improved, and if so, how.

\ifshort\clearpage\fi

\ack The authors are supported by the Vienna Science and Technology Fund (WWTF)~[10.47379/ VRG18012]. We would like to thank the referees for their helpful comments.

\bibliography{bib}

\ifshort
\clearpage
\appendix
\appendixtext
\fi

\end{document}